\pdfoutput=1
\documentclass[cernpreprint,atlasdraft=true,texlive=2019,UKenglish,texmf]{atlasdoc}
 
\usepackage{atlaspackage}
\usepackage{atlasbiblatex}
 
\usepackage{atlasphysics}
 
\graphicspath{{logos/}{figures/}}
 
\usepackage{ANA-MDET-2019-01-PAPER-defs}
\usepackage{afterpage}
 
\AtlasTitle{Performance of the ATLAS RPC detector and Level-1 muon barrel trigger at $\sqrt{s}=13$~TeV}
 
\AtlasAbstract{
The ATLAS experiment at the Large Hadron Collider (LHC) employs a trigger system consisting of a first-level hardware trigger (L1) and a software-based high-level trigger. The L1 muon trigger system selects muon candidates, assigns them to the correct LHC bunch crossing and classifies them into one of six transverse-momentum threshold classes. The L1 muon trigger system uses resistive-plate chambers (RPCs) to generate the muon-induced trigger signals in the central (barrel) region of the ATLAS detector.  The ATLAS RPCs are arranged in six concentric layers and operate in a toroidal magnetic field with a bending power of 1.5 to 5.5~Tm. The RPC detector consists of about 3700 gas volumes with a total surface area of more than 4000~m$^2$. This paper reports on the performance of the RPC detector and L1 muon barrel trigger using 60.8~fb$^{-1}$ of proton--proton collision data recorded by the ATLAS experiment in 2018 at a centre-of-mass energy of 13~TeV. Detector and trigger performance are studied using $Z$ boson decays into a muon pair. Measurements of the RPC detector response, efficiency, and time resolution are reported. Measurements of the L1 muon barrel trigger efficiencies and rates are presented, along with measurements of the properties of the selected sample of muon candidates. Measurements of the RPC currents, counting rates and mean avalanche charge are performed using zero-bias collisions. Finally, RPC detector response and efficiency are studied at different high voltage and front-end discriminator threshold settings in order to extrapolate detector response to the higher luminosity expected for the High Luminosity LHC.
}
 
\PreprintIdNumber{CERN-EP-2020-245}
 
\author{The ATLAS Collaboration}
 
\AtlasRefCode{MDET-2019-01}
 
\AtlasDate{\today}
 
\AtlasJournal{JINST}
 
\AtlasJournalRef{JINST 16 (2021) P07029}
 
\AtlasDOI{10.1088/1748-0221/16/07/P07029}
 
\hypersetup{pdftitle={ATLAS document},pdfauthor={The ATLAS Collaboration}}

\begin{document}
 
\maketitle
 
\tableofcontents
 
\clearpage
 
\section{Introduction}
ATLAS~\cite{Aad:2008zzm,ATLAS-TDR-2010-19,PIX-2018-001} is a general-purpose detector that records high-energy collisions of protons and heavy ions delivered by the Large Hadron Collider (LHC). The detector has been taking data since its completion in 2008 and is scheduled to operate until approximately 2040, following extensive accelerator and detector upgrades. These data have been used by the ATLAS Collaboration to publish a diverse set of results that include the discovery of the Higgs boson~\cite{Aad:2012tfa} and measurements of its properties~\cite{Aaboud:2017jvq,ATLAS:2020wny,Aad:2020xfq}, searches for new phenomena~\cite{Aaboud:2018spl,Aad:2019vnb,Aad:2019fac}, and precision measurements of the Standard Model (SM) properties~\cite{Aaboud:2017svj,Aaboud:2019nmv,Aad:2019wmn,Aad:2020jvx,Aad:2020tmz}. Efficient online selection of events containing muons~\cite{Aad:2020uyd} produced in collisions is essential for many of these measurements.
 
The first-level (L1) hardware-based trigger system of ATLAS~\cite{Aad:2012xs,Aaboud:2016leb} uses resistive-plate chambers (RPCs) to identify muon candidates in the central (barrel) region of the muon spectrometer (MS)~\cite{ATLAS-TDR-10}. RPCs are gaseous detectors with nanosecond-level time resolution~\cite{Santonico:1981sc,Santonico:1988qi}. Excellent performance of the RPC detector and its associated trigger system are therefore fundamental for the ATLAS physics programme. This L1 muon barrel trigger system~\cite{Anulli:2009zz} selects muon candidates produced in LHC collisions; it assigns the selected candidates to the correct LHC bunch crossing and measures the transverse momentum ($\pT$) of the muon candidates using six predetermined programmable thresholds.
 
This paper reports measurements of the performance of the ATLAS RPC detector and L1 muon barrel trigger system using \ilumi of proton--proton collision data recorded by the ATLAS experiment in 2018 at a centre-of-mass energy of 13~\TeV. The paper is organised as follows. Section~\ref{sec:detector} briefly describes the RPC detector and L1 muon barrel trigger system. Section~\ref{sec:analysis} details the dataset and tools used for the measurements presented in the subsequent sections. Section~\ref{sec:result} presents measurements of RPC detector efficiency and time resolution. This section also describes the RPC response to the passage of a muon. Section~\ref{sec:trigger} presents measurements of the L1 muon barrel trigger response, including measurements of the efficiency, rates and composition of the selected muon candidates. Section~\ref{sec:current_rate} presents measurements of the currents in the gas volumes as a function of the operating voltage, temperature and instantaneous luminosity. Measurements of RPC counting rates are also reported in this section. The current and counting rate measurements are combined to determine an average RPC avalanche charge using zero-bias collisions. Section~\ref{sec:HL-LHC} presents studies of expected RPC detector response at the High Luminosity LHC (HL-LHC). For these studies, the measurements reported in Section~\ref{sec:current_rate} are extrapolated to the HL-LHC design luminosity in order to predict RPC detector response. In addition, studies of the RPC detector response at different voltage working points and front-end (FE) electronics discriminator thresholds, corresponding to expected RPC operational parameters for the HL-LHC, are presented.
 
\section{ATLAS detector and resistive-plate chambers}
\label{sec:detector}
ATLAS is a general-purpose detector at the LHC with a cylindrical geometry\footnote{ATLAS uses a right-handed coordinate system with its origin at the nominal interaction point (IP) in the centre of the detector and the $z$-axis along the beam direction. The $x$-axis points from the IP to the centre of the LHC ring, and the $y$-axis points upward. Cylindrical coordinates $(r,\phi)$ are used in the $(x, y)$ plane, $\phi$ being the azimuthal angle around the $z$-axis. The pseudorapidity is defined in terms of the polar angle $\theta$ as $\eta=-\ln\tan(\theta/2)$. The distance $\Delta R$ is defined as $\Delta R=\sqrt{(\Delta \eta)^2+(\Delta \phi)^2}$.} that provides nearly full solid angle coverage around the collision point located at the centre of the detector. The detector consists of an inner tracking detector (ID), electromagnetic and hadronic calorimeters, and a muon spectrometer. The detector is subdivided into a barrel and two endcap sections, and provides complete azimuthal angle coverage. The ID covers the pseudorapidity range of $|\eta| < 2.5$ and is surrounded by a thin superconducting solenoid that provides a 2~T axial magnetic field. The ID consists of silicon pixel, silicon microstrip, and transition radiation tracking detectors. The calorimeter system covers the pseudorapidity range $|\eta| < 4.9$. High-granularity lead and liquid-argon (LAr) sampling calorimeters provide electromagnetic calorimetry within the pseudorapidity range $|\eta| < 3.2$. An additional thin LAr presampler is used to correct for energy losses in the material upstream of the calorimeters in the pseudorapidity range $|\eta| < 1.8$. A steel and scintillator-tile sampling calorimeter provides hadronic calorimetry within the pseudorapidity range $|\eta| < 1.7$. Two copper/LAr endcap hadronic calorimeters cover the pseudorapidity range of 1.5 < $|\eta|$ < 3.2. The forward coverage is extended up to $|\eta| = 4.9$ with copper/LAr and tungsten/LAr calorimeters, which are optimised for electromagnetic and hadronic measurements, respectively.

\begin{figure}[h]
\begin{center}
\subfloat[]{\includegraphics[width=0.480\textwidth]{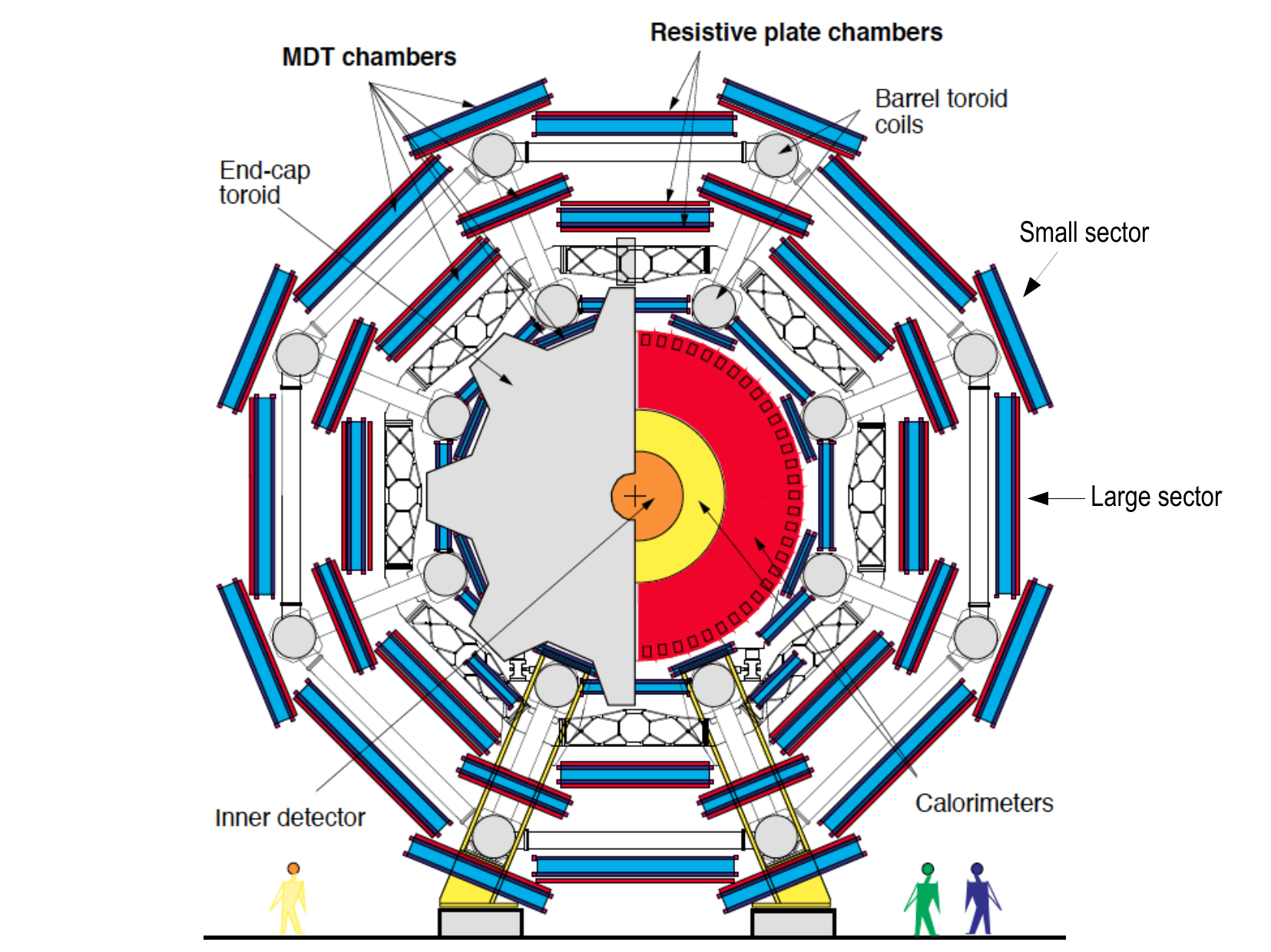}\label{subfig:msxy}}
\subfloat[]{\includegraphics[width=0.519\textwidth]{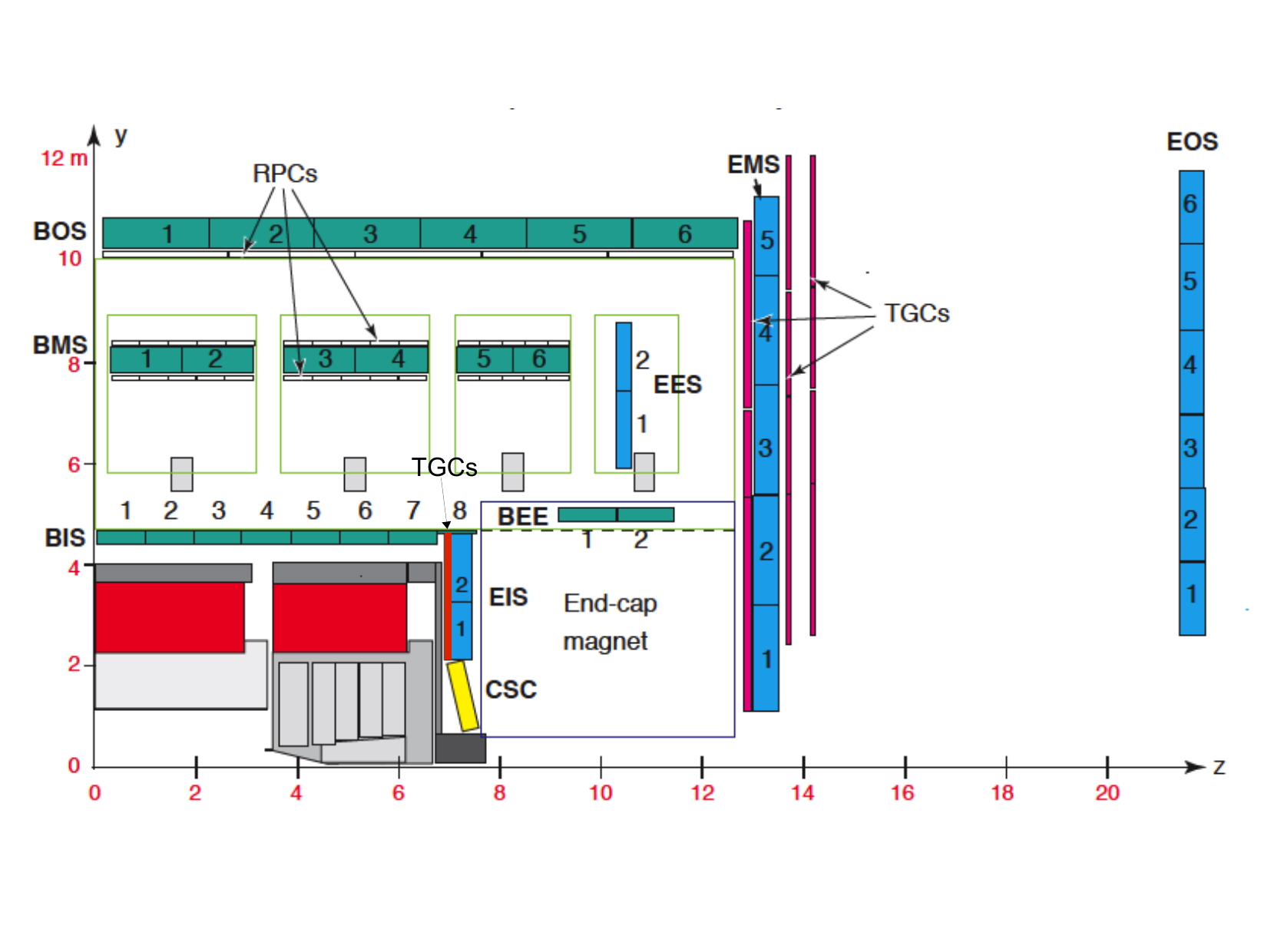}\label{subfig:msrz}}
\caption{(a) View of the ATLAS MS barrel detectors in the transverse $(x, y)$ plane. (b) View of the ATLAS MS layout in the $(z, y)$ plane for a small azimuthal sector containing the barrel toroid coils. The green (blue) chambers are MDT chambers in the barrel (endcap) regions of the spectrometer. The TGCs, RPCs, and CSCs are shown in red, white, and yellow, respectively.}
\label{fig:msdet}
\end{center}
\end{figure}
 
\subsection{ATLAS muon spectrometer}
The MS is designed to identify muon candidates and to measure the muon momentum and charge independently from the ID. The MS is the outermost system of the ATLAS detector and it consists of one barrel and two endcap sections. Each section incorporates an air-core magnet, with each magnet consisting of eight superconducting coils. The barrel toroid magnet provides 1.5 to 5.5~Tm of bending power for muon tracks in the pseudorapidity range $|\eta| < 1.4$. The toroidal field causes muon track deflections primarily in the $(r, z)$ plane, which are measured using the $\eta$ coordinate. The MS is subdivided into eight large and eight small azimuthal sectors, with the small barrel sectors containing the magnet coils. The muon barrel chambers are arranged in six concentric cylindrical layers around the beam axis at radii of approximately 5~m, 7.5~m, and 10~m. In the endcap regions, muon chambers form wheels positioned perpendicular to the beam axis at distances of approximately 7.4~m, 10.8~m, 14~m, and 21.5~m from the interaction point.  The MS barrel geometry is illustrated in Figure~\ref{subfig:msxy} in the $(x, y)$ plane and in Figure~\ref{subfig:msrz} in the $(r, z)$ plane for the small sectors.
 
The MS contains fast trigger detectors and precision tracking detectors that cover the pseudorapidity ranges $ |\eta| < 2.4$ and $|\eta| < 2.7$, respectively. The RPCs and thin-gap chambers (TGCs) are used for triggering in the barrel and endcap regions, respectively. The RPC and TGC detectors are also used to measure the muon $\phi$ direction in the non-bending $(x, y)$ plane. The precision muon tracking chambers are constructed from monitored drift tubes (MDTs). The MDT detector provides an average resolution of about $80~{\mu}$m per tube and $35~{\mu}$m per chamber, when measured in the bending $(r, z)$ plane, which is approximately perpendicular to the magnetic field lines. Cathode-strip chambers (CSCs) are used for precision tracking in the inner endcap region, $2 < |\eta| < 2.7$.
 
The following nomenclature is used to identify different MS elements. The ATLAS detector is divided into two halves along the $z$-axis, called side A ($z$ > 0) and side C ($z$ < 0). The barrel and endcap regions are labelled as `B' and `E', respectively. Letters `I' (inner), `M' (middle) and `O' (outer) are used to identify the corresponding MS layers. Letters `S' and `L' specify whether a chamber belongs to a small or large sector, respectively. For example, the ATLAS nomenclature for a chamber located in a large sector of the middle barrel layer is `BML', followed by: the chamber position index along the \textit{z}-direction, ranging from 1 to 8; detector side A or C; and the azimuthal sector number, ranging from 1 to 16. In addition, special muon chambers were added in the barrel--endcap transition region $1.05 < |\eta| < 1.3$ in order to close gaps and increase the overall detector coverage. These chambers are referred to as `barrel endcap extra' (BEE) and `extended endcap' (EE) chambers.
 
\subsection{ATLAS resistive-plate chambers}
\label{sec:detrpc}
The ATLAS RPC detector~\cite{ATLAS-TDR-10,Aad:2012xs} provides up to six position measurements along the muon trajectory in the MS, with a space--time resolution of the order of $2~\text{cm} \times 2$~ns. The RPC detector covers the pseudorapidity range $|\eta|<1.05$. It consists of approximately 3700 gas volumes, with a total surface area of more than 4000~m$^2$, and operates in a toroidal magnetic field of about 0.5~T. Each RPC consists of two independent detector layers (referred to as a doublet), separated by about 2~cm. The RPCs are arranged in three concentric cylindrical doublet layers at radii of approximately 7.8~m (6.8~m), 8.4~m (7.5~m), and 10.2~m (9.8~m) for the small (large) azimuthal sectors. These three doublet layers are referred to as the `middle confirm layer' (RPC1), `middle pivot layer' (RPC2) and `outer confirm layer' (RPC3). Following the initial detector operations from late 2009 through early 2013, additional so-called feet and elevator chambers were installed during the first long shutdown to increase the system acceptance before the resumption of operations in 2015~\cite{Aaboud:2016leb,Corradi:2016moz}.
 
\begin{figure}[!h]
\begin{center}
\includegraphics[width=0.70\textwidth]{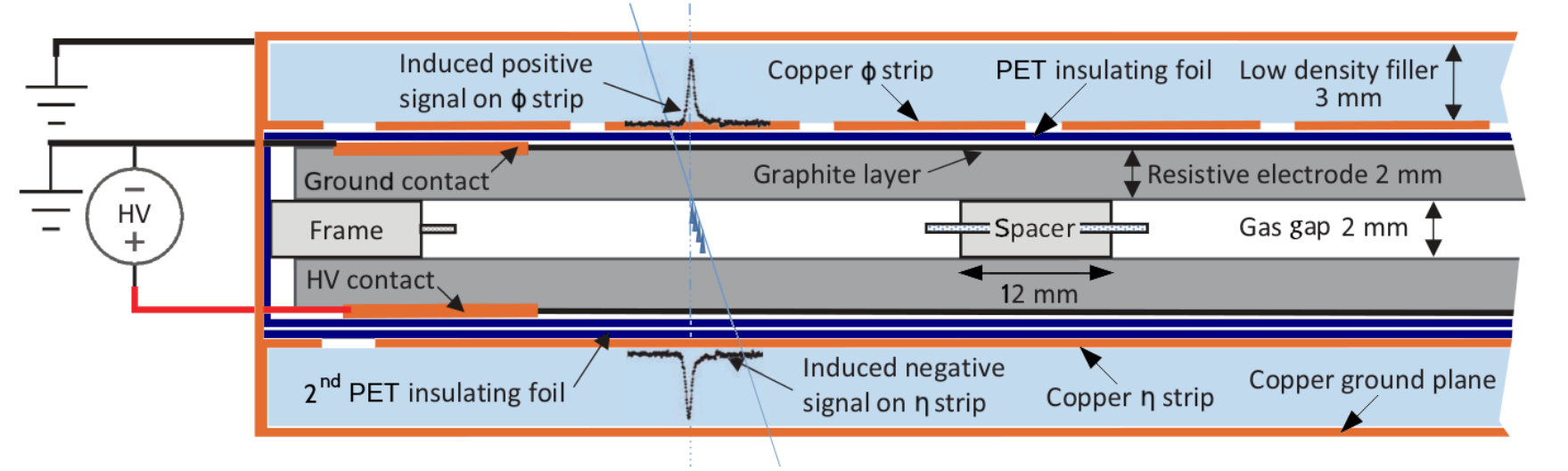}
\caption{A schematic drawing of an ATLAS RPC detector module.}
\label{fig:rpcpanel}
\end{center}
\end{figure}

Each single RPC detector layer is constructed from two parallel resistive electrodes which are made of high-pressure phenolic--melaminic laminate (bakelite) with a high resistivity of approximately $10^{10}\ \Omega~\textrm{cm}$, as illustrated in Figure~\ref{fig:rpcpanel}. The bakelite sheet prevents self-sustaining discharges and limits the amount of charge produced in an ionisation event, thus allowing the high-rate operation of RPCs. A thin coat of linseed oil is applied to the inner surfaces of the electrodes in order to ensure their smoothness. The two electrodes are separated by a distance of 2~mm using insulating polycarbonate spacers, with a diameter of approximately 12~mm, which are placed every 10~cm. The spacers cover approximately 1\% of the detector surface and therefore reduce the RPC efficiency by the equivalent amount.
 
The external sides of the resistive electrodes are coated with a graphite paint. A reference voltage of 9.6~kV is typically applied across the two electrodes. The actual applied voltage is automatically adjusted according to continuous readings of environmental temperature and pressure parameters in order to provide a muon detection efficiency equivalent to that obtained at a temperature of $24^\circ$C and a pressure of 970~mbar~\cite{Aielli:2013nea}. The average applied voltage across all detector chambers was 9.6~kV, with the standard deviation of about 26~V. These variations were due to changes in environmental temperature and pressure with respect to the standard conditions.
 
The RPCs are continuously flushed with a gas mixture of $\mathrm{C_2H_2F_4}(94.7\%)\text{--}\mathrm{C_4H_{10}}(5\%)\text{--}\mathrm{SF_6}(0.3\%)$. This mixture includes a quencher component ($\mathrm{C_4H_{10}}$) that helps to avoid propagation of the discharge and an electronegative component $(\mathrm{SF_6})$ that helps to limit the growth of avalanches. This gas mixture has a strong greenhouse effect and it is currently being phased down in the European Union, thereby also leading to rising cost. For these reasons, new gas mixtures are under investigation for future RPC operation~\cite{Rigoletti:2020exj,Guida:2020yqs}.
 
The ATLAS RPCs are operated in avalanche mode, with the probability to produce streamers during the gas multiplication of the primary ionisation kept below 2\%~\cite{Guida:2016qjq}. The reduced pulse charge in avalanche mode provides a high rate capability and enables stable operation over the required detector lifetime. The lower gas amplification of the avalanche mode is compensated for by a high signal amplification gain of the FE electronics~\cite{abbrescia2018resistive}.

Each single RPC layer measures $\eta$ and $\phi$ coordinates using orthogonal copper strips placed on opposite sides of the electrodes, with the strip widths varying in a range between 24.5 and 33.3~mm. Muon sagittae due to the magnetic field are measured by $\eta$ strips, aligned perpendicularly to the bending $(r, z)$ plane. Strip foils are glued on 3~mm-thick low-density polyester foam plates. The strips are isolated from the electrode with insulating polyethylene terephthalate (PET) foil. The signal is recorded via capacitive coupling to the copper strips, which are connected to FE electronics.
 
The RPC layout and readout functionality are not fully symmetric between $\eta$ and $\phi$ views. A second PET foil is placed between the $\eta$ strips and the electrode where the positive voltage is applied, resulting in the charge collected by the $\eta$ strips being slightly smaller than that collected by the $\phi$ strips for the same avalanche. Moreover, FE boards for $\phi$ strips include an additional circuit which inverts the polarity of the positively charged signal induced in $\phi$ strips. After that, the same circuit is used for $\eta$ and $\phi$ strip readout.
 
The following nomenclature is used to refer to the RPC detector elements. One RPC gas volume (gas gap) together with the $\eta$ and $\phi$ readout strips is called `module' and is shown in Figure~\ref{fig:rpcpanel}. A group of strips in one view ($\eta$ or $\phi$) belonging to one module is thereafter referred to as a strip readout panel or simply a panel. Two layers, consisting of two or four modules each, assembled into a common mechanical structure are referred to as a unit (or chamber). Each RPC unit therefore forms one doublet layer. One or two RPC units, which are integrated with the corresponding MDT chambers, make a (muon) station. One station in the MS middle layer integrates the RPC1 and RPC2 doublet layers, while one station in the MS outer layer includes only the RPC3 doublet layer.

There are 384 muon stations that contain RPCs, separated into 2 MS layers (middle and outer), 2 detector sides, 16 sectors and 6 locations along the $z$-axis for each side. In addition, there are a small number of special RPCs that are described later. The position of each station is determined by a coordinate along the $z$-axis (\ieta ranging from 1 to 6, for $|\eta|$ values from 0.05 to 1.05) and a coordinate along the $\phi$ direction (\iphi ranging from 1 to 16). Modules within one station are represented in figures using half-integer values of the \ieta and \iphi.
 
A signal recorded by the FE electronics in one strip is referred to as a hit. All RPC hits within a $200$~ns window, centred on the bunch crossing selected by the trigger system, are recorded for offline analysis. After registering a hit in a channel, the FE electronics imposes the dead-time window of 100~ns for that channel. LHC proton--proton collisions at an instantaneous luminosity of $10^{34}~\text{cm}^{-2}\text{s}^{-1}$ result in RPC counting rates in the range 10 to 30~$\text{Hz}/\text{cm}^2$, depending on the chamber location, as discussed in Section~\ref{sec:current_rate}.
 
The ATLAS detector control system (DCS)~\cite{BarriusoPoy:2008zz,Polini:2012zz} is used to operate and monitor the RPC detector. The DCS is used to ramp up/down the applied voltage and to set the thresholds of the FE discriminators. The DCS also provides continuous monitoring of configuration parameters (such as threshold and voltage settings) and of operational detector parameters (such as temperature, pressure and chamber currents). This monitoring information is available in real time in the ATLAS control room during data-taking and is also recorded in a central database for subsequent offline analysis.
 
The parameter {\vset} is used by the DCS to control the threshold of the FE discriminators of readout channels in a panel. The absolute threshold applied at the discriminator level is determined by the FE electronics using the following expression: $\vfe = \vref - \vset$, where {\vref} is a reference voltage which is set to 2~V. The {\vfe} corresponds to the physical threshold applied to the RPC signal after the amplification stage~\cite{Aielli:2003bx}. Nearly all {\vfe} parameters are typically in the range 0.8~V to 1.5~V, in steps of 0.1~V, and approximately 90\% of panels use the nominal value of 1~V. This physical {\vfe} threshold parameter is used for studies presented later.
 
\subsection{L1 muon barrel trigger}
\label{subsec:L1MuonBarrelTrigger}
Interesting collision events are selected using a two-level trigger system~\cite{Aad:2012xs,Aaboud:2016leb,Aad:2020wji}. The L1 trigger processes events at a rate of 40~MHz, set by the LHC beam structure consisting of bunches separated by 25~ns. The L1 trigger selects events at a rate of 100~kHz using data from the calorimeters and muon trigger detectors. The high-level trigger (HLT) employs software algorithms with access to the full detector information to analyse the accepted L1 events. The HLT selects events for offline analysis at a rate of approximately 1~kHz.
 
The L1 muon barrel trigger system~\cite{Anulli:2009zz} uses the RPCs to identify a region of interest (RoI) containing a muon candidate in the pseudorapidity range $|\eta| < 1.05$. A typical RoI has $\Delta \eta\times \Delta \phi$ dimensions of approximately $0.1\times 0.1$. The L1 trigger assigns muon candidates to the correct LHC bunch crossing and determines the muon transverse momentum ($\pT$) using six programmable thresholds. To compensate for different signal propagation times due to the different lengths of readout cables, timing response of RPC electronics channels are calibrated using programmable delays in steps of 3.125~ns, corresponding to an eighth of the LHC bunch spacing~\cite{Anulli:2009zz}. One calibration constant is used for each group of eight channels.
 
\begin{figure}[h]
\begin{center}
\includegraphics[width=0.55\textwidth]{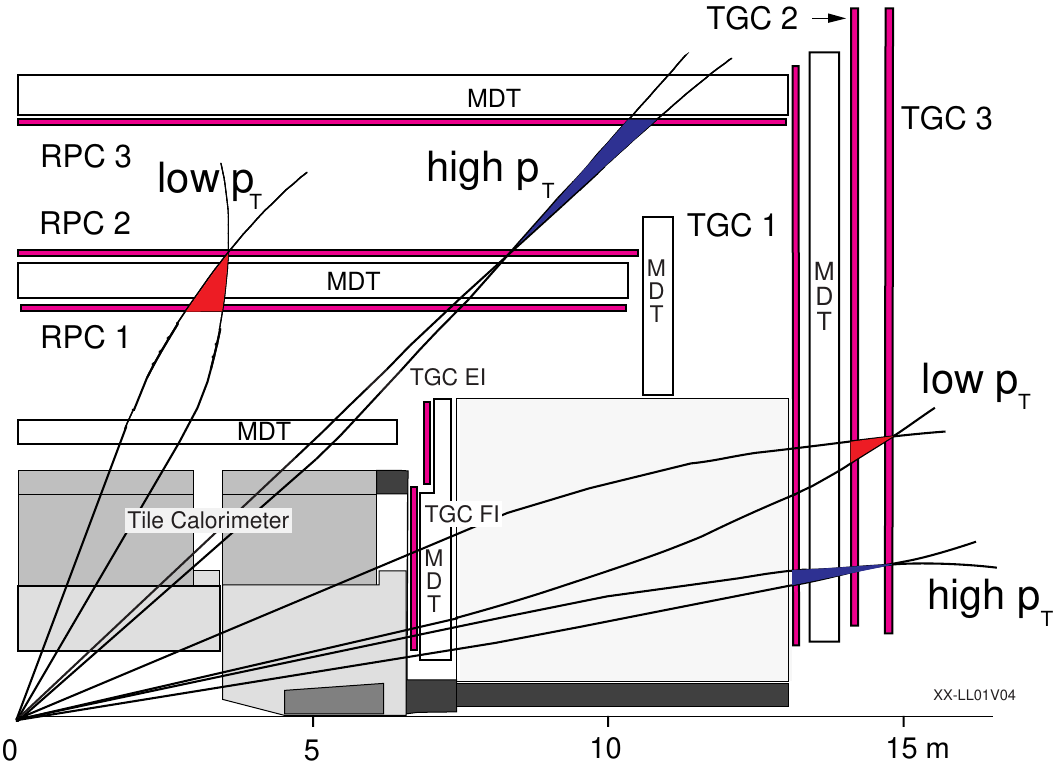}
\caption{Illustration of the low-$\pT$ and high-$\pT$ L1 muon trigger algorithms in the barrel and endcap regions.}
\label{fig:trigroads}
\end{center}
\end{figure}

The transverse momentum of muon candidates is measured by the L1 muon barrel trigger using different algorithms for low-$\pT$ and high-$\pT$ triggers~\cite{Anulli:2009zz}, as illustrated in Figure~\ref{fig:trigroads}. The low-$\pT$ algorithm starts with a signal in an RPC2 (pivot) strip and then checks for matching signals in RPC1 (confirm) strips within a narrow cone pointing back to the collision point, shown in red in Figure~\ref{fig:trigroads}. The low-$\pT$ algorithm requires signals to be present in three out of four detector layers, which results in a significant suppression of random coincidences due to background events. The high-$\pT$ algorithm starts with a muon candidate identified by the low-$\pT$ algorithm and then checks for the presence of matching signals in one of the two RPC3 (confirm) layers within a narrower cone pointing back to the collision point, shown in blue in Figure~\ref{fig:trigroads}. Each trigger $\pT$ threshold always satisfies the conditions of the lower ones. Therefore, only the highest $\pT$ threshold passed by a muon candidate is reported by the L1 muon barrel trigger system.
 
Three low-$\pT$ thresholds and three high-$\pT$ thresholds were defined for the L1 muon barrel trigger system~\cite{Aad:2020uyd}. In 2015--2018, the low-$\pT$ trigger thresholds were $\pT=4$, 6 and 10~\GeV, which are referred to as MU4, MU6 and MU10 triggers, respectively. In 2015--2016, the high-$\pT$ trigger thresholds were $\pT=10$, 15 and 20~\GeV, which are referred to as MU11, MU15 and MU20 triggers, respectively. The MU11 nomenclature was used to distinguish this trigger from the low-$\pT$ MU10 trigger. In 2017--2018, the MU15 trigger was removed, and the MU21 trigger was introduced. The MU21 trigger was identical to the MU20 trigger except that the so-called new feet RPCs were not included in its trigger logic. These new feet chambers were installed as a fourth RPC doublet layer (RPC4) in $\phi$ sectors 12 and 14, which contain the ATLAS detector support structures (feet).
 
These new feet chambers increase the geometrical acceptance of the L1 muon barrel trigger system for detecting muons with $\pT>20$~\GeV\ from approximately 67\% to around 70\%, as described in Section~\ref{subsec:trigEff}. In these two sectors, some RPCs are absent in the middle MS layer (RPC1 and RPC2) in order to make room for the detector feet. Therefore, muons travelling from the collision point encounter only the two outermost RPC doublet layers in these sectors (RPC3 and RPC4). For this reason, the trigger logic of the new feet trigger requires a geometrical matching only in the two outermost doublet layers, for both the low- and high-$\pT$ trigger thresholds. This leads to a larger expected rate of background events for the high-$\pT$ triggers, as discussed in Section~\ref{subsec:trigComp}. The MU21 trigger was introduced as a backup trigger in case the acceptance rate from the new feet trigger exceeded the allowed limit.
 
\section{Dataset and event selection}
\label{sec:analysis}
The measurements presented here were performed using proton--proton collision data recorded by the ATLAS experiment in 2018 at a centre-of-mass energy of 13~TeV, with 25~ns spacing between LHC bunches. Only LHC fills with an integrated luminosity greater than 50~pb$^{-1}$ were used for these measurements. The total integrated luminosity of the analysed dataset amounts to \ilumi.
 
Muon candidates are reconstructed by combining ID and MS information~\cite{Aad:2016jkr,Aad:2020gmm}. The precision MDT chambers are used to reconstruct muon trajectories in the $(r, z)$ plane. This plane is approximately orthogonal to the magnetic field lines, thus allowing precise measurements of the sagitta of the muon tracks. In the barrel region, the RPC detector is used to reconstruct the muon trajectory in the non-bending azimuthal $(x, y)$ plane which is transverse to the beam direction.
 
RPC $\phi$ hits are used to build track patterns in the $\phi$ view. For the pattern finding procedure in the $\eta$ view, both RPC and MDT hits are used. Three-dimensional track segments are reconstructed in each station, then combined among different stations and fitted to form a final MS track candidate. The fitting procedure takes into account effects of multiple scattering, magnetic field inhomogeneities and inter-chamber misalignments. At least three MDT $\eta$ hits are required to make a track segment along the precision $\eta$ view. In the $\phi$ view, at least two RPC hits are required to form a track candidate.
 
Events were selected using several different trigger criteria which were based on the presence of a muon candidate, a high-$\pT$ hadronic jet, or significant missing transverse momentum~\cite{Aaboud:2018tkc}. The majority of the selected events come from the Drell--Yan production of $W$ and $Z$ bosons decaying into a muon and neutrino or into a muon pair, respectively. A smaller fraction of events are due to production of top quark pairs, electroweak vector-boson pairs and decays of hadrons containing bottom or charm quarks.
 
Selected events were recorded in a dedicated data stream during prompt data reconstruction at CERN. This data includes the reconstructed muon candidates, the L1 muon trigger information and the full information related to muon detector system, including all hits recorded by the RPC detector in the $200$~ns window centred on the selected bunch crossing. The offline reconstruction selects only the first hit in time in each RPC strip, resulting in at most one hit associated with a single strip. Selected events were required to contain at least one reconstructed muon candidate with $\pt>10$~GeV that satisfied the \textit{loose} identification criteria~\cite{Aad:2016jkr,Aad:2020gmm}. Muon candidates were also required to originate from the reconstructed vertex with the highest sum of $\pT^2$ of the associated reconstructed ID tracks. Requirements on the significance of the transverse impact parameter $d_{0}$ (|$d_{0}|/\sigma(d_{0}) < 3.0$) and on the longitudinal impact parameter $z_0$ ($|z_0 \sin\theta|<0.5$~mm) of the muon are also imposed.
 
Detector and trigger performance were evaluated using events containing a $Z$ boson decay into two muons. Events containing a $Z$ boson candidate were first selected using the primary single-muon trigger~\cite{Aad:2020uyd} with $\pT>26$~GeV. At least two reconstructed muon candidates were then required to be present in each selected event. Among them, one muon (referred to as a tag) was required to be matched with the muon candidate selected by the trigger system. The other muon (referred to as a probe) was required to form an opposite-charge pair with the tag muon. Finally, the invariant mass of each dimuon pair was required to be in the range 50~GeV $< m_{\mu \mu} <$ 150~GeV, consistent with the $Z$ boson mass. Probe muons selected in this way are unbiased by the trigger system and were used for the measurements presented in Sections~\ref{sec:result} and~\ref{sec:trigger}.
 
\section{RPC detector performance measurements}
\label{sec:result}
This section presents studies of the RPC detector response to the passage of probe muons produced in Z boson decays.  A dedicated analysis algorithm was developed to study the response of RPCs to the passage of muons. This algorithm extrapolates the trajectory of the muon track through the MS and computes the expected position of the muon impact point on the surface of the RPC detector modules near the trajectory. The algorithm takes into account the magnetic field configuration, detector geometry and material density distribution. RPC modules with expected muon-induced signals are selected for analysis following a two-step procedure. First, modules positioned within a distance of $\Delta R < 0.5$ are retained, where $\Delta R$ is computed between the muon track and the vector pointing from the detector centre to the geometric centre of the module. Second, the extrapolated muon track position on the module surface is required to be within $20$~mm of the centre of at least one strip belonging to that module. Each selected RPC module is therefore expected to contain muon-induced hits because a muon trajectory is predicted to pass through the active surface of the module.
 
This section is organised as follows. Section~\ref{subsec:RPC_response} presents the response of one representative RPC module in order to introduce the analysis technique. Section~\ref{subsec:RPC_clusters} reports the response of the entire RPC detector, which is evaluated using probe muons. Section~\ref{subsec:det_eff} shows the measurement of the efficiency of all RPC detector modules, as well as the stability of the RPC detector response as a function of time. Finally, Section~\ref{subsec:time_res} presents measurements of the time resolution of the RPC detector and the readout system.
 
\subsection{RPC single-module response}
\label{subsec:RPC_response}
 
\begin{figure}[!h]
\centering
\subfloat[]{\includegraphics[scale=0.4]{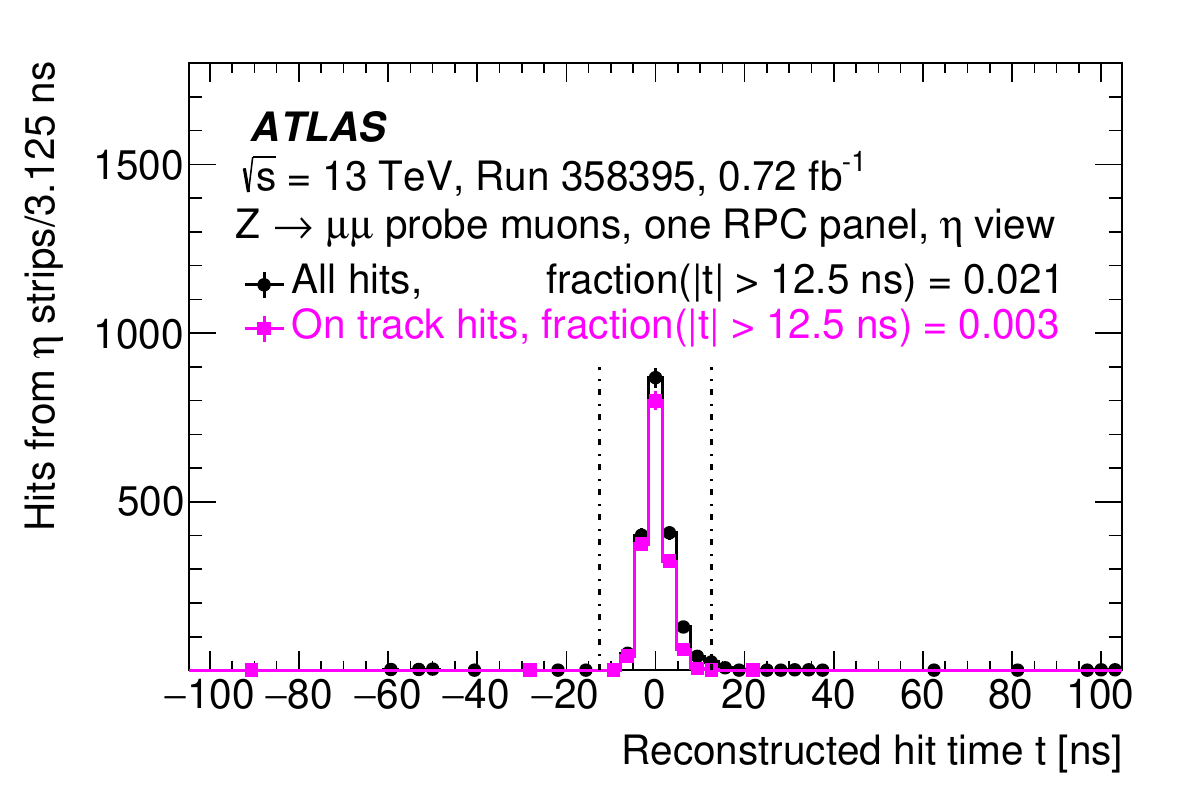}}
\subfloat[]{\includegraphics[scale=0.4]{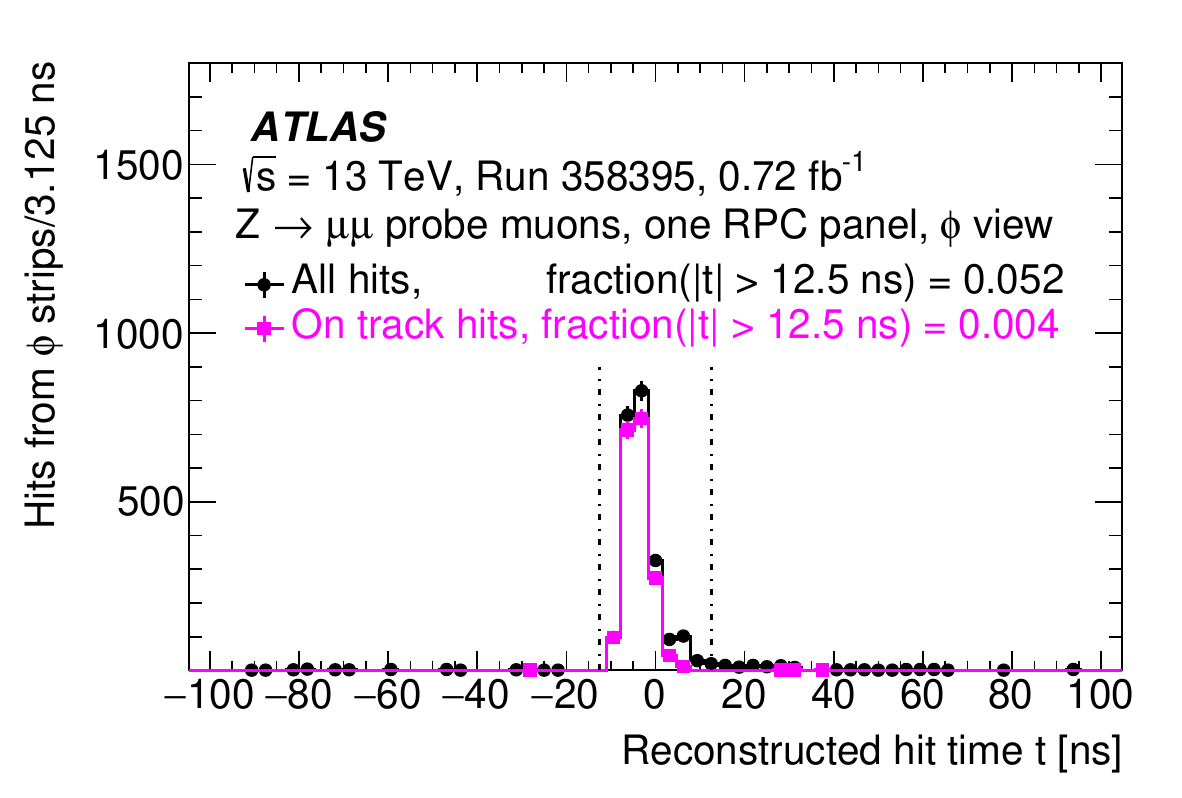}}
\caption{Time distributions of the calibrated hits recorded by the (a) $\eta$ panels and (b) $\phi$ panels belonging to a representative RPC module. Only events that contain a muon that is expected to pass through this module are used. Dashed vertical lines correspond to the 25~ns time window centred at zero. The fractions of hits with $|t| > 12.5$~ns are also reported.}
\label{pic:time_distribution}
\end{figure}

The response of one representative RPC module is evaluated using data recorded in one ATLAS run. This module is located on side C in the RPC1 doublet layer in large sector 11, in the station with $\ieta$ 1. In this module, $\eta$ and $\phi$ strip widths are 27~mm and 25~mm, respectively. Time distributions of $\eta$ and $\phi$ hits in this module are shown in Figure~\ref{pic:time_distribution} for events containing muon candidates that are predicted to pass through this module. Hit times are calibrated according to the procedure described in Section~\ref{subsec:L1MuonBarrelTrigger}. The zero time of the $x$-axis corresponds to the arrival time of an ultra-relativistic particle produced at the interaction point in the bunch crossing selected by the trigger system. The small timing offsets observed in Figure~\ref{pic:time_distribution} are partly due to imperfect timing calibrations which are nevertheless still within required levels of precision. When considering all hits recorded by this module, about 2\% (5\%) of $\eta$ ($\phi$) hits lie outside the 25~ns window centred at zero, corresponding to the time interval that separates the selected bunch crossing from the previous and subsequent bunch crossings. The fraction of hits outside this time window is reduced to less than 0.5\% when only the hits belonging to the strip containing the expected muon impact point are considered (therefore selecting only those hits that lie on the muon trajectory). This small fraction illustrates the typically good performance of the RPC time calibration for assigning muon-induced hits to the correct bunch crossing. The timing response of the full RPC detector is presented in Section~\ref{subsec:RPC_clusters}.
 
\begin{figure}[!h]
\centering
\subfloat[]{\includegraphics[scale=0.4]{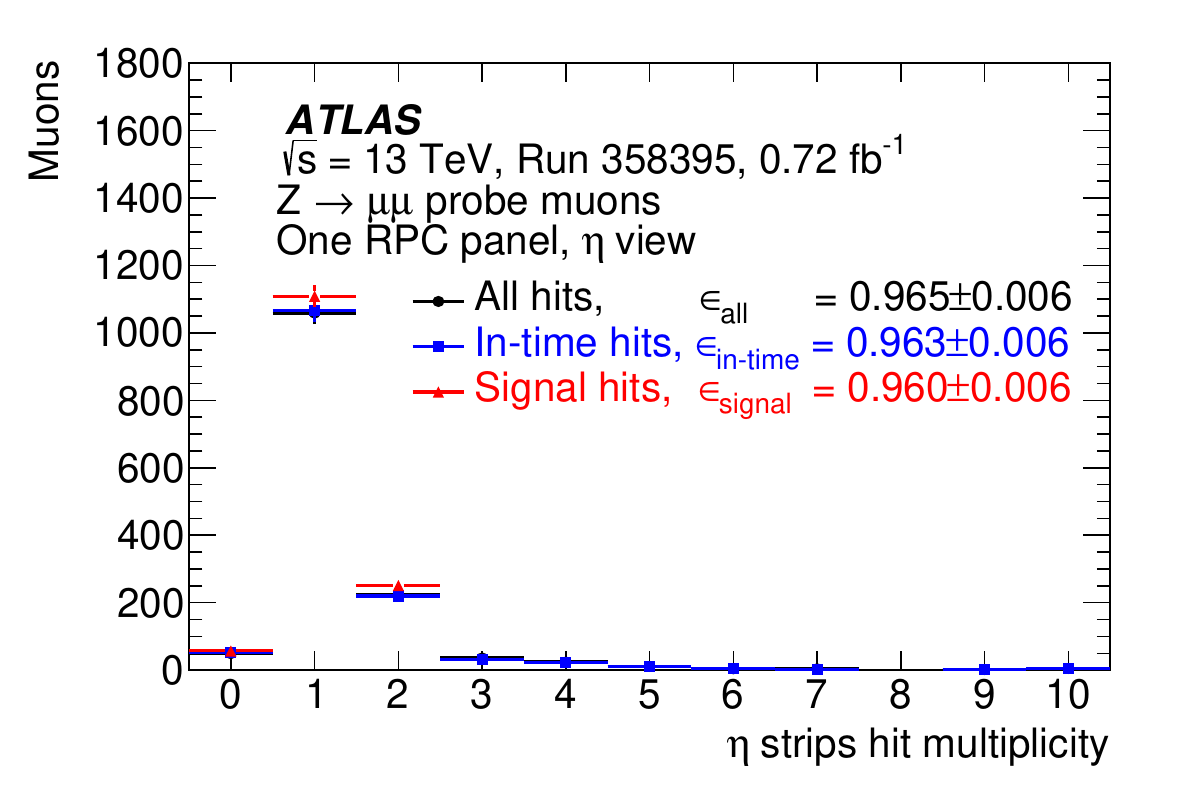}}
\subfloat[]{\includegraphics[scale=0.4]{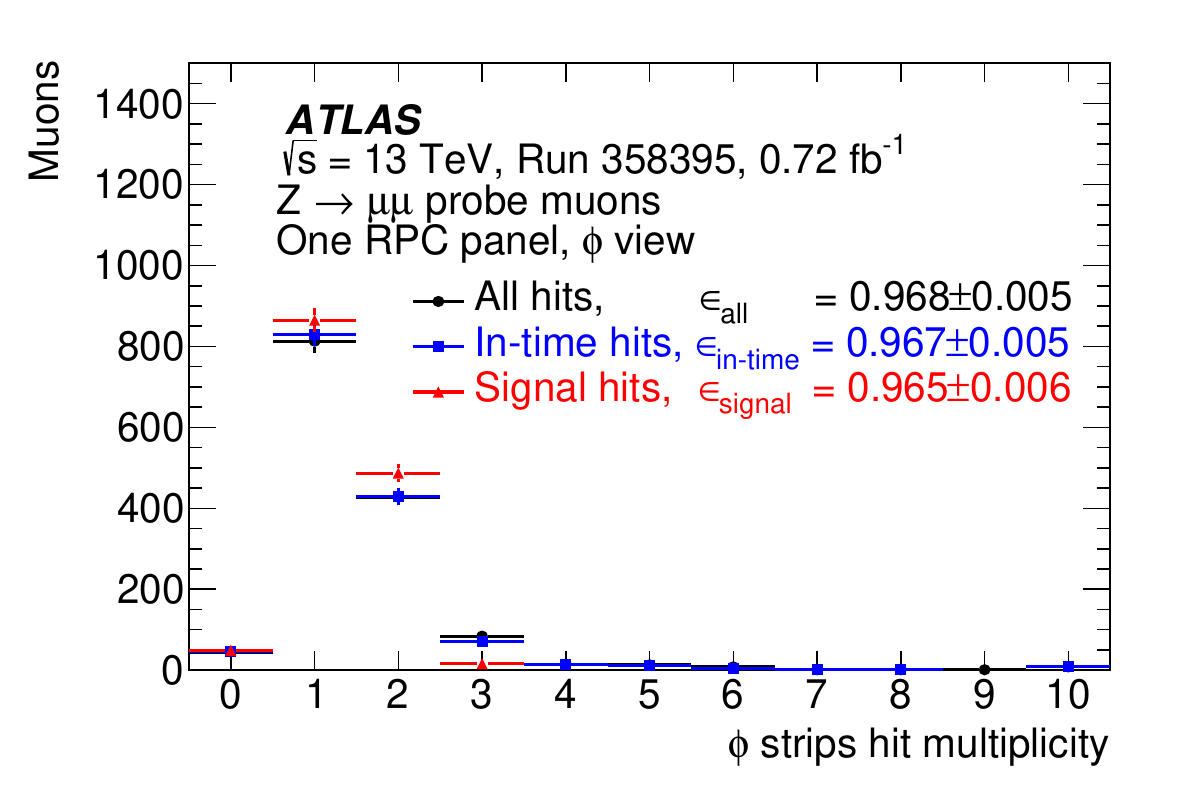}}
\caption{Hit multiplicity and detector efficiency for the (a) $\eta$ panels and (b) $\phi$ panels belonging to the same representative RPC module. Only events where a muon is expected to pass through this module are used.}
\label{pic:hit_multiplicity}
\end{figure}

The efficiency of detecting a muon-induced ionisation signal is evaluated by counting the hit multiplicity in a module using events that contain muons that pass through that module. Three sets of selection criteria are studied for counting RPC hits. Considering all hits recorded in a given module results in the maximum possible efficiency but also includes noise contributions. Hits within the 25~ns window centred at zero are referred to as in-time hits. In-time hits belonging to strips with their centre within $30$~mm of the extrapolated muon track position are referred to as signal hits. The hit multiplicity distributions obtained using these three selection criteria are shown in Figure~\ref{pic:hit_multiplicity} for the $\eta$ and $\phi$ panels belonging to the representative module. The RPC module efficiency, $\epsilon$, is computed as the number of muon candidates that generate at least one selected hit divided by the total number of muon candidates that are predicted to pass through that module. This representative module detects muon-induced ionisation signals with an efficiency of approximately 96\%. The measured efficiencies calculated using the three hit-selection criteria differ by less than $1\%$.
 
As shown in Figure~\ref{pic:hit_multiplicity}, a muon passing through an RPC module typically generates one or two hits. A cluster is defined as a set of hits in contiguous strips, with hit times that are separated by less than 12.5~ns. The size of the muon-induced clusters is an important parameter of RPC performance and is studied in Section~\ref{subsec:RPC_clusters}. Muon clusters are typically located close to the impact point of a muon track.
 
\begin{figure}[!h]
\centering
\subfloat[]{\includegraphics[width=0.49\textwidth]{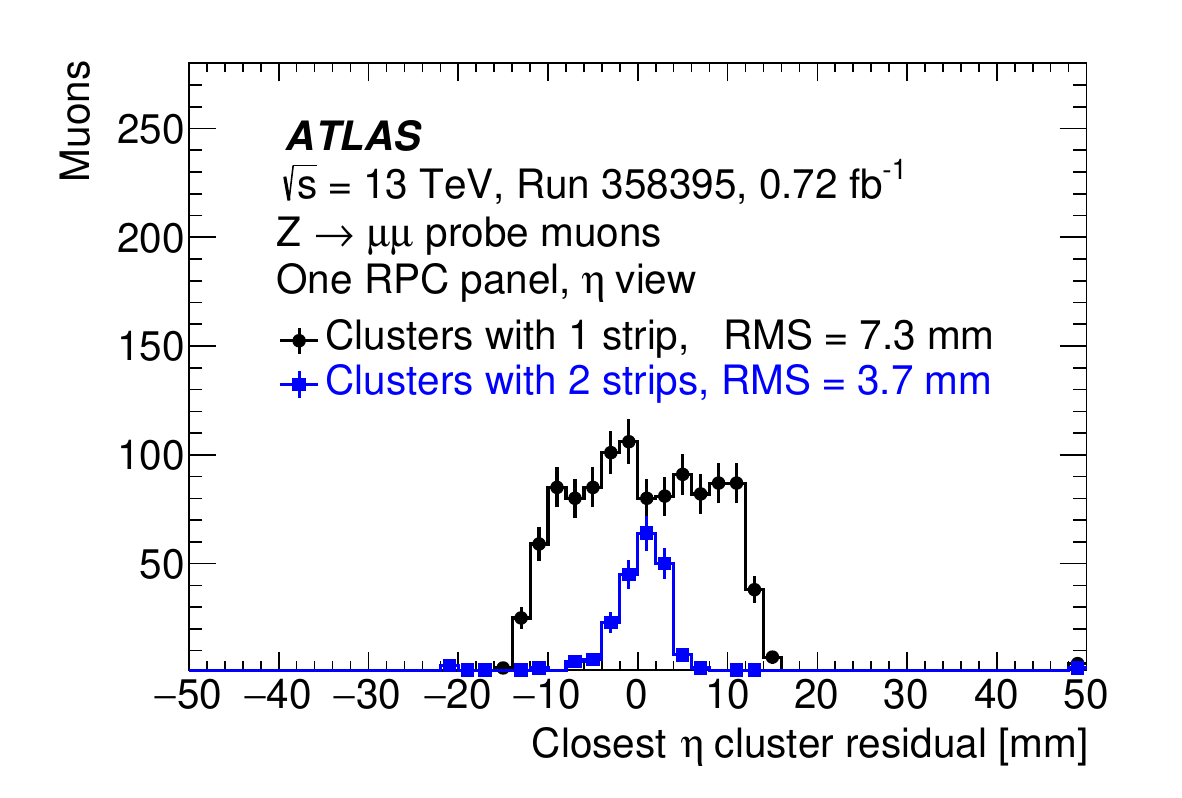}}
\subfloat[]{\includegraphics[width=0.49\textwidth]{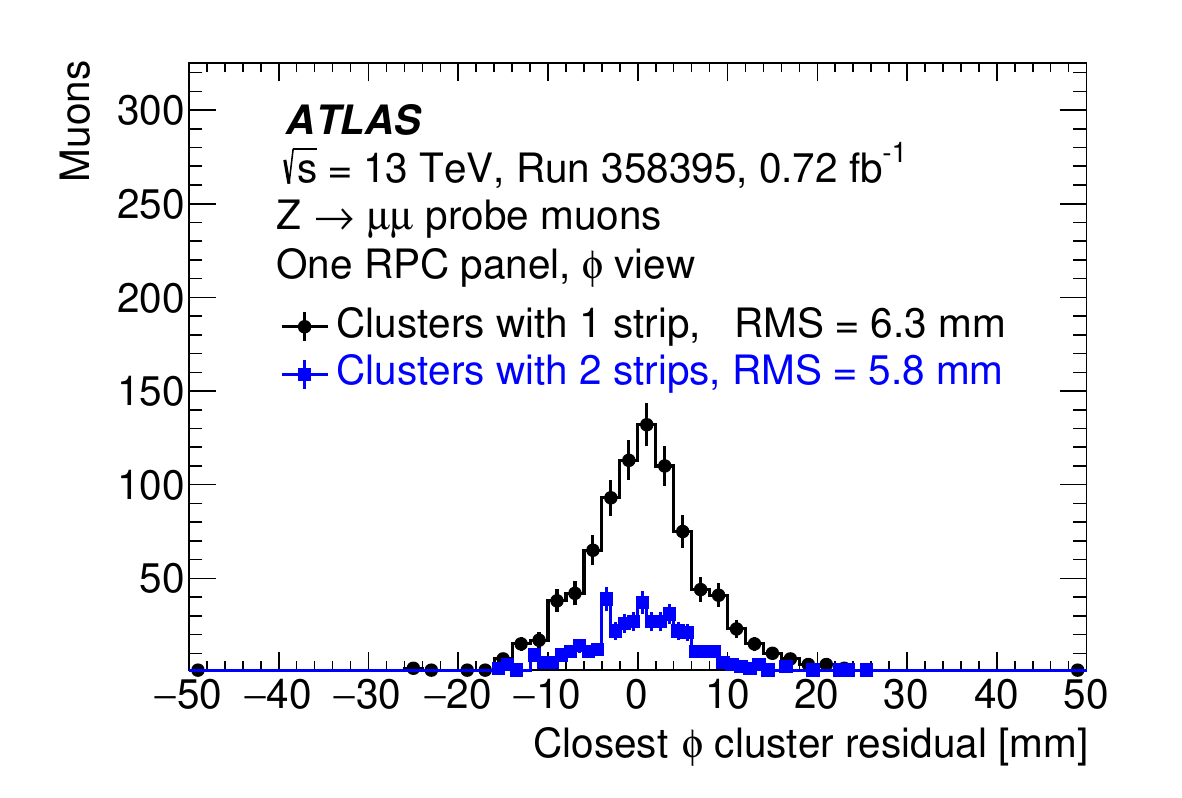}}
\caption{Cluster residuals for one representative RPC module for (a) $\eta$ strips and (b) $\phi$ strips. The residuals were computed by subtracting the expected muon position from the centre of the cluster. Clusters with one strip and with two strips are shown separately. The RMS values are computed using entries with residuals between $\pm 20$~mm. No time requirements are used to select the individual RPC hits. For clusters made of two strips, the hit times are required to be within 12.5~ns of each other.}
\label{pic:cluster_resolution}
\end{figure}

The cluster position residuals are defined as the difference between the cluster centre and the extrapolated impact point of a reconstructed muon track on the module surface. These residuals are shown in Figure~\ref{pic:cluster_resolution} for the representative RPC module for clusters with one or two strips. For clusters with one strip, the cluster centre is given by the strip centre and the width of the $\eta$ residual distribution is therefore determined by the strip width. This is because the $\eta$ coordinate of a muon track was measured using the MDT detector, which has much more precise $\eta$ position resolution than that of the RPC detector. For clusters with two strips, the cluster centre is defined as the midpoint between two strips. For these clusters the muon is more likely to pass between the two strips and therefore the width of the $\eta$ residual distribution is smaller because of the slightly better estimate of the muon impact point compared to the clusters with a single hit. Muon tracks are more likely to pass close to the $\phi$ cluster centre because the $\phi$ coordinate of the muon tracks in the MS was measured using the RPC detector. Therefore, muon tracks are biased to be close to recorded $\phi$ hits, resulting in the peak near zero for the $\phi$ residual distributions. The $\phi$ residual distribution widths are similar for the clusters with one hit and with two hits because the muon track $\phi$ position uncertainty is dominated by the RPC hit position uncertainty.
 
\FloatBarrier
 
\subsection{RPC detector response}
\label{subsec:RPC_clusters}
\begin{figure}[!h]
\centering
\subfloat[]{\includegraphics[scale=0.4]{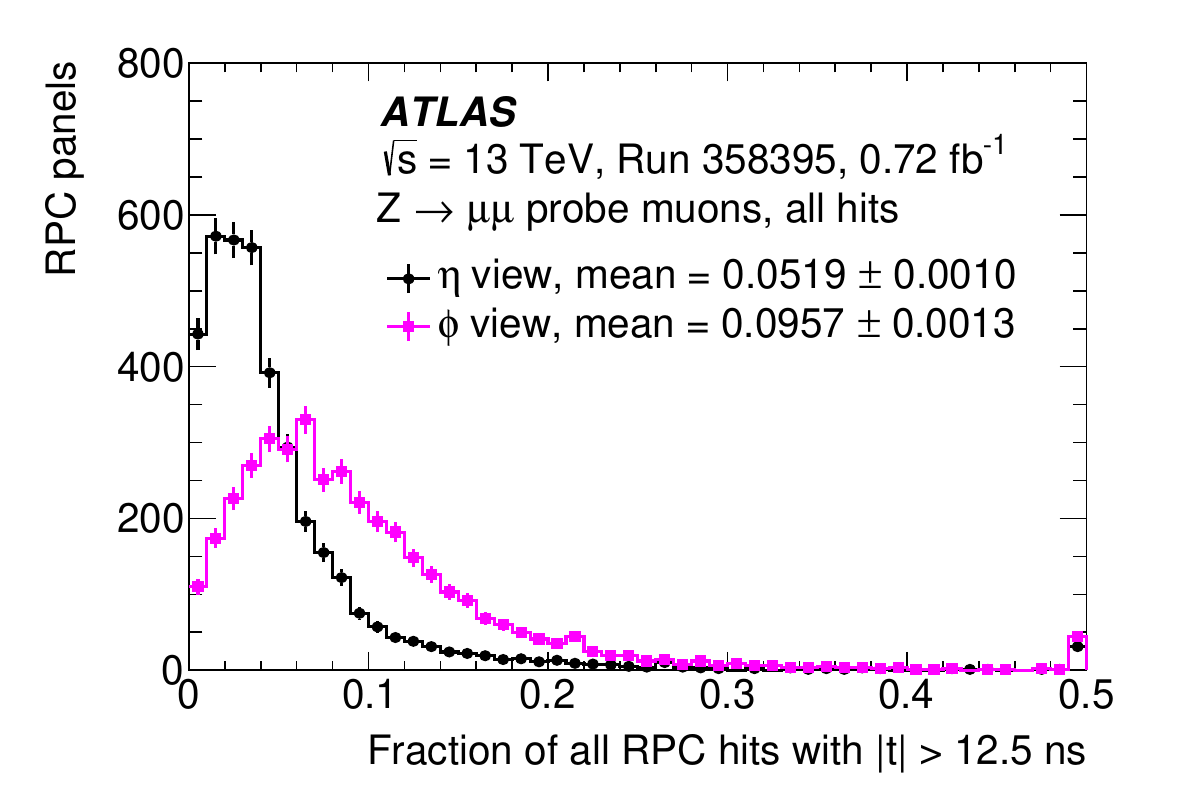} \label{pic:f_out_of_time_all}}
\subfloat[]{\includegraphics[scale=0.4]{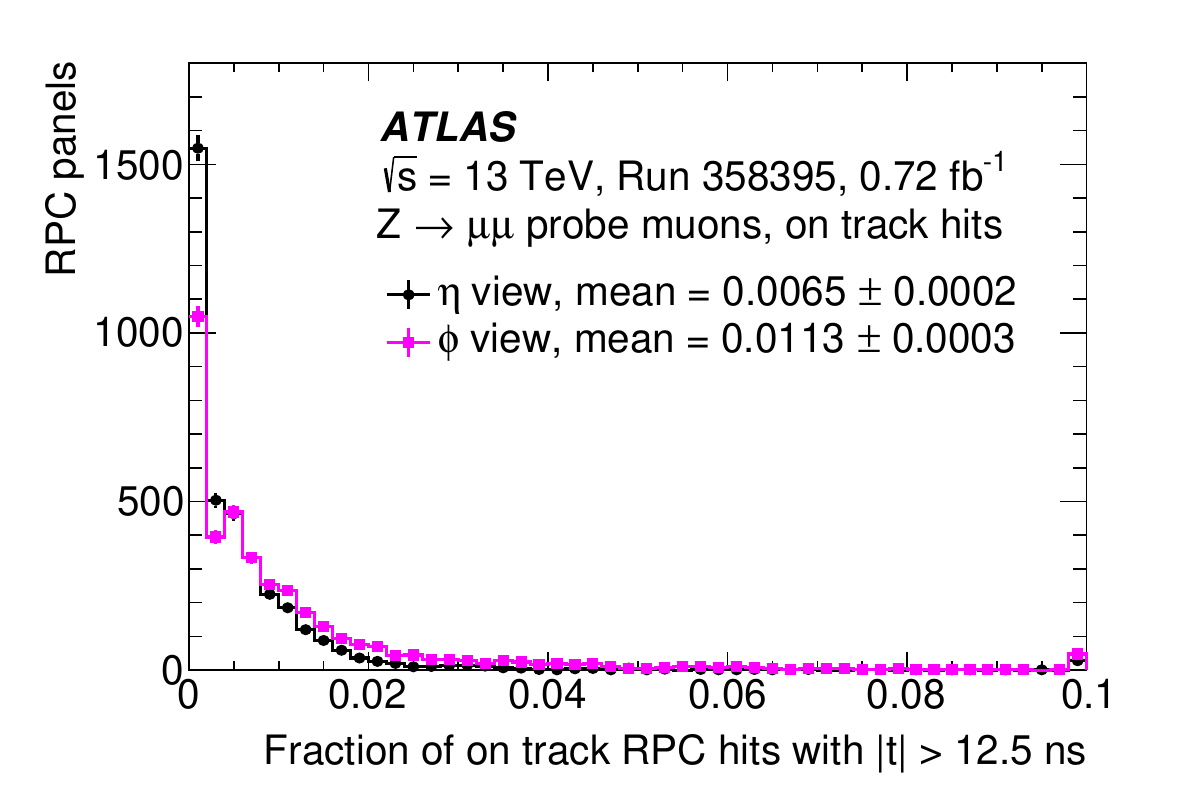} \label{pic:f_out_of_time_track}}
\caption{Fraction of RPC hits with a time $|t| > 12.5$~ns, for (a) all hits and (b) only for hits belonging to strips with an expected muon hit. Only active RPC modules were used for these plots.}
\label{pic:f_out_of_time}
\end{figure}

As discussed in Section~\ref{subsec:RPC_response}, a majority of the hits recorded in a module containing the muon impact point are expected to be closely associated in time with the arrival time of that muon. RPC hit times are calibrated in order to equalise the RPC timing response for ultra-relativistic particles produced at the interaction point. A small number of out-of-time hits, defined as hits with time $|t| > 12.5$~ns, are also recorded, primarily due to noise and delayed signals, with a small contribution from imperfect RPC timing calibration. Figure~\ref{pic:f_out_of_time_all} shows the fractions of out-of-time hits for all RPC modules. These fractions were computed for each module using only those events that include a muon candidate predicted to pass through that module. The larger mean out-of-time fraction for $\phi$ hits is due to the differences between $\eta$ and $\phi$ panels described in Section~\ref{sec:detrpc}. The mean out-of-time fraction is reduced to about 1\% when only the hits lying on the extrapolated muon track are included in the computation, as shown in Figure~\ref{pic:f_out_of_time_track}. This result demonstrates that only a small fraction of the out-of-time muon hits originate from imperfect RPC timing calibrations.
 
The passage of a muon through an RPC module can produce several hits and form a cluster. Two quantities are used to study cluster properties: the cluster hit multiplicity and the cluster size. The cluster hit multiplicity is an integer distribution, with one entry corresponding to one muon candidate, where a given multiplicity bin is filled with the number of hits recorded in one event for a muon candidate passing through a given panel. Distributions for individual panels are added up to produce the cluster hit multiplicity distribution averaged over a collection of panels or over the entire detector. These distributions can be computed using muon candidates recorded in one run or in the entire dataset. The cluster size is defined as the mean value of the cluster hit multiplicity distribution for one panel.
 
\begin{figure}[!h]
\centering
\subfloat[]{\includegraphics[width=0.49\textwidth]{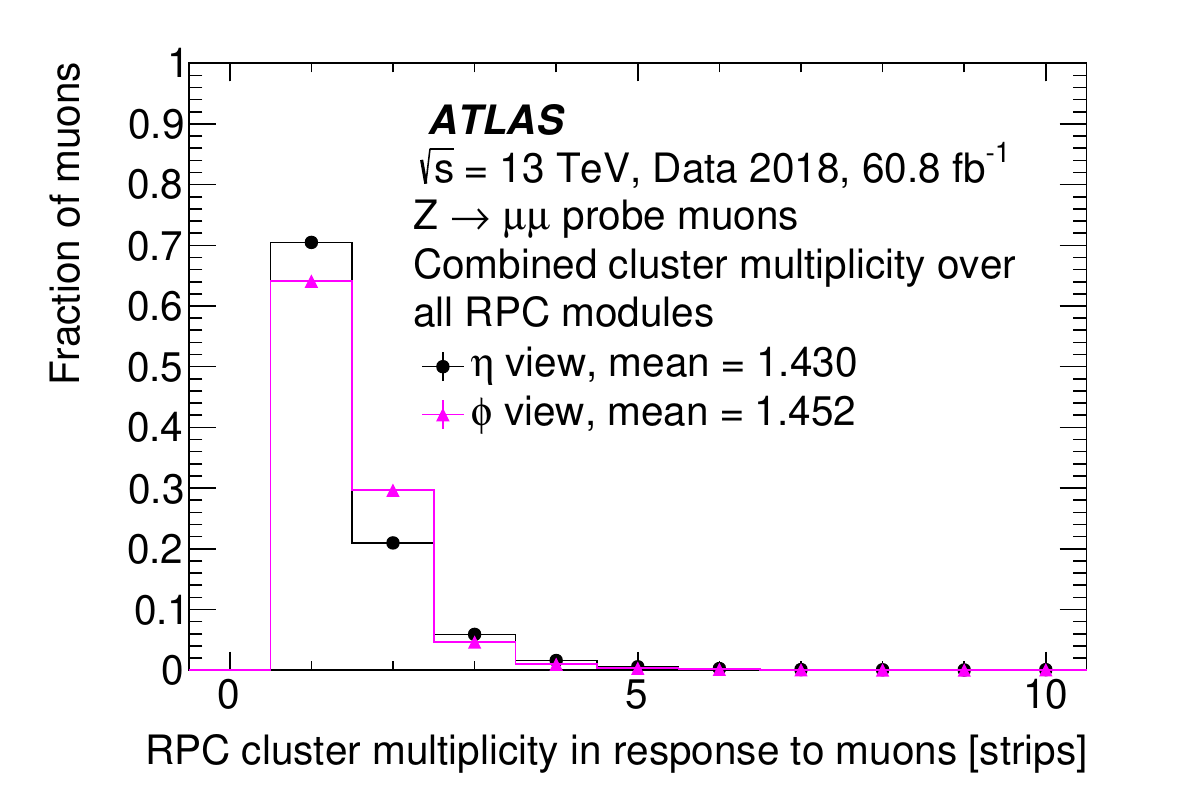}\label{pic:cluster_shape}}
\subfloat[]{\includegraphics[width=0.49\textwidth]{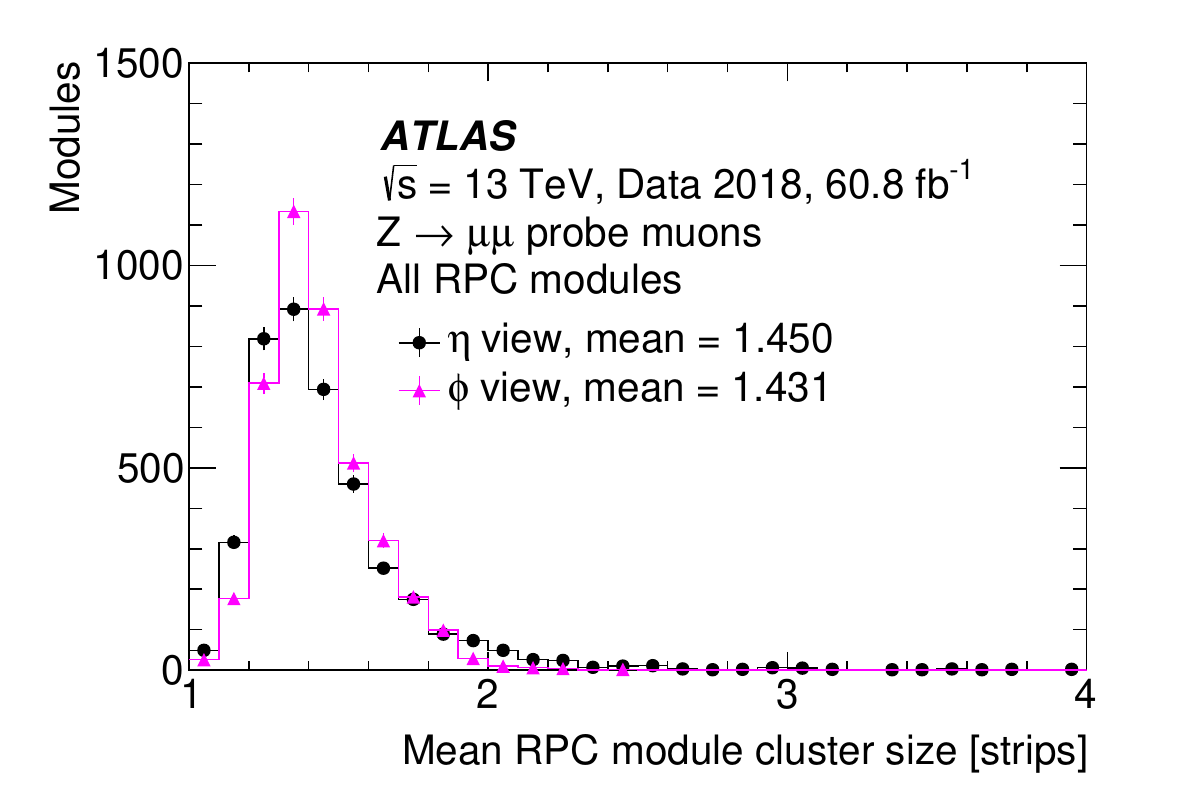}\label{pic:cluster_size}}
\caption{(a) RPC cluster hit multiplicity distribution and (b) RPC cluster size distribution, both averaged over all modules, in response to the passage of a muon through a module, with $\eta$ and $\phi$ panels shown separately.}
\label{pic:cluster_size_shape}
\end{figure}

The cluster hit multiplicity distribution is averaged over all RPC panels and shown in Figure~\ref{pic:cluster_shape}, computed using the full 2018 dataset. The average RPC cluster is approximately 1.5~strips wide, with the average taken separately over $\eta$ and $\phi$ panels and using all active RPC modules. The average $\eta$ panel has a cluster multiplicity which is slightly smaller than that of the average $\phi$ panel, due to the differences in the construction and readout as discussed in Section~\ref{sec:detrpc}. Figure~\ref{pic:cluster_size} shows the average cluster size distributions for all active RPC modules, which is computed as the mean of the cluster hit multiplicity distribution for each $\eta$ and $\phi$ panel. There are some $\eta$ panels that have larger-than-typical cluster sizes, seen as the tail of the $\eta$ distribution in Figure~\ref{pic:cluster_size}.
 
The cluster hit multiplicity distribution averaged over the entire RPC detector is given in Table~\ref{tab:cluster_size_overall}. Muons produce clusters with two or more hits in 30\% and 36\% of events for the $\eta$ and $\phi$ panels, respectively. Larger clusters containing four or more hits are produced in 2.7\% and 1.6\% of events for the $\eta$ and $\phi$ panels, respectively. Variations in the cluster hit multiplicity are due to the differences between $\eta$ and $\phi$ panels described in Section~\ref{sec:detrpc} and to small differences in strip widths.
 
\begin{table}[!h]
\centering
\caption{
Cluster hit multiplicity distribution in response to a passage of a muon, shown separately for $\eta$ and $\phi$ panels.
These multiplicity distributions are summed over all active RPC modules,
separately for $\eta$ and $\phi$ panels. The statistical uncertainty for all table entries is smaller than 0.1\%.}
\vspace{0.2cm}
\begin{tabular}{cccccccc}
\hline
Cluster hit multiplicity & 1 & 2 & 3 & 4 & 5 & > 5 \\
\hline
$\eta$ panels & $70.5\%$ & $21.0\%$ & $5.9\%$ & $1.6\%$ & $0.6\%$ & $0.5\%$ \\
$\phi$ panels & $64.1\%$ & $29.6\%$ & $4.6\%$ & $1.0\%$ & $0.3\%$ & $0.3\%$ \\
\hline
\end{tabular}
\label{tab:cluster_size_overall}
\end{table}
 
\begin{figure}[!h]
\centering
\subfloat[]{\includegraphics[width=0.49\textwidth]{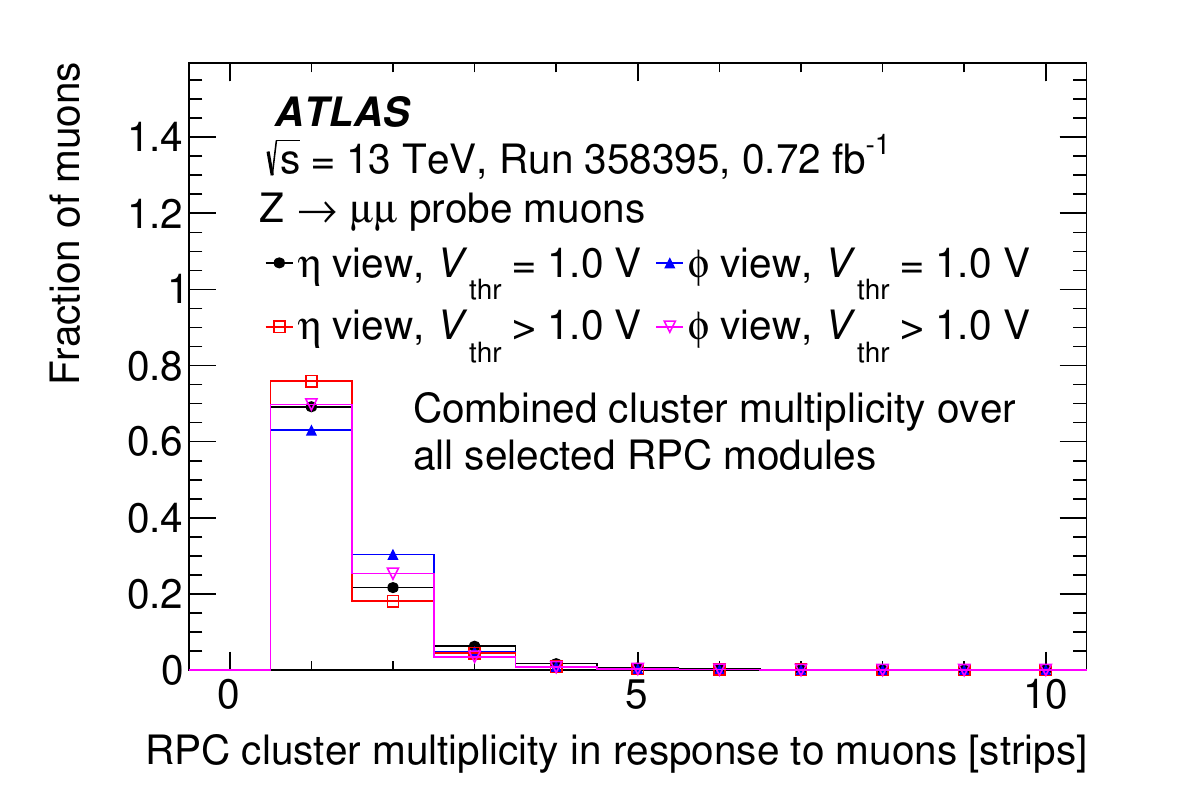}}
\subfloat[]{\includegraphics[width=0.49\textwidth]{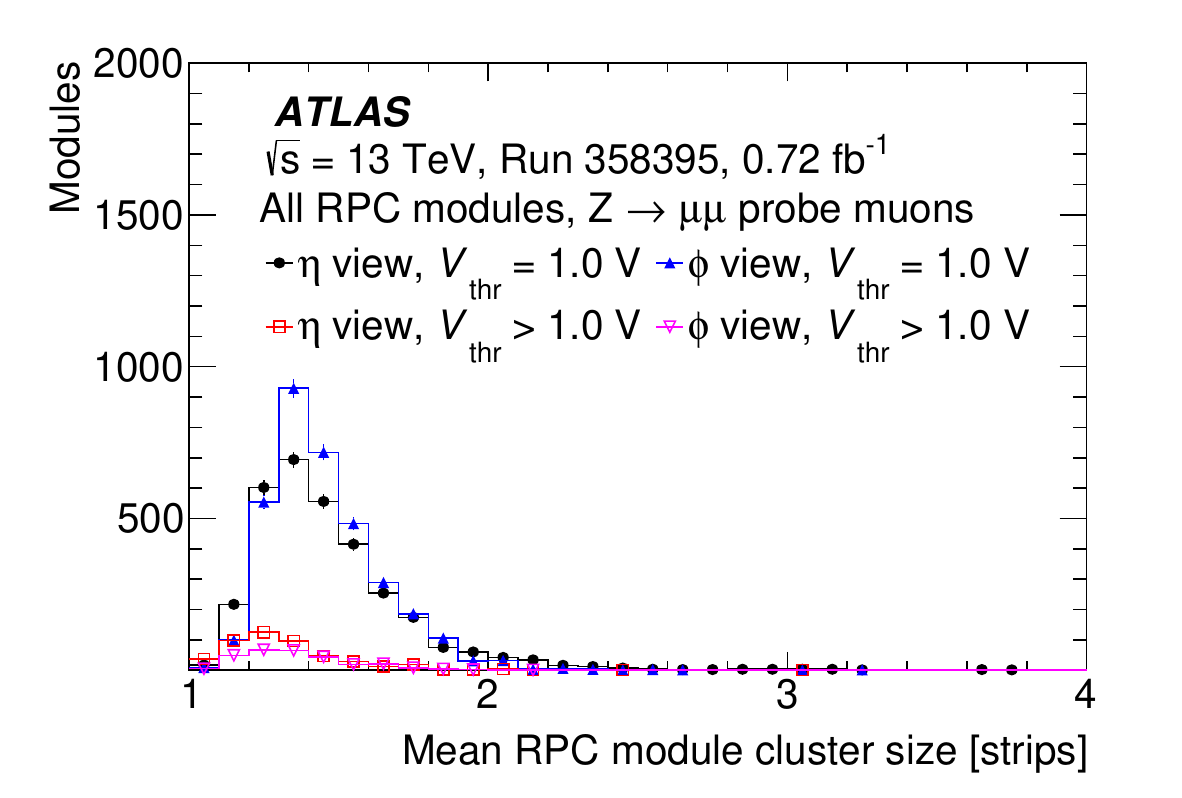}}
\caption{(a) Cluster hit multiplicity and (b) cluster size distributions for all selected modules shown separately for $\eta$ and $\phi$ panels using nominal ($\vfe = 1.0$~V) and higher than nominal ($\vfe > 1.0$~V) FE thresholds.}
\label{pic:cluster_size_shape_vth}
\end{figure}
 
To illustrate the effect of the FE thresholds on the RPC cluster size, the cluster hit multiplicity distribution and cluster size were measured separately for panels with nominal FE thresholds and for panels with FE thresholds higher than their nominal values. One typical ATLAS run was analysed for this study. In this run, the majority of the panels utilised the default FE threshold setting of $\vfe = 1.0$~V. A small fraction of the panels utilised higher-than-nominal FE thresholds that varied in the range from $\vfe = 1.1$~V to $\vfe = 1.5$~V in steps of 0.1~V. These results are shown in Figure~\ref{pic:cluster_size_shape_vth} separately for the $\eta$ and $\phi$ panels. As expected, higher FE thresholds result in smaller cluster hit multiplicity and smaller cluster sizes.
 
\begin{figure}[!h]
\centering
\subfloat[]{\includegraphics[width=0.49\textwidth]{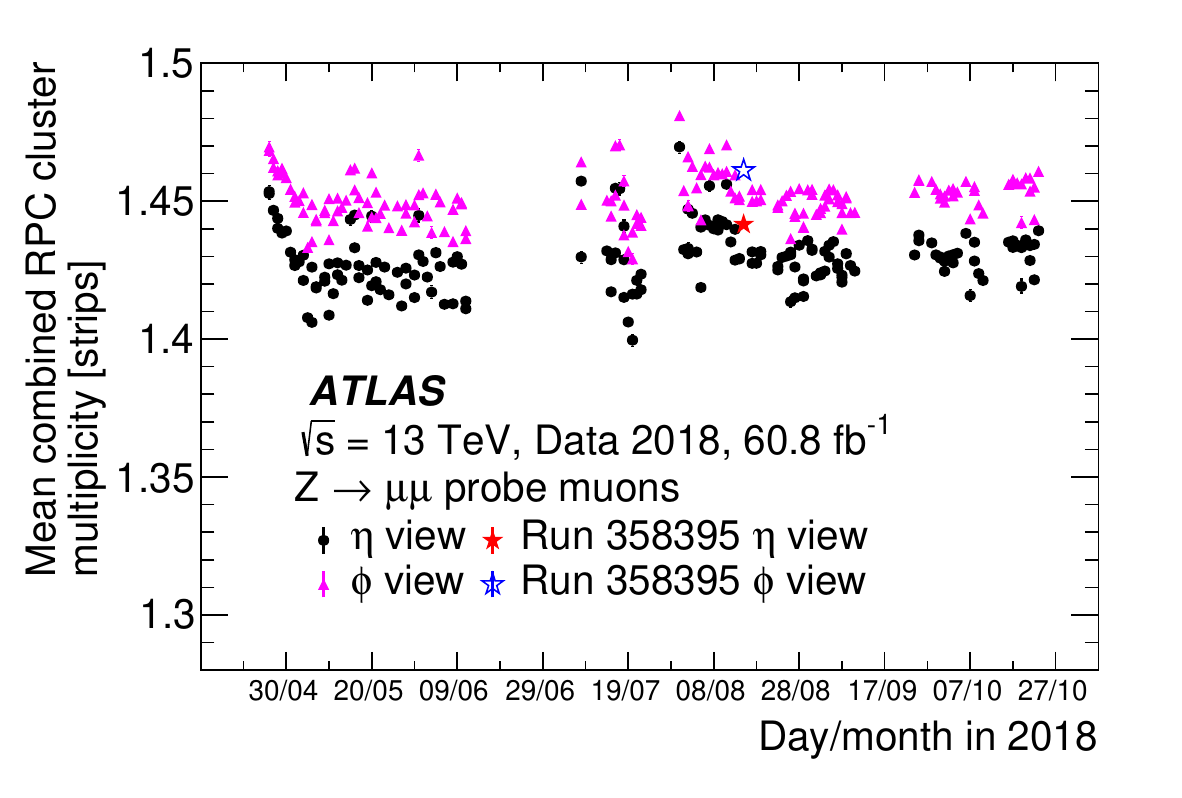}\label{pic:cluster_vs_run}}
\subfloat[]{\includegraphics[width=0.49\textwidth]{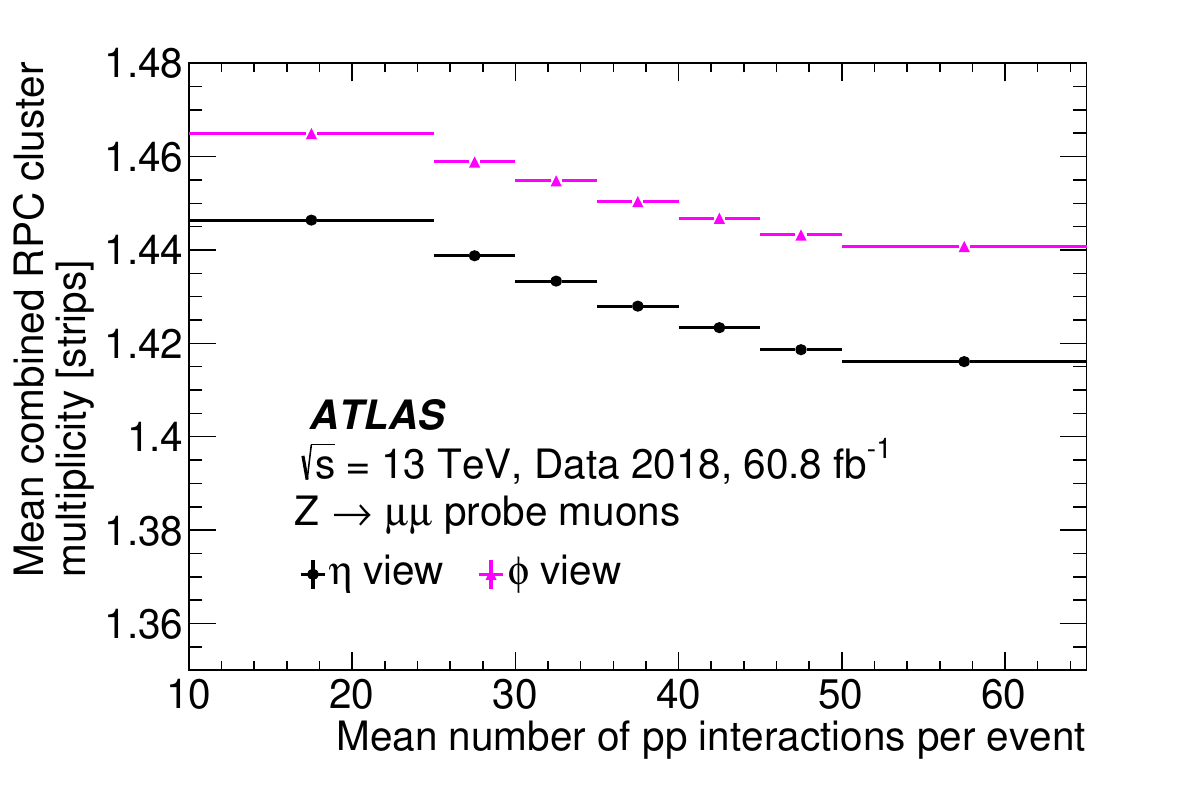}\label{pic:cluster_vs_mu}}
\caption{Mean of the RPC detector cluster hit multiplicity distribution plotted (a) as a function of time and (b) as a function of average number of proton--proton collisions per event. Mean values for $\eta$ and $\phi$ panels are shown separately. The blue and red stars refer to the representative run used in most of the performance studies shown in this paper.}
\end{figure}
 
The average cluster hit multiplicity was monitored as a function of time, as shown in Figure~\ref{pic:cluster_vs_run} where each entry corresponds to one ATLAS run. The small variations among runs are due to changes in the FE thresholds and changes in the detector conditions, such as automatic adjustments of the applied voltage. Overall, the mean RPC cluster multiplicity was stable in 2018 to within a few percent. The small decrease in the cluster multiplicity in late April and early May was due to operating the RPCs at nominal voltage with collisions following the winter LHC shutdown when the RPC voltage was off. The small increase in August was due to adjustments of the FE thresholds, leading to the slightly larger detector efficiency described in Section~\ref{subsec:det_eff}.
 
The average cluster hit multiplicity is plotted in Figure~\ref{pic:cluster_vs_mu} as a function of the number of proton--proton interactions per event. An instantaneous luminosity of $2 \times \silumi{34}$ corresponds on average to about 56 collisions per event. A reduction of approximately 1\% to 2\% between the first and last bins was observed. This small reduction is due to dead time imposed by the readout system and to offline reconstruction in each individual strip, in addition to a small chamber inefficiency at higher detector occupancy~\cite{Acitelli:1995ft,Aielli:2002th}. If a strip records a hit due to a background event, then a later muon hit within the dead-time window will be discarded, thereby introducing a small inefficiency in that channel. These background hits would decrease both the mean cluster size and also the muon detection efficiency, as shown in Section~\ref{subsec:det_eff}.
\FloatBarrier
 
\subsection{RPC detector performance}
\label{subsec:det_eff}
\begin{figure}[!h]
\centering
\subfloat[]{\includegraphics[width=0.49\textwidth]{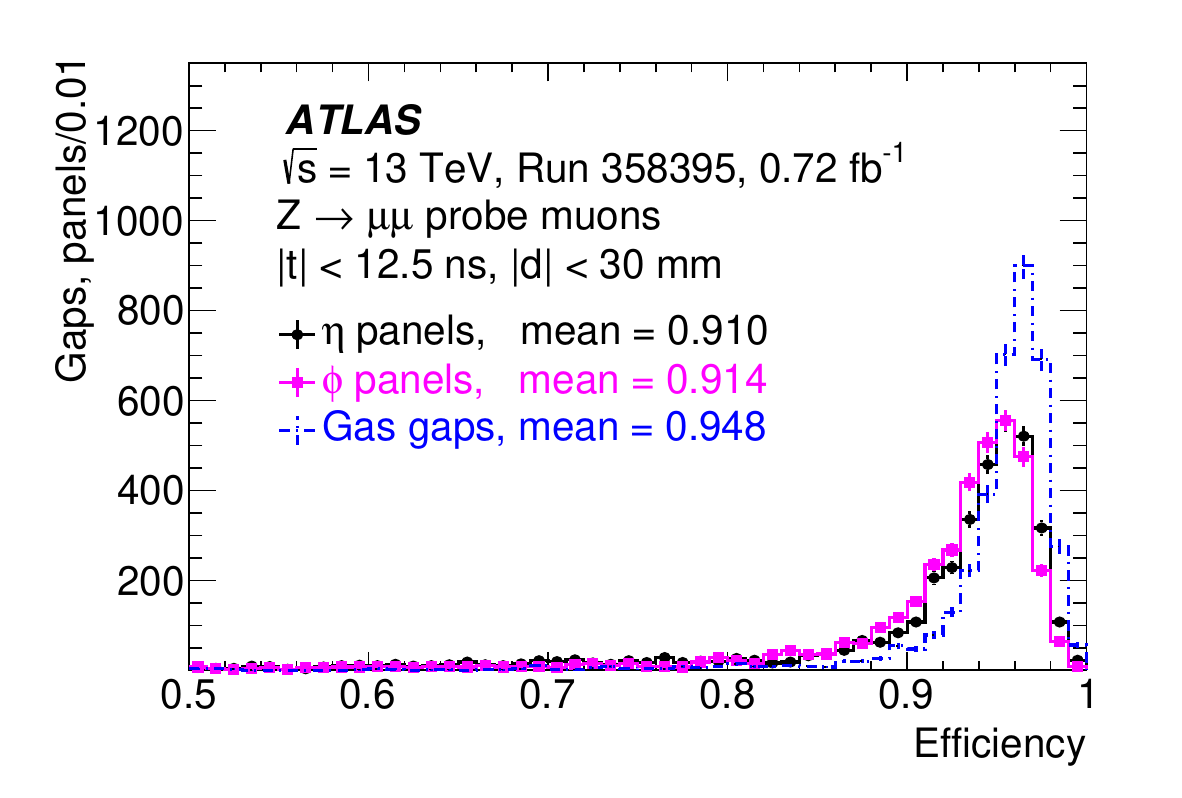}\label{pic:PanelEffDist}}
\subfloat[]{\includegraphics[width=0.49\textwidth]{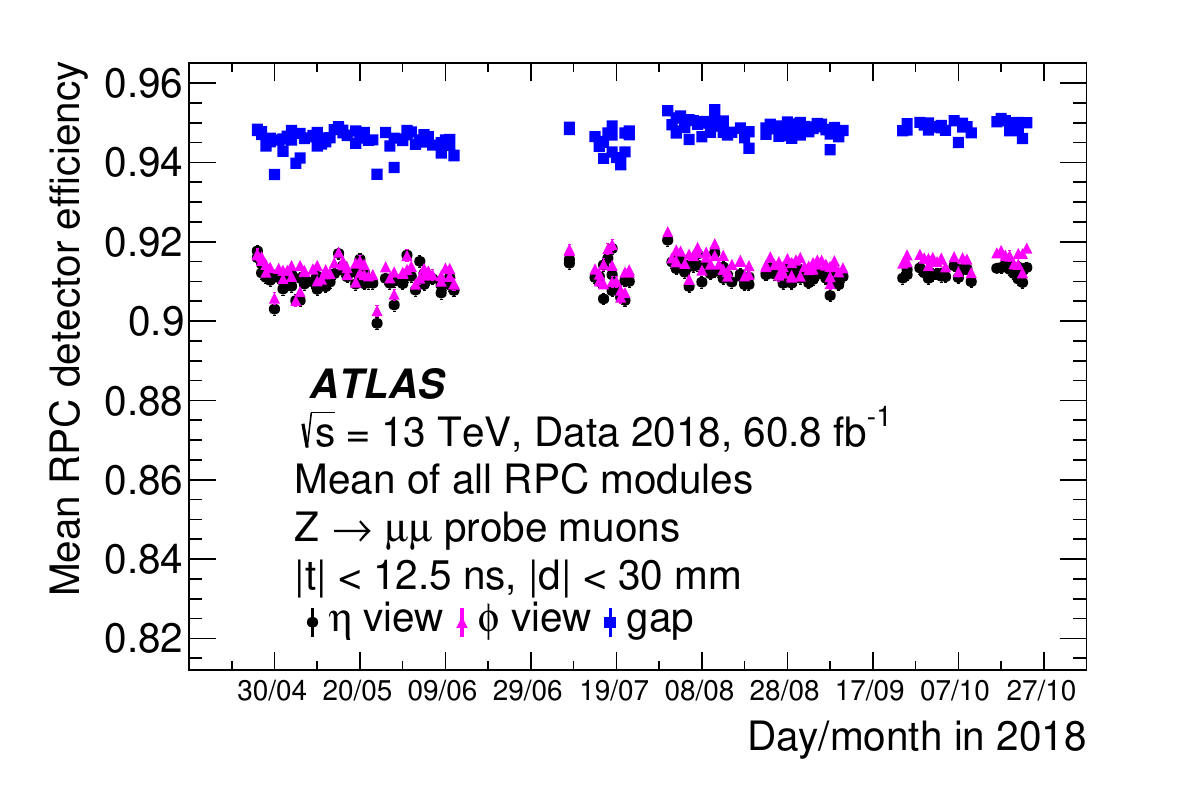}\label{pic:GapPanelMeanEffTime}}
\caption{(a) Distribution of the muon detection efficiency for all active RPC modules. (b) Muon detection efficiency for $\eta$ panels, $\phi$ panels and gaps averaged over all active RPC modules plotted as a function of time.}
\label{pic:RPCDetEff}
\end{figure}

The RPC efficiency for muon-induced signals is shown in Figure~\ref{pic:PanelEffDist} for all RPC modules without known defects using data from one representative ATLAS run. The efficiency is computed using signal hits, defined in Section~\ref{subsec:RPC_response}. The efficiency for $\eta$ and $\phi$ panels belonging to the same module is computed independently by requiring at least one hit in $\eta$ or $\phi$ strips, respectively. The module (gas gap) efficiency is computed by requiring at least one hit either in $\eta$ or $\phi$ strips belonging to the same module. The gap efficiency corresponds to the intrinsic RPC efficiency to detect a muon-induced avalanche. The efficiency of the individual $\eta$ or $\phi$ panel within a module includes the additional effect due to intrinsic efficiency of the FE electronics to register the avalanche signal. This FE efficiency is estimated as the difference between the module efficiency and individual $\eta$ or $\phi$ panel efficiency. The average efficiency of the FE electronics to record the avalanche signal is measured to be approximately 97\%. This number also includes small contributions from a few faulty strips in some of the panels.
 
The RPC detector efficiency is plotted as a function of time in Figure~\ref{pic:GapPanelMeanEffTime}, which shows the $\eta$/$\phi$ panel efficiency and gas gap efficiency. The mean detector efficiency in each ATLAS run is computed by averaging over all active RPC modules with efficiency greater than 50\%, which represent about 90\% of the total number of modules. A small increase in the average efficiency of about 0.3\% in August 2018 was due to adjustments of FE thresholds. No significant ageing effects have been observed during 2018 while the ATLAS detector recorded approximately {\ilumi} of proton--proton collision data.
 
\begin{figure}[!h]
\centering
\subfloat[]{\includegraphics[width=0.49\textwidth]{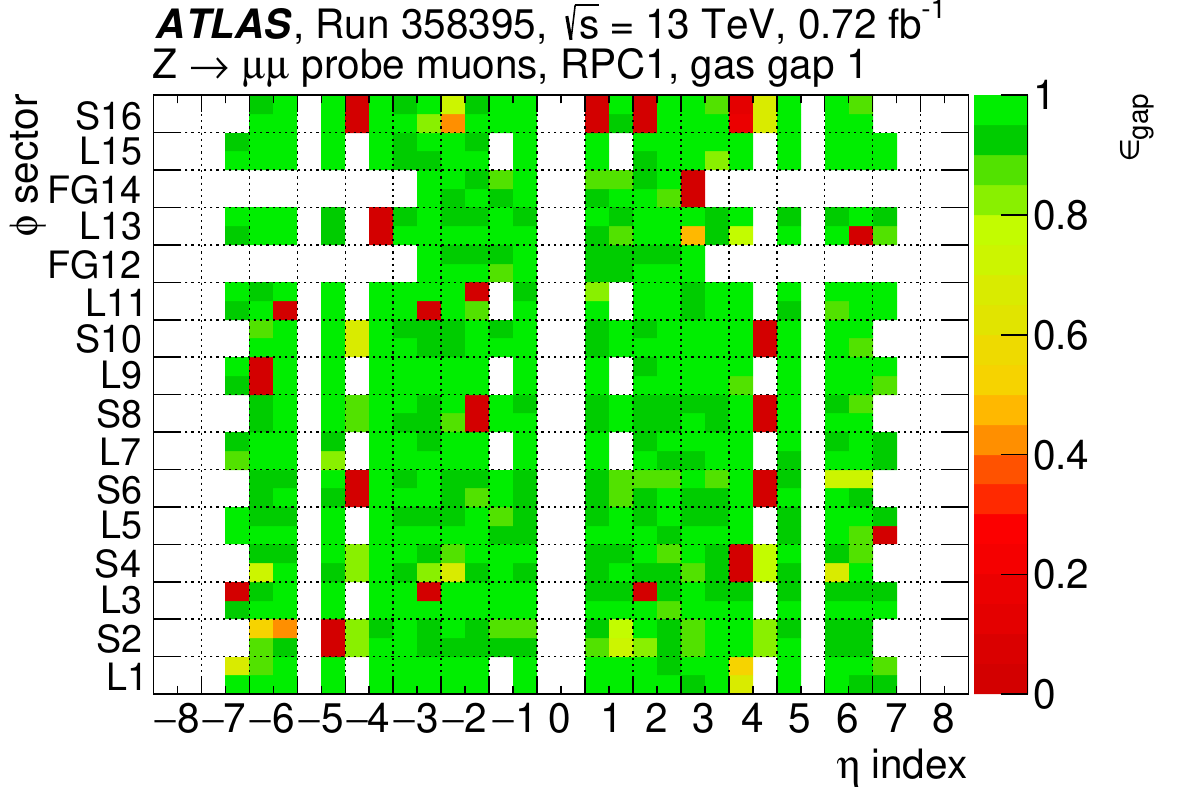}\label{pic:BMDR1GG1GapEffMap}}
\subfloat[]{\includegraphics[width=0.49\textwidth]{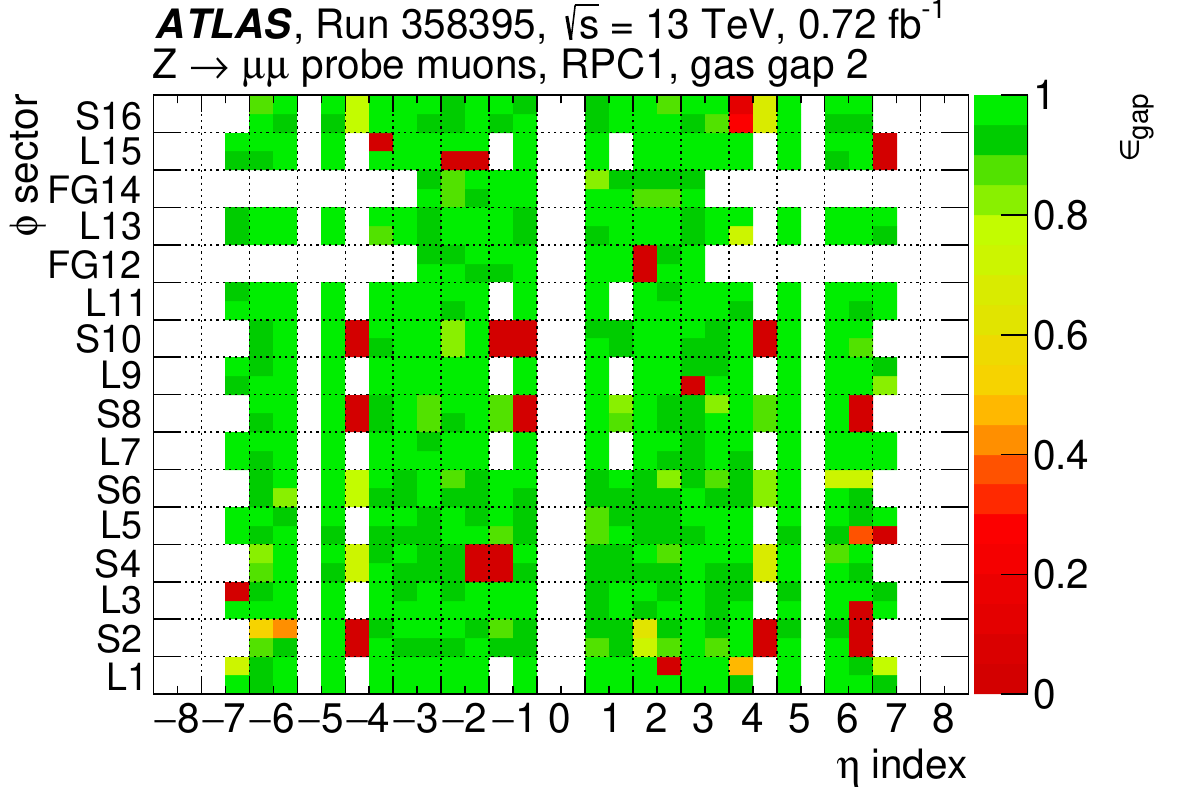}\label{pic:BMDR1GG2GapEffMap}}
\caption{Muon detection efficiency in the $\eta$--$\phi$ plane for all modules in (a) the first layer and (b) the second layer of the RPC1 doublet layer. Modules within one station are represented using half-integer values of \ieta and \iphi. Empty bins correspond to logical combinations of indices that do not represent installed RPCs.}
\label{pic:EtaPhiMapRPC1}
\end{figure}

The module efficiency is shown in Figure~\ref{pic:EtaPhiMapRPC1} for all modules in the first and second layers of the RPC1 doublet layer. The red bins indicate modules with low efficiency or modules with the operating voltage off due to gas leaks. The fraction of the chambers with gas leaks was around 5\% in 2015 and increased to about 10\% in 2018 at the end of the most recent data-taking period. These less efficient modules had only a small impact on the overall muon trigger efficiency because muons producing a single hit in the two layers can still satisfy the trigger logic requirements, as discussed in Section~\ref{subsec:trigRoad}. Approximately two thirds of the chambers with gas leaks are expected to be repaired during the 2019-2021 LHC shutdown.
 
\begin{figure}[!h]
\centering
\includegraphics[width=0.49\textwidth]{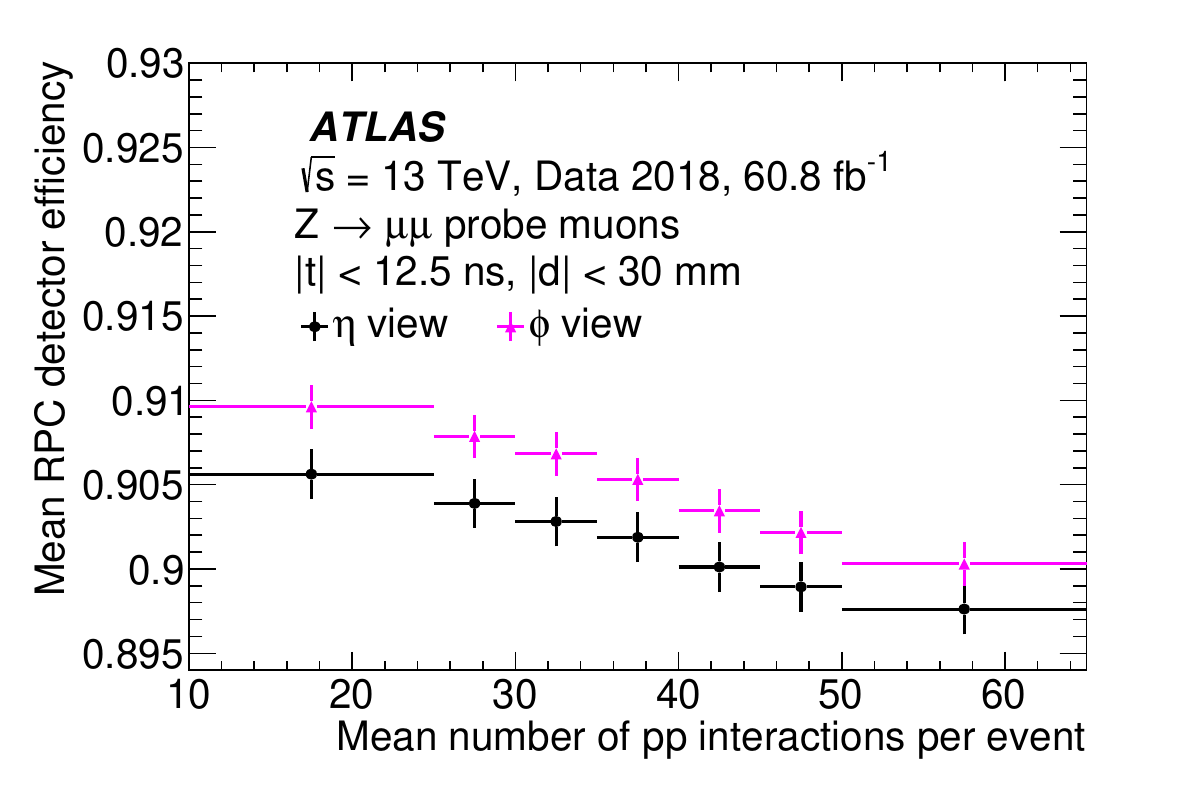}
\caption{Muon detection efficiency for $\eta$ and $\phi$ panels, averaged over all active RPC modules, plotted as a function of average number of proton--proton collisions per event.}
\label{pic:PanelMeanEffAveMu}
\end{figure}
 
The muon detection efficiency is shown in Figure~\ref{pic:PanelMeanEffAveMu} as a function of the mean number of proton--proton interactions per event. The detector efficiency is reduced by approximately 1\% over the range considered for the reasons discussed in Section~\ref{subsec:RPC_clusters}. The detector efficiency linearly extrapolated to an instantaneous luminosity of $7.5 \times \silumi{34}$ is approximately 87\%. The dependence of the detector efficiency on detector occupancy is investigated further in Section~\ref{subsec:pileup}.
 
One potential bias in measuring the efficiency of $\phi$ panels can arise from the offline reconstruction of muon candidates, which uses RPCs to measure the $\phi$ coordinates of muon tracks in the MS. The RPC detector provides up to six $\eta$ and $\phi$ position measurements for each muon track. Therefore, the impact of a single RPC detector layer on the muon detection efficiency is expected to be negligible. This hypothesis was first checked by extrapolating ID tracks to the RPC detector surfaces. No detectable biases were observed for the measurements reported in this section when using this alternative extrapolation algorithm. This effect was also checked by comparing the efficiencies of the $\eta$ panel and $\phi$ panel belonging to the same RPC module. The average difference between the $\eta$ and $\phi$ efficiencies is consistent with zero, confirming that no significant bias was introduced by using RPCs for reconstructing $\phi$ positions of muon tracks in the MS.
 
\subsection{Time resolution of RPC detector and readout system}
\label{subsec:time_res}
The RPC technology was chosen by the ATLAS experiment for the L1 muon barrel trigger because of its fast response, good time and position resolution, and relatively low cost~\cite{ATLAS-TDR-10}. Before the construction of the ATLAS detector, the RPC time resolution was measured to be between 1.1~ns and 1.4~ns, using full-size prototype units of the ATLAS RPCs equipped with the final version of the FE electronics~\cite{Aielli:2000rz}. The time resolution of the installed RPC detectors was measured using muons produced in proton--proton collisions recorded in 2011 at a centre-of-mass energy of 7~\TeV. A time resolution value of around $2$~ns was obtained~\cite{Chiodini:2013rta}. This higher value, compared to the results from test-beam facilities, was due to combined effects of the RPC intrinsic resolution and the readout system resolution.

In this section, the intrinsic time resolution of the RPC detector and the time resolution of RPC electronics are evaluated separately using newly developed procedures for the analysis of the RPC timing response to probe muons produced in the decays of $Z$ bosons, with the selection criteria detailed in Section~\ref{sec:analysis}. These results aim to quantify the RPC intrinsic time resolution for detecting a signal induced by an ionising particle, excluding effects related to the resolution of its absolute arrival time. The intrinsic RPC time resolution depends on fluctuations in the location of the first ionisation event along the muon path through the RPC gas gap and on the statistical fluctuations inherent in the subsequent avalanche amplification process~\cite{Riegler:2002vg}. The electronics time resolution component includes several sources, such as the $320$~MHz sampling frequency and the associated jitter~\cite{Anulli:2009zz}. The total time resolution (\sigtot) of an RPC is defined as:
\begin{equation}
\sigtot^2 = \sigin^2+\sigel^2
\label{equation:total_time_resolution}
\end{equation}
where \sigin is the component corresponding to the intrinsic RPC time resolution and \sigel is the electronics component corresponding to the time resolution of the readout system.

The total RPC time resolution was estimated using time differences between detector signals generated by the same muon passing through two parallel single RPC layers belonging to the same module. These two layers are separated by a distance of approximately 20~mm. The time difference was computed using hits produced in the two parallel strips (either a pair of $\eta$ strips or a pair of $\phi$ strips) that are closest to the muon track in each of the two detector layers. A relativistic muon travelling nearly at the speed of light will produce a time-of-flight difference of about 0.07~ns when travelling perpendicular to the two detector layers. Since the contribution from the muon time-of-flight component is negligible, the time differences between signals produced in the two layers are dominated by the total RPC time resolution.
 
\begin{figure}[!h]
\centering
\subfloat[]{\includegraphics[width=0.49\textwidth]{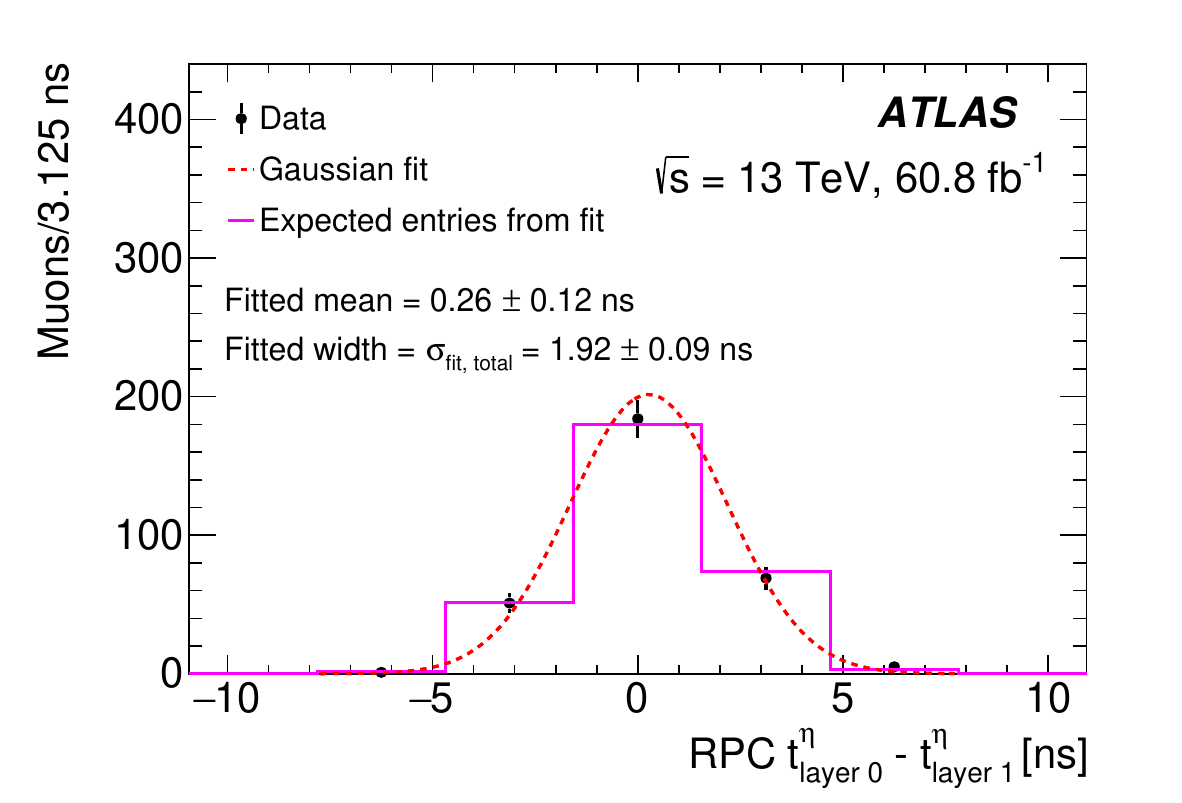}\label{pic:onetimefit}}
\subfloat[]{\includegraphics[width=0.49\textwidth]{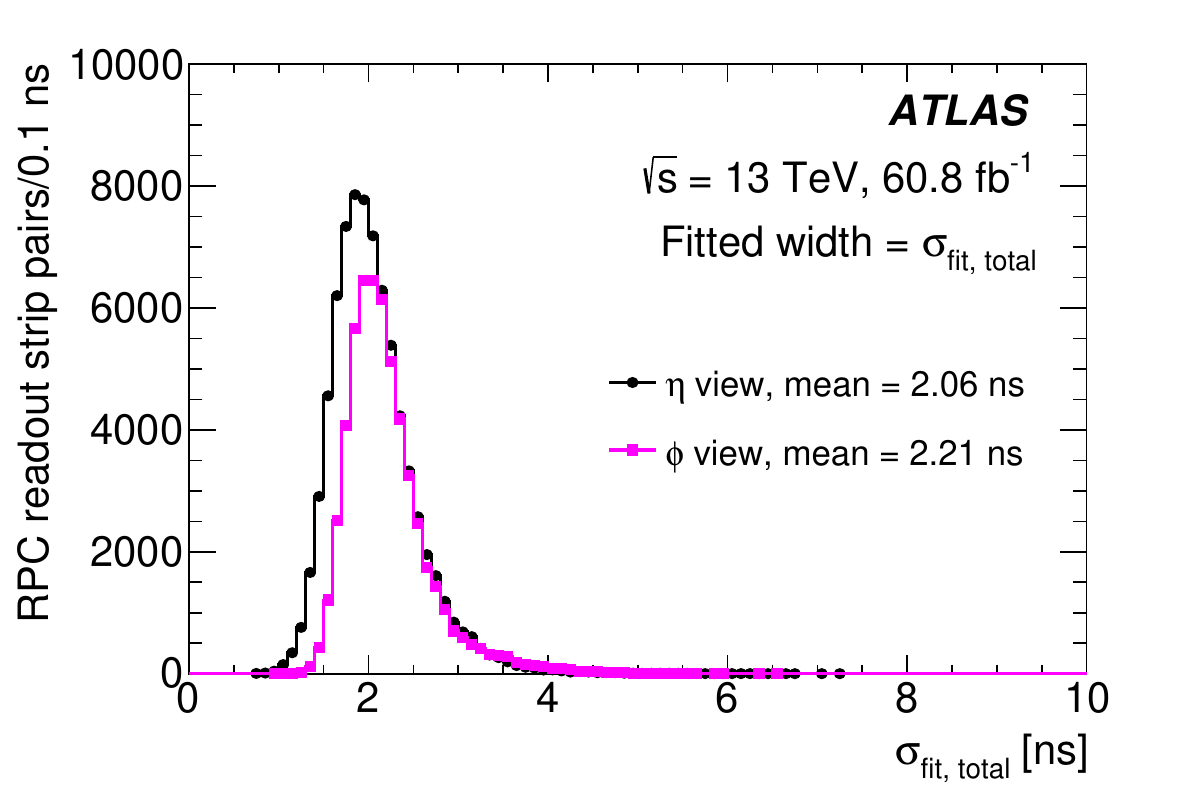}\label{pic:sigtotal}}
\caption{(a) Time-difference distribution between signals generated by the same muon in a single pair of $\eta$ strips, matched with the muon track, in two parallel RPC detector layers. The bin width corresponds to the 3.125~ns sampling time. (b) Distribution of the total time differences for the all selected RPC strip pairs.}
\label{pic:time1}
\end{figure}

The time-difference distributions were obtained for all geometrically possible combinations of $\eta$ strip pairs and $\phi$ strip pairs in each doublet layer of the RPC detector. The same timing circuit was used to measure these two signals so no differences are introduced due to clock synchronisation. Only distributions with at least 100 entries were selected for the analysis in order to remove strips with low efficiency. The width of the time-difference distribution was determined from a binned maximum-likelihood fit of a Gaussian function to the observed distribution of time differences. An example fit is shown in Figure~\ref{pic:onetimefit} for a pair of $\eta$ strips from the two parallel layers of one RPC chamber. The Kolmogorov--Smirnov (KS) test was performed to assess the goodness of fit. To perform this test, a new histogram was generated with the number of entries in each bin equal to the area of the Gaussian curve in that bin divided by the bin width. Only strip pair distributions with a KS probability greater than 0.1 and with at least three non-empty bins were retained for the final analysis. These selection criteria have a combined efficiency of about 98\%.

The best-fit Gaussian widths for all selected $\eta$ strip pairs and $\phi$ strip pairs are shown in Figure~\ref{pic:sigtotal} for the full RPC system. The mean width of the time differences is approximately 2.1~ns and 2.2~ns for $\eta$ and $\phi$ strips, respectively. The small difference between the $\eta$ and $\phi$ strips is due to the different cluster size distributions for $\eta$ and $\phi$ panels, which is discussed in Section~\ref{subsec:RPC_clusters}. When the time differences were measured using only those muons that produce clusters with a single hit in both layers, no difference between $\eta$ and $\phi$ strips was observed for the time difference widths.
 
\begin{figure}[!h]
\centering
\subfloat[]{\includegraphics[width=0.49\textwidth]{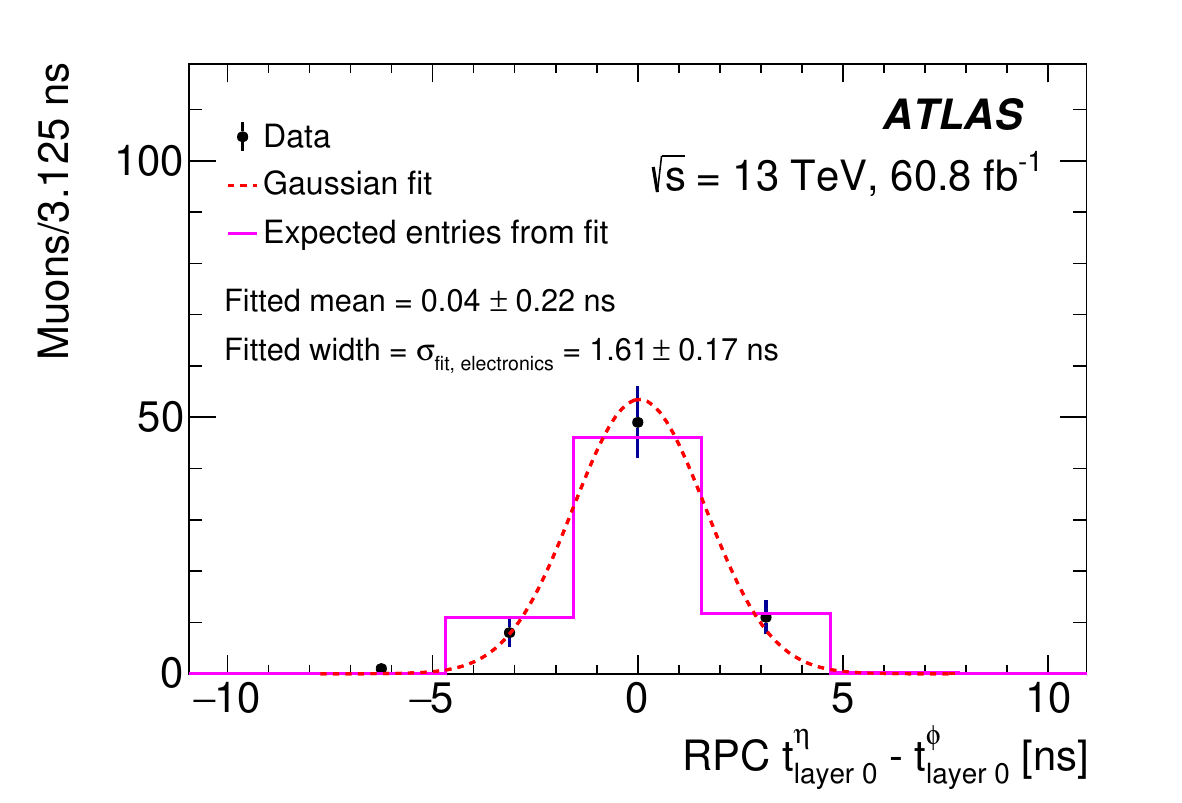}\label{pic:OneElecFit}}
\subfloat[]{\includegraphics[width=0.49\textwidth]{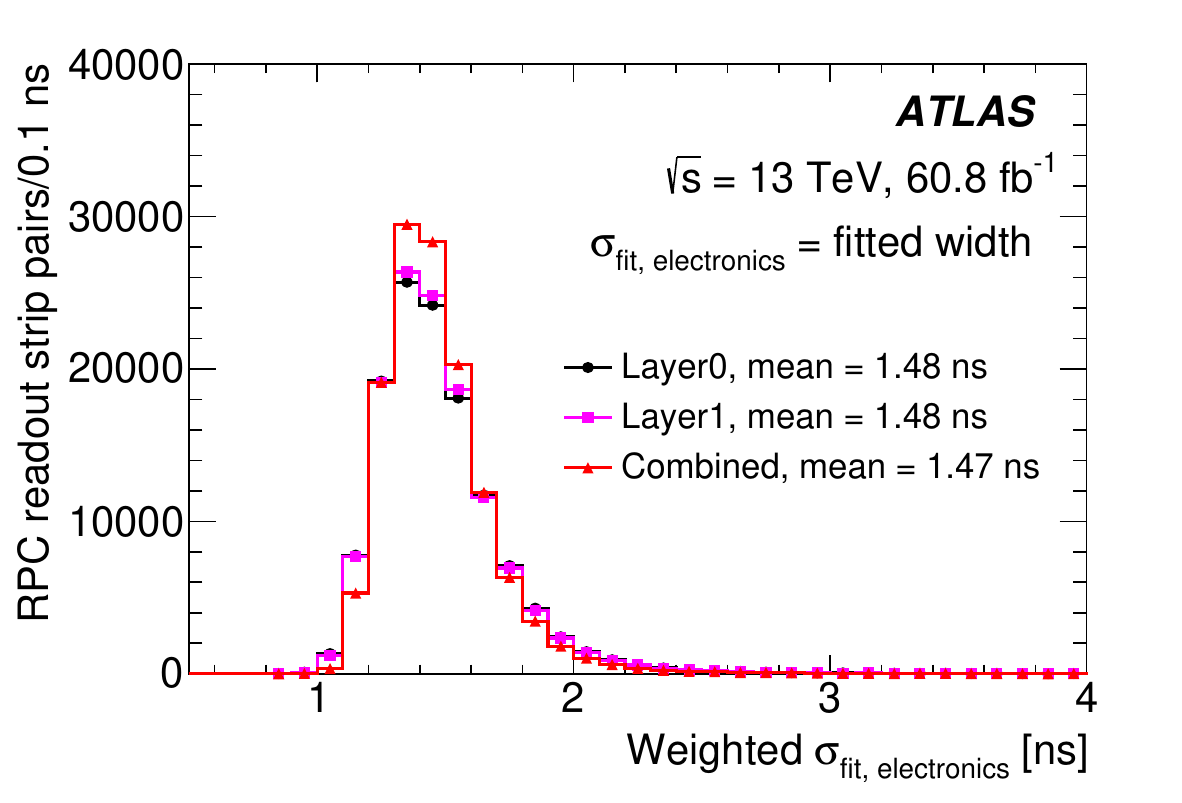}\label{pic:sigelec}}
\caption{(a) Time-difference distribution between signals generated by the same muon in a single pair of $\eta$ and $\phi$ strips, matched with the muon track, belonging to one detector layer. (b) Distribution of the electronic component of the time resolution.}
\end{figure}
 
The electronic component of the time resolution was estimated by taking the difference between time measurements of simultaneous $\eta$ and $\phi$ hits from the same avalanche event induced by a muon passing through a single detector layer. The intrinsic RPC resolution cancels out in this measurement because the same avalanche event is observed by a pair of $\eta$ and $\phi$ strips. Two different timing circuits were used to measure these two signals but since their clocks are synchronised to the same LHC reference clock, any differences in clock synchronisation would produce a shift of the mean of the time-difference distribution but leave its width unchanged.
 
The time-difference distribution was computed for each pair of orthogonal $\eta$ and $\phi$ strips observing the same avalanche event in the common gas volume. Delays due to different strip and cable lengths result in a constant offset of the mean value of this time-difference distribution. The width of this time-difference distribution is then proportional to the electronics time resolution. A binned maximum-likelihood fit to a Gaussian function was performed for each time-difference distribution. Only distributions with at least 20 entries were selected in order to remove strip pairs with low efficiency. An example fit is shown in Figure~\ref{pic:OneElecFit}. The KS test was performed to assess the goodness of fit. Only strip pair distributions with a KS probability greater than 0.4 and with at least three non-empty bins were retained for the final analysis, with the later criterion removing the majority of strip pairs. These selection criteria have a combined efficiency of about 75\%.
 
\begin{figure}[h]
\centering
\includegraphics[width=0.49\textwidth]{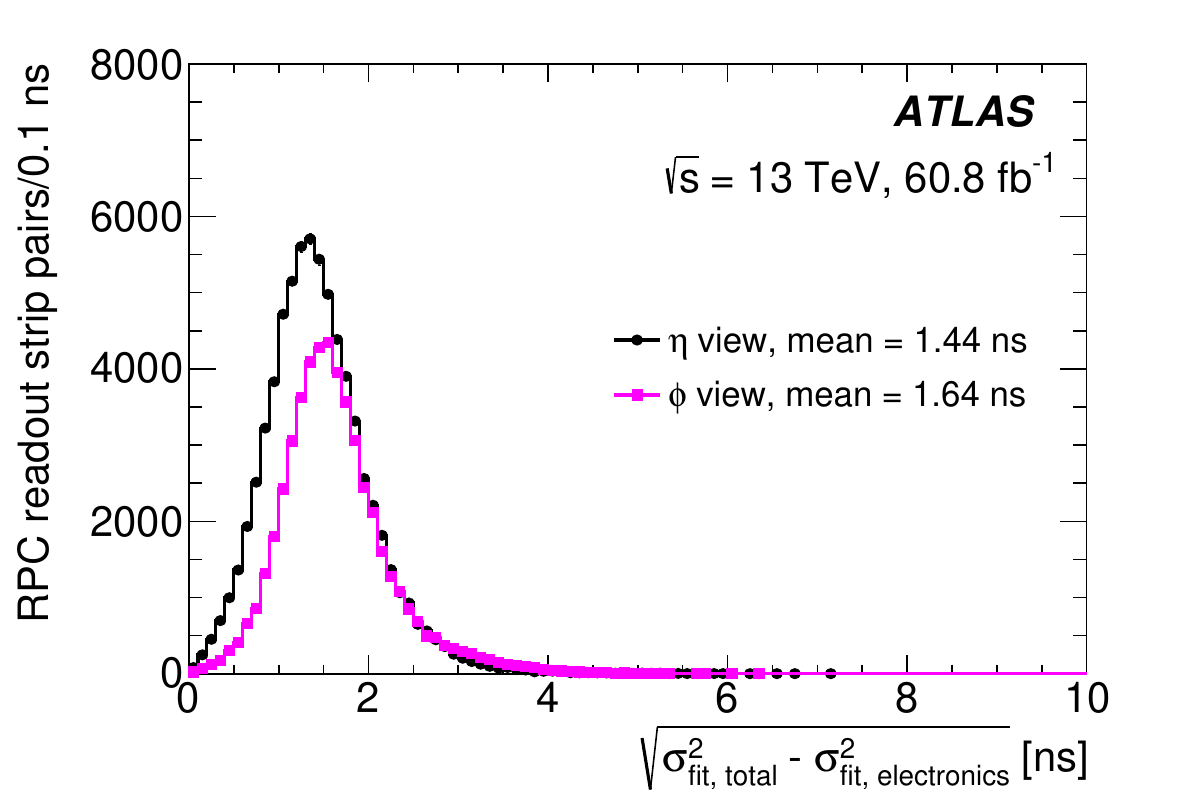}
\caption{Distribution of the intrinsic component of the time resolution for $\eta$ and $\phi$ panels.}
\label{pic:sigin}
\end{figure}
 
All possible combinations of $\eta$ and $\phi$ strip pairs were considered in each layer of each module for the full RPC system. For each $\eta$ ($\phi$) strip, several combinations with orthogonal $\phi$ ($\eta$) strips can be made. The electronics time resolution component associated with each single strip is therefore estimated as the statistically weighted average among all possible combinations with the orthogonal strips, where each individual measurement was weighted by the inverse of the square of the uncertainty in the fitted value of the Gaussian width parameter. Figure~\ref{pic:sigelec} shows the distribution of the best-fit Gaussian widths of the time-difference distributions for the full RPC system. The results for two layers for each module were combined for the estimate of the intrinsic time resolution component using the same weighting procedure, since both layers were used in the total time resolution measurements.
 
Finally, the intrinsic component of the time resolution was estimated using Eq.~(\ref{equation:total_time_resolution}). The resulting distributions are shown in Figure~\ref{pic:sigin} for all selected $\eta$ and $\phi$ strip pairs. The mean values of the distributions should be divided by $\sqrt{2}$ to extract the intrinsic time resolution of one RPC module. The mean value of the resulting intrinsic time resolution is consistent with the previous measurements obtained at test-beam facilities.

\FloatBarrier
 
\section{Performance of L1 muon barrel trigger}
\label{sec:trigger}
This section reports measurements of the L1 muon barrel trigger performance obtained using proton--proton collision data. Section~\ref{subsec:trigRoad} studies the performance of individual trigger towers. Section~\ref{subsec:trigEff} presents measurements of the L1 muon barrel trigger efficiency as a function of several quantities. The measurements in these two sections were performed using events containing a $Z$ boson decay into a pair of muons. Section~\ref{subsec:trigRate} presents measurements of the L1 muon trigger's event selection rate as a function of the instantaneous luminosity. Finally, Section~\ref{subsec:trigComp} studies the composition of the events selected by the L1 muon barrel trigger.
 
\subsection{Trigger roads}
\label{subsec:trigRoad}
Processing of signals (hits) produced by the RPC FE electronics is first performed by the on-detector electronics contained in the processor box (PAD)~\cite{Anulli:2009zz}. Each PAD contains four coincidence matrices, with each matrix implemented in one application-specific integrated circuit, referred to as CMA. These custom-designed circuits perform digital signal shaping, set a programmable dead time, mask channels and perform trigger logic operations. The CMA trigger logic aligns the FE signals in time, checks the time coincidence of RPC hits, and applies the geometrical matching criteria for selecting one of the three programmable $\pT$ thresholds. Two types of PADs are deployed: one is responsible for the low-$\pT$ trigger and another is responsible for the high-$\pT$ trigger. In each PAD, two CMAs collect the signals from the $\eta$ view strips and another two CMAs collect the signals from the $\phi$ view strips. Each RPC PAD covers an $\eta \times \phi$ detector region of approximately $0.2 \times 0.2$, with the overlap of one $\phi$ CMA and one $\eta$ CMA corresponding to a single RoI of approximately $0.1 \times 0.1$. Two PADs, one for the low-$\pT$ trigger and one for the high-$\pT$ trigger, make one trigger tower. A set of six, seven or eight trigger towers, placed along the $z$-axis at a fixed $\phi$ position, makes one trigger sector. There are in total 432 trigger towers, divided in 64 azimuthal trigger sectors, with 32 sectors on each ATLAS side.
 
CMAs identify muon candidates and measure their momentum by searching for geometrical matching of RPC hits inside programmable windows, called trigger roads, defined using the detector strips, as illustrated in Figure~3 of Ref.~\cite{Ciapetti:2007zz}. A muon candidate satisfies the L1 barrel trigger logic conditions if it generates RPC hits inside the trigger roads for a matching pair of $\eta$ and $\phi$ CMAs. The low-$\pT$ (high-$\pT$) trigger checks for coincident RPC signals between the pivot RPC2 and RPC1 (RPC3) doublet layers within the corresponding trigger road. The low-$\pT$ (high-$\pT$) trigger also requires signals in three out of four (one out of two) RPC detector layers, in both the $\eta$ and $\phi$ views. The trigger roads were defined to contain 95\% of  positively and negatively charged muons, simulated using the ATLAS simulation infrastructure~\cite{SOFT-2010-01} with a fixed $\pT$ value equal to the trigger $\pT$ threshold. Trigger roads encode the RPC detector layout, magnetic field configuration, and geometric relationships among strips in the different layers, as seen by a muon travelling from the interaction region.
 
\begin{table}[!h]
\caption{List of the hit selection criteria used to evaluate the performance of each individual CMA.}
\vspace{0.25cm}
\begin{center}
\begin{tabular}{lclcl}
\hline
Type        & CMA type & Selection criteria  \\
\hline
\specialrule{0em}{2pt}{1pt}
Muon kinematics    & Low $\pT$    & $\pT \geq 10$~GeV, $\left|\eta\right|\leq 1.05$\\
& High $\pT$   & $\pT \geq 20$~GeV, $\left|\eta\right|\leq 1.05$\\
\specialrule{0em}{2pt}{1pt}
\hline
\specialrule{0em}{2pt}{1pt}
Time               &                   & | $t_{\text{pivot layer channel}} - t_{\text{confirm layer channel}} | \leq 12.5$~ns \\
\specialrule{0em}{2pt}{1pt}
\hline
\specialrule{0em}{2pt}{1pt}
Layer              & Low $\pT$    & $N_{\text{pivot layers with hits}} + N_{\text{confirm layers with hits}} \geq 3$ \\
\specialrule{0em}{2pt}{1pt}
& High $\pT$   & $N_{\text{pivot layers with hits}} \geq 1$, $N_{\text{confirm layers with hits}} \geq 1$ \\
\specialrule{0em}{2pt}{1pt}
\hline
\specialrule{0em}{2pt}{1pt}
Hit multiplicity   & Low $\pT$    & $N_{\text{hits in pivot layer}} \leq 4$, $N_{\text{hits in confirm layer}} \leq 4$ \\
\specialrule{0em}{2pt}{1pt}
& High $\pT$   & $N_{\text{hits in pivot layer}} \leq 2$, $N_{\text{hits in confirm layer}} \leq 4$ \\
\specialrule{0em}{2pt}{1pt}
\hline
\end{tabular}
\end{center}
\label{table:TriggerRoadCut}
\end{table}
 
The performance of each individual CMA was evaluated using probe muons produced in decays of the $Z$ bosons. Probe muons were matched with the four closest CMAs by requiring the angular $\Delta R$ distance between the muon track and the centre of the CMA to be less than 0.15. The MU10 (MU20) trigger roads were evaluated using offline muon candidates with $\pT > 10~(20)$~GeV. Signals in the confirm layer were required to be within the 25~ns time window centred at the pivot signal time, thereby requiring the selected signals to belong to the same bunch crossing. To reduce contributions from background events due to random coincidences, at least three out of four layers of RPC1 and RPC2 were required to contain signals in the selected low-$\pT$ CMA. Similarly, at least one of the two layers of RPC3 was required to contain signals in the selected high-$\pT$ CMA. These two selection criteria approximate similar conditions applied by the on-detector PAD electronics~\cite{Anulli:2009zz}. Finally, events with more than four signals in one detector layer were removed to reduce the number of combinations between the pivot and confirm layers. These selection criteria are summarised in Table~\ref{table:TriggerRoadCut}.
 
\begin{figure}[!h]
\centering
\subfloat[]{\includegraphics[width=0.49\textwidth]{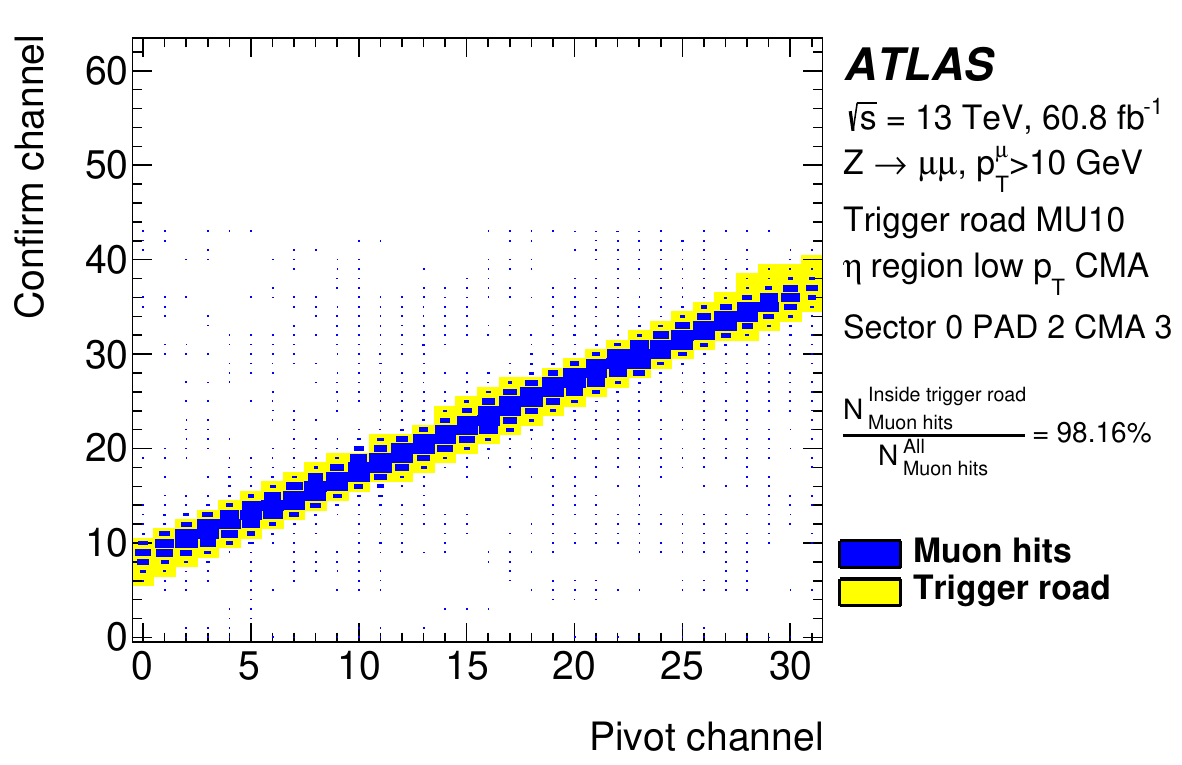}}
\subfloat[]{\includegraphics[width=0.49\textwidth]{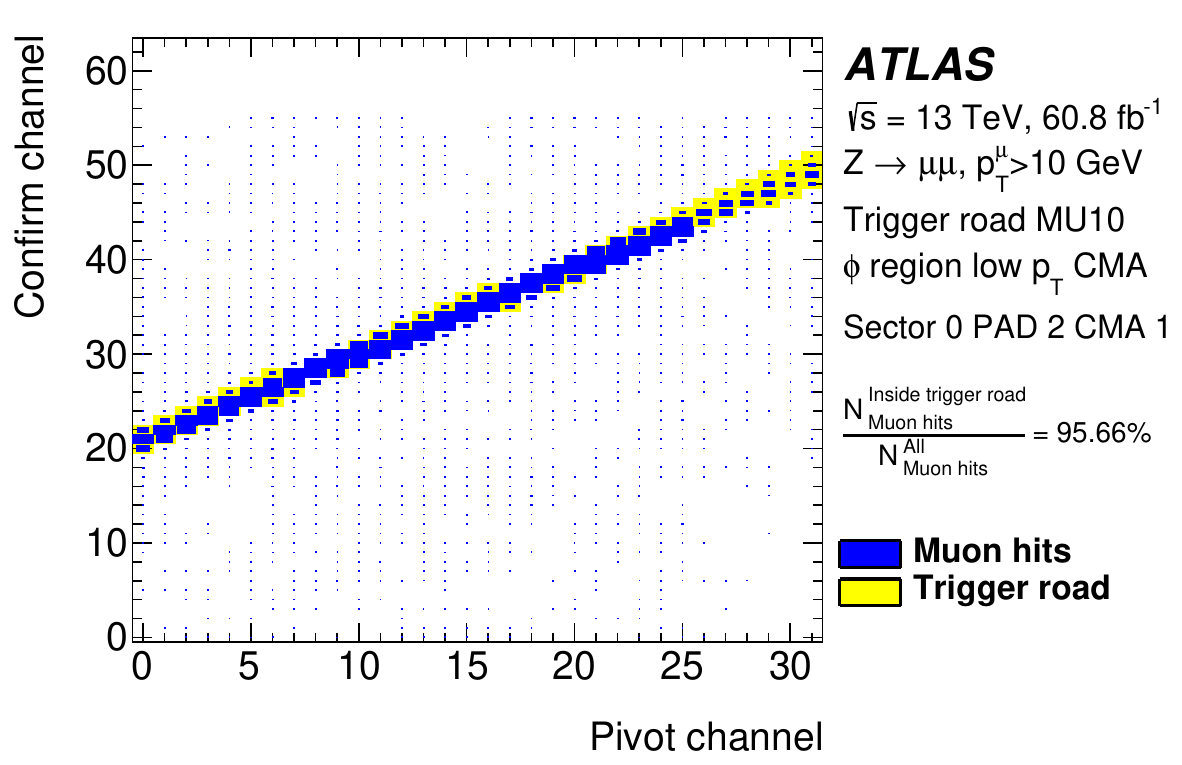}}
\\
\subfloat[]{\includegraphics[width=0.49\textwidth]{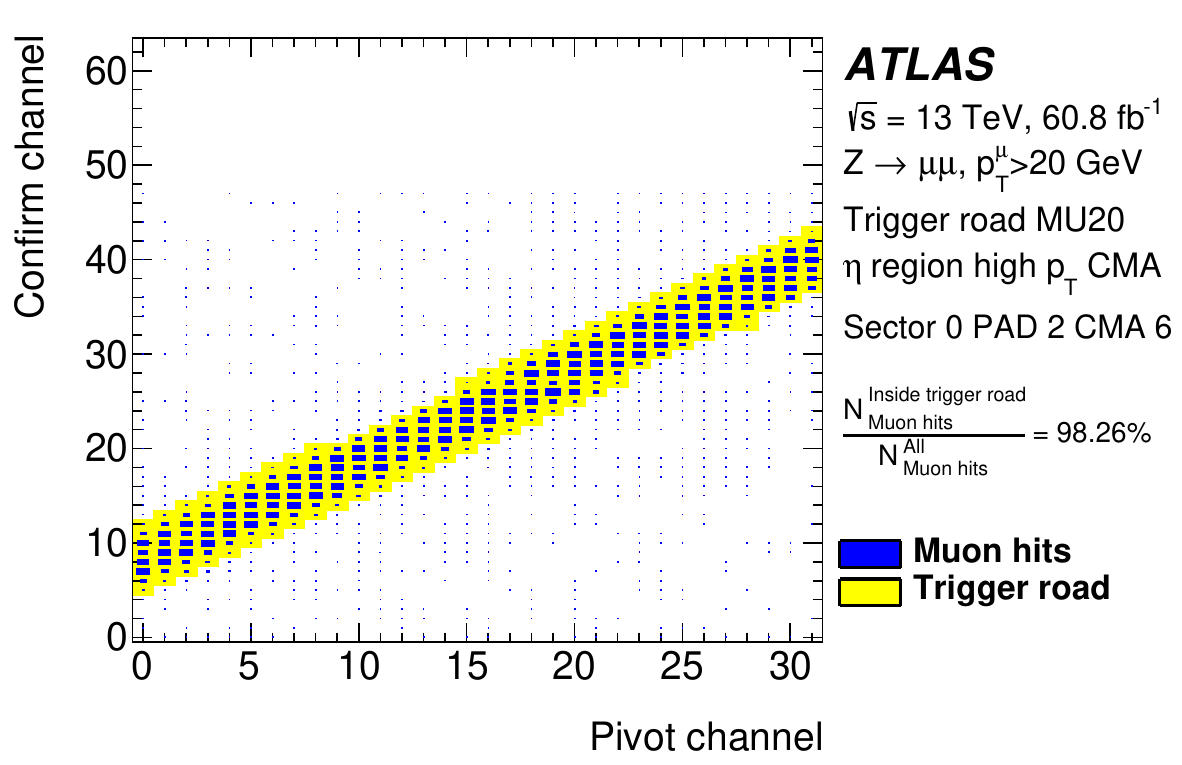}}
\subfloat[]{\includegraphics[width=0.49\textwidth]{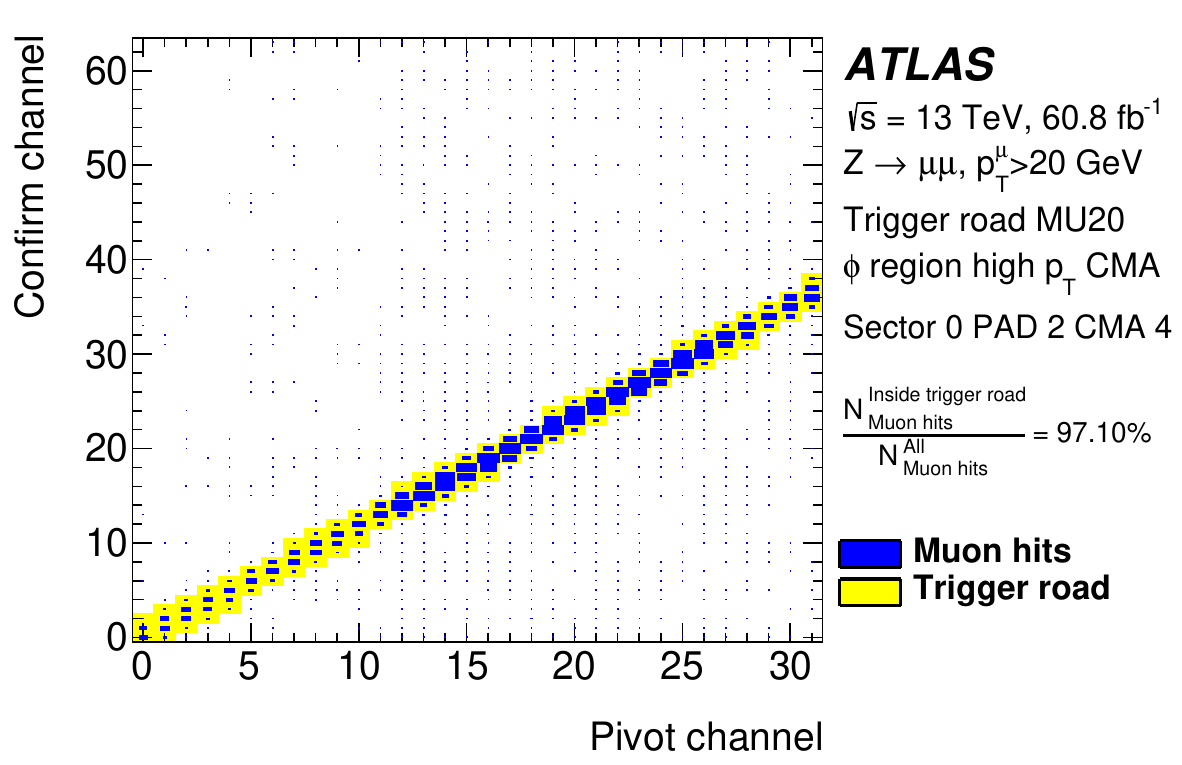}}
\caption{L1 muon barrel trigger roads for four example CMAs: (a) MU10 low-$\pT$ $\eta$ CMA, (b) MU10 low-$\pT$ $\phi$ CMA, (c) MU20 high-$\pT$ $\eta$ CMA, (d) MU20 high-$\pT$ $\phi$ CMA.}
\label{pic:triggerRoadCMA}
\end{figure}
 
Figure~\ref{pic:triggerRoadCMA} shows the selected RPC hits in the pivot and confirm layers for four representative CMAs belonging to the same PAD: low- and high-$\pT$ CMAs in both the $\eta$ and $\phi$ views. For each hit in the pivot layer, all possible combinations with the selected hits in the confirm layer are reported. The trigger roads used for 2018 data-taking are shown in yellow. As expected, the majority of hits are produced by the probe muons and are contained within the trigger roads.
 
\begin{figure}[!h]
\centering
\subfloat[]{\includegraphics[width=0.49\textwidth]{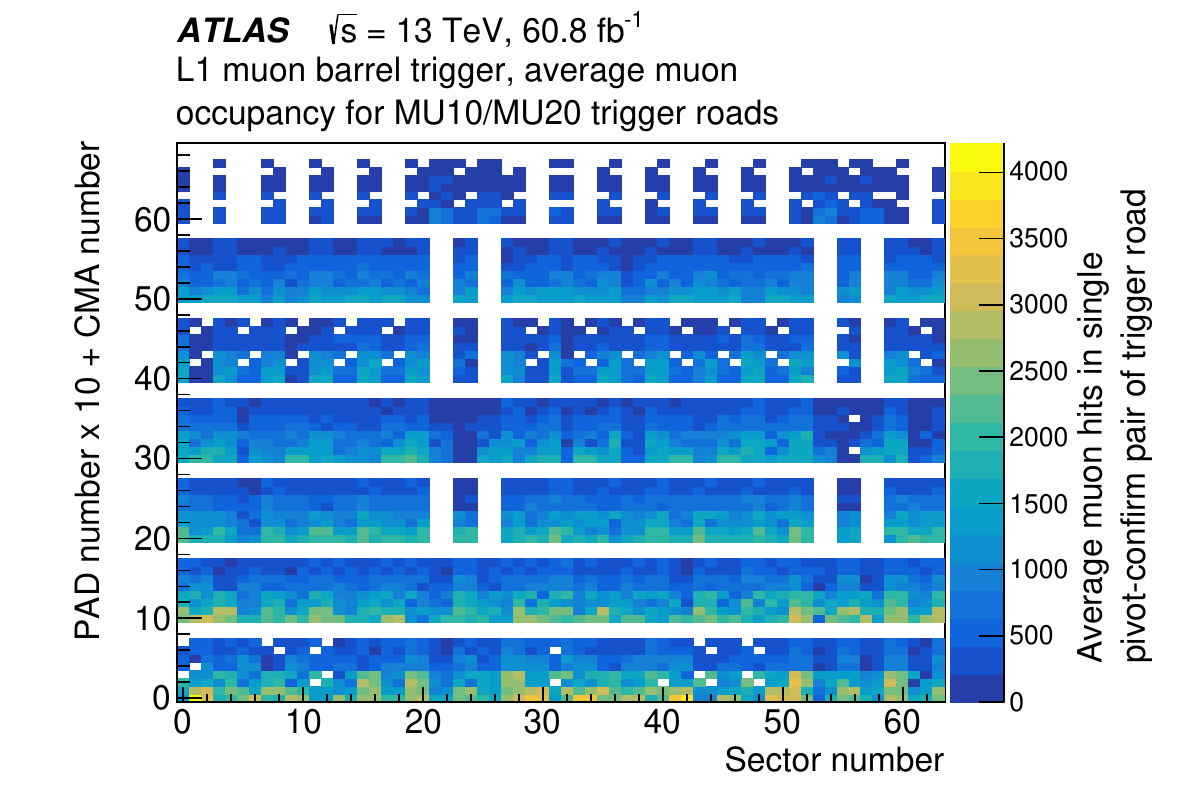}\label{pic:triggerRoadAvgHit2D}}
\subfloat[]{\includegraphics[width=0.49\textwidth]{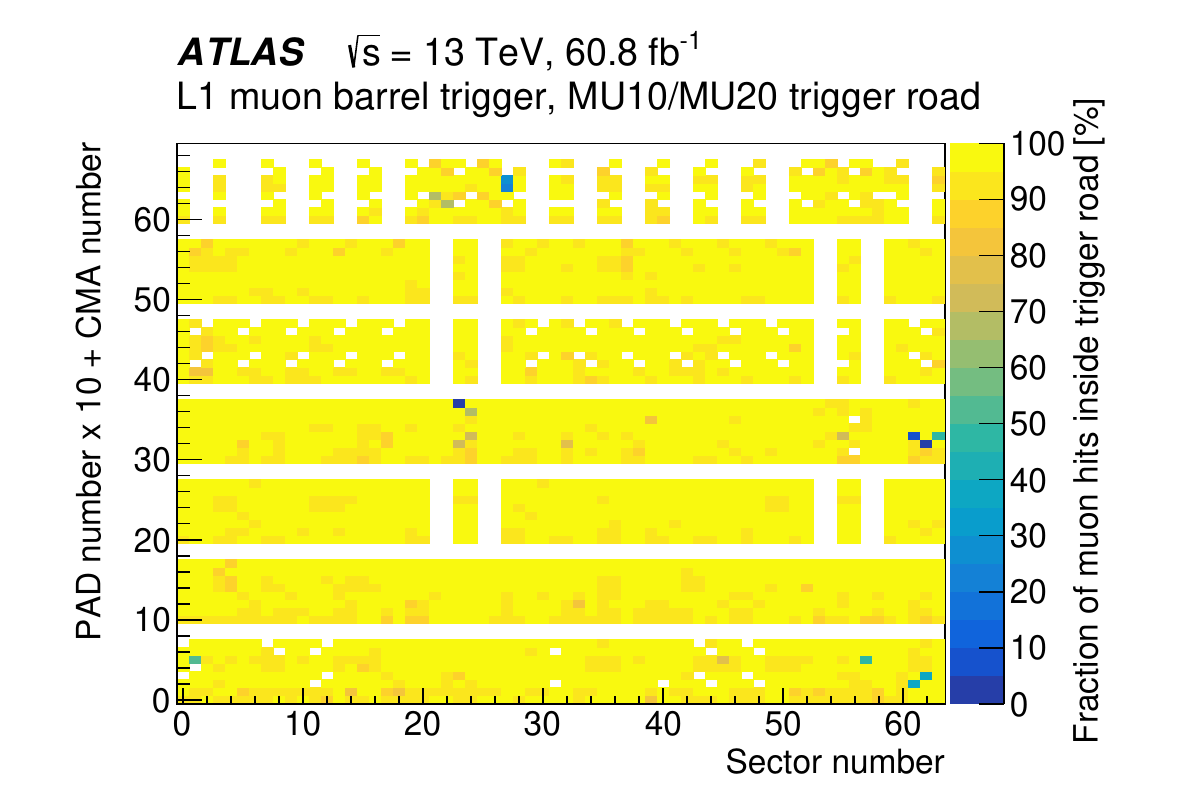}\label{pic:triggerRoadFrac2D}}
\caption{(a) Average number of the probe muons passing through each CMA and (b) fraction of the selected RPC hits inside the trigger road for each CMA. The horizontal axis corresponds to the trigger sector number. The vertical axis shows the PAD number multiplied by 10 plus the CMA number. Low-$\pT$ CMA numbers range from 0 to 3 and high-$\pT$ CMA numbers range from 4 to 7. The trigger roads employed for this study correspond to the MU10 and MU20 triggers for the low-$\pT$ and high-$\pT$ CMAs, respectively.}
\label{pic:triggerRoadWorkStatus}
\end{figure}
 
The fractions of the selected RPC hit pairs inside the trigger roads (referred to as trigger road hit fractions) were computed for all CMAs in order to study the trigger road performance using the actual detector. Figures~\ref{pic:triggerRoadAvgHit2D} and \ref{pic:triggerRoadFrac2D} show the number of probe muons passing through each CMA and the trigger road hit fractions for each CMA, respectively. Typically, several hundred probe muons pass through each CMA, providing sufficient precision to study the performance of the trigger roads.
 
\begin{figure}[!h]
\centering
\subfloat[]{\includegraphics[width=0.49\textwidth]{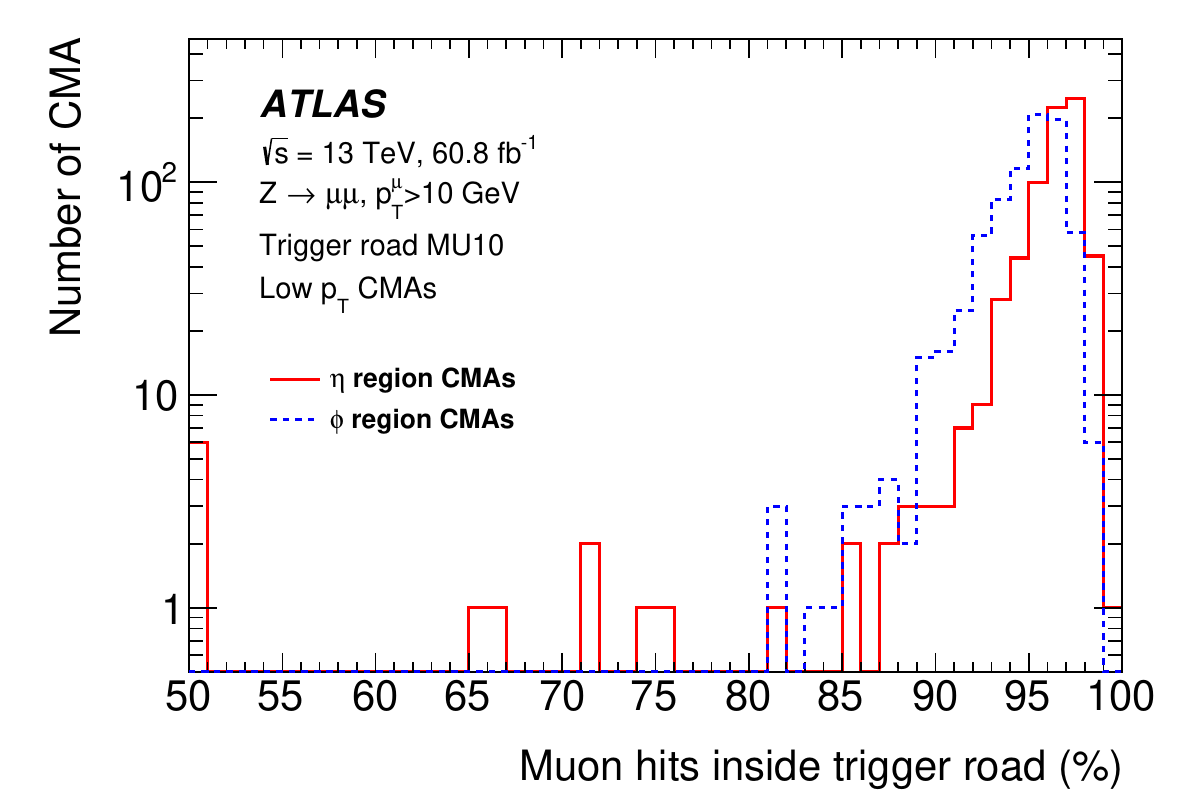}\label{pic:triggerRoadFracMU10}}
\subfloat[]{\includegraphics[width=0.49\textwidth]{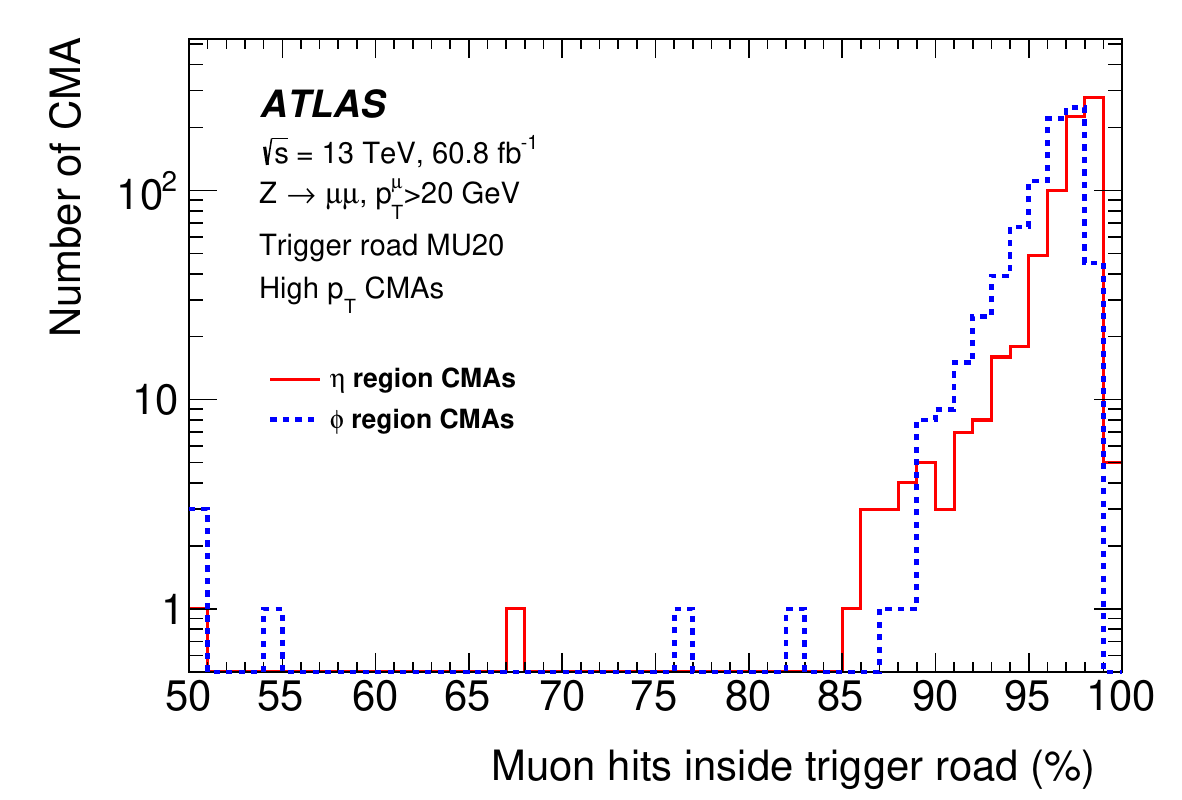}\label{pic:triggerRoadFracMU20}}
\caption{Distributions of the fraction of selected RPC hits inside the trigger road for the (a) MU10 and (b) MU20 triggers, shown separately for $\eta$ and $\phi$ CMAs.}
\label{pic:triggerRoadFrac1D}
\end{figure}
 
Figure~\ref{pic:triggerRoadFrac1D} shows the distributions of the trigger road hit fractions for the MU10 and MU20 triggers for all CMAs. In events with probe muons passing through the CMA, more than $95\%$ of the selected pivot and confirm hit pairs are contained within the trigger road. The trigger road hit fractions are larger than 95\% for the majority of CMAs, indicating good performance of the MU10 and MU20 triggers. These fractions are slightly larger for the $\eta$ CMAs than for the $\phi$ CMAs because the trigger roads were defined to contain 95\% of muons with $\pT=20$~\GeV\ while the majority of the probe muons have $\pT$ between 30 and 50~\GeV. Therefore, probe muon trajectories in the bending $\eta$ view have a slighter higher probability to fall within the trigger road than in the non-bending $\phi$ view. Approximately 2\% of $\eta$ CMAs and 4\% of $\phi$ CMAs have trigger road hit fractions below 90\%. These lower fractions are due to residual differences between the actual and simulated detector geometries.
 
\FloatBarrier
 
\subsection{Trigger efficiency and timing}
\label{subsec:trigEff}
\begin{figure}[!h]
\centering
\subfloat[]{\includegraphics[width=0.49\textwidth]{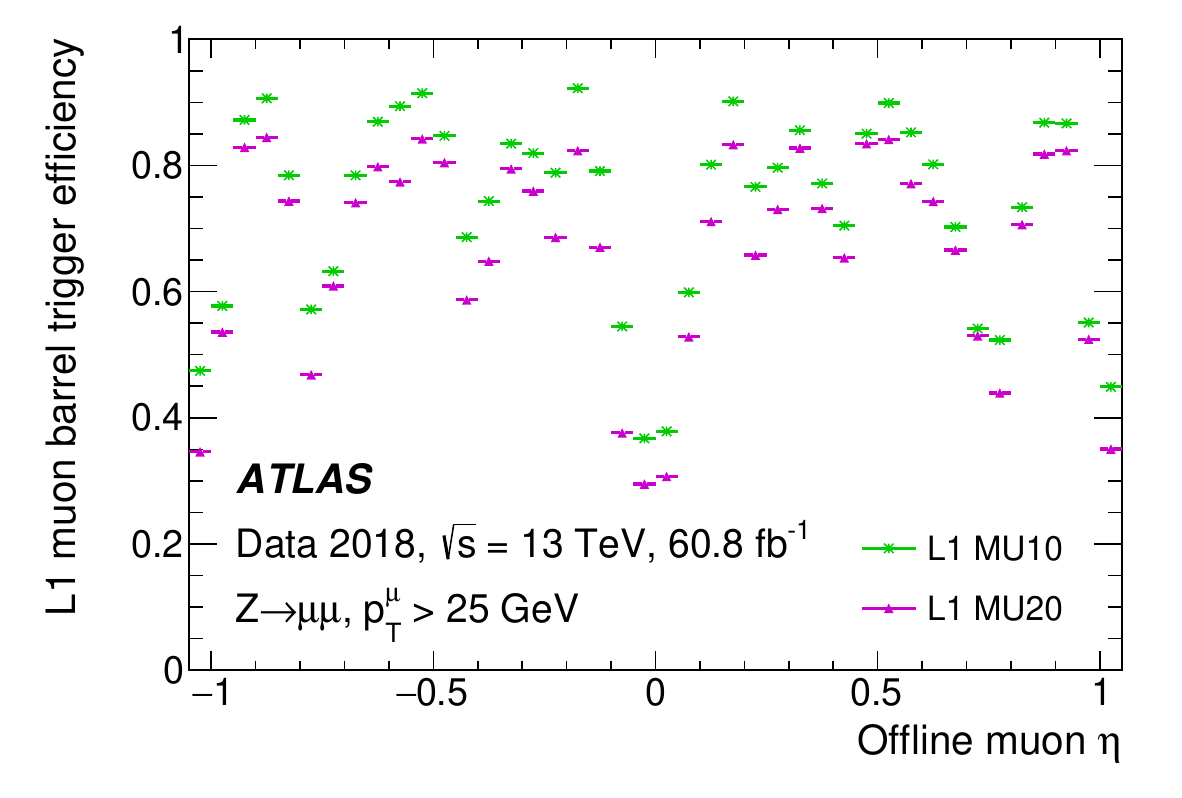}\label{pic:triggerEffEta}}
\subfloat[]{\includegraphics[width=0.49\textwidth]{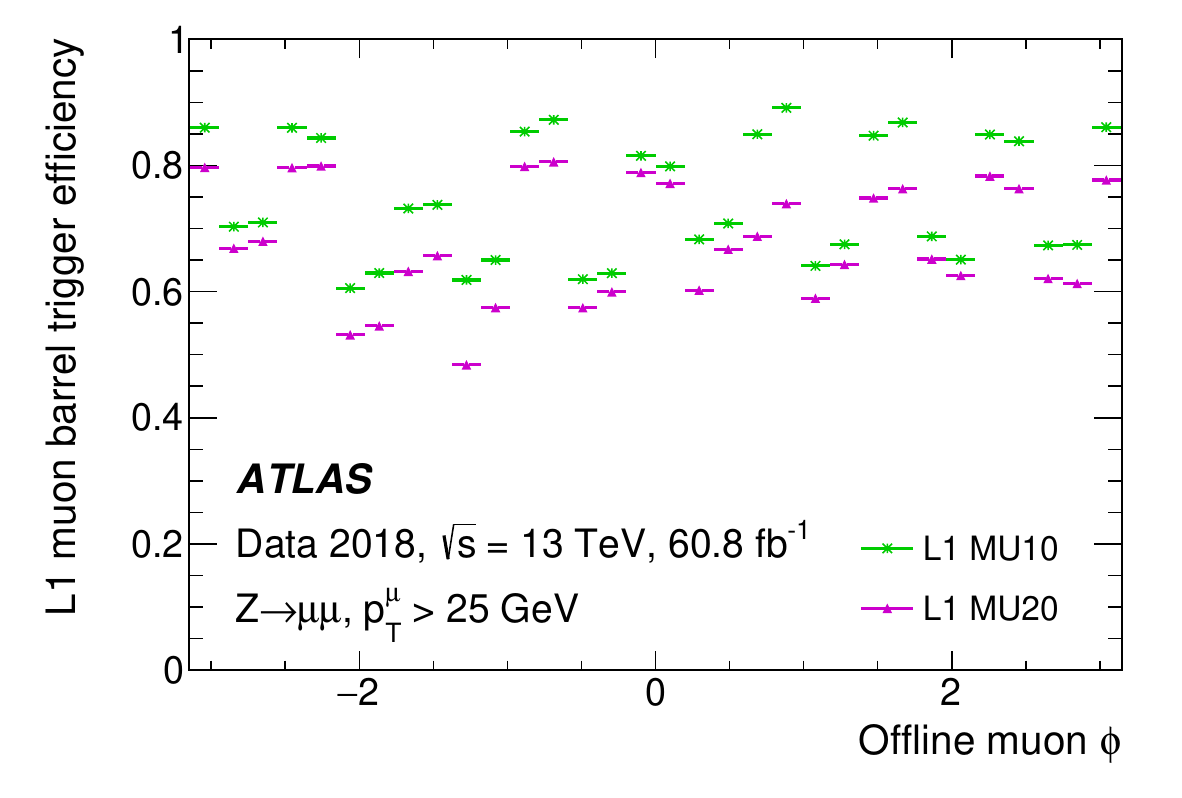}\label{pic:triggerEffPhi}}
\caption{L1 muon barrel trigger efficiency as a function of (a) muon pseudorapidity and (b) muon azimuthal angle.}
\label{pic:triggerEff1d}
\end{figure}

The efficiency of the L1 muon barrel trigger to detect a probe muon was evaluated as a function of several parameters. Figure~\ref{pic:triggerEff1d} shows the trigger efficiency as a function of probe muon $\eta$ and $\phi$ for the MU10 and MU20 triggers. Only muons with $\pT > 25$~\GeV\ were used for these measurements; this removes the dependence of the trigger efficiency on the muon $\pT$. Regular features in the trigger efficiency distributions correspond to the ATLAS detector support structures and service elements. The drop in the trigger efficiency at $\eta \sim 0$ in Figure~\ref{pic:triggerEffEta} is due to the presence of detector services, such as gas pipes, water pipes, cryogenic lines, and cables, which cause the region $|\eta|<0.1$ to be only sparsely instrumented with RPCs. Other, smaller efficiency drops, mainly at $|\eta|\sim0.4$ and $|\eta|\sim0.8$, are due to the presence of the detector support and service structures. The eight barrel magnet coils are located in the small RPC sectors and their presence is visible in Figure~\ref{pic:triggerEffPhi}, where a regular eightfold structure can be observed.
 
\begin{figure}[!h]
\centering
\subfloat[]{\includegraphics[width=0.49\textwidth]{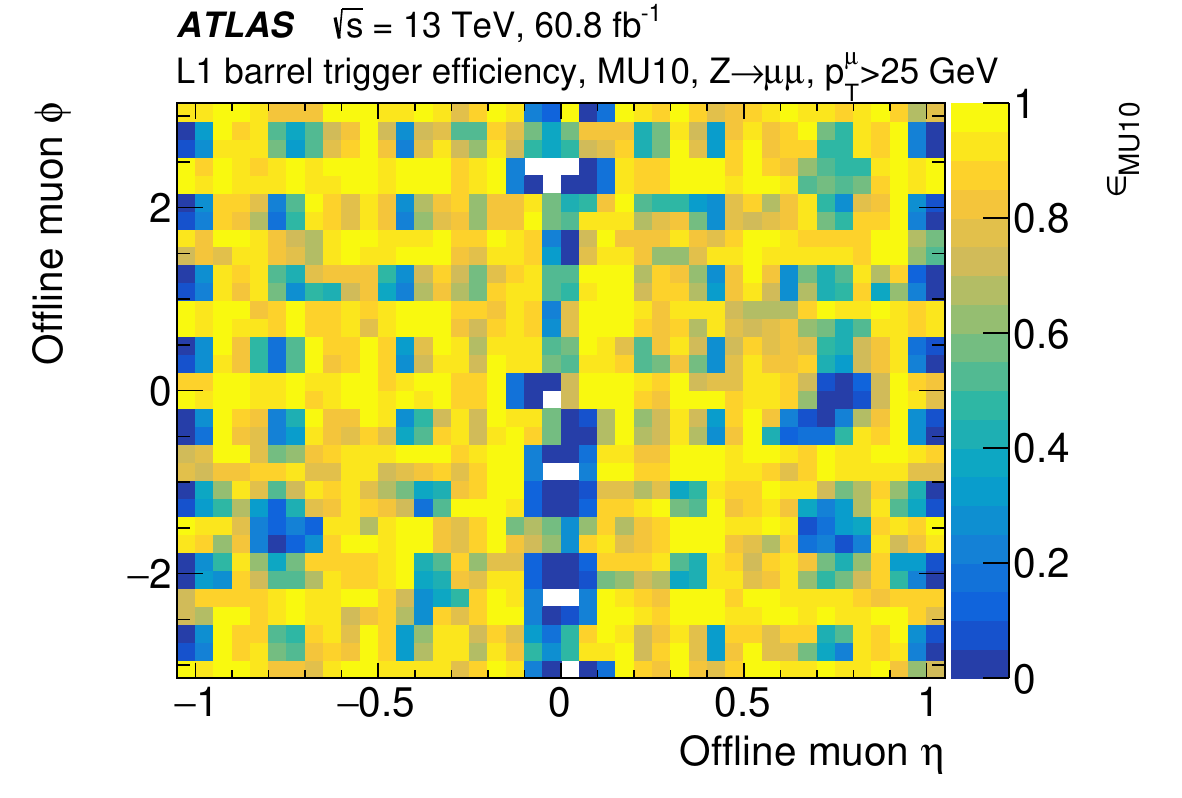}\label{pic:triggerEff2dMU10}}
\subfloat[]{\includegraphics[width=0.49\textwidth]{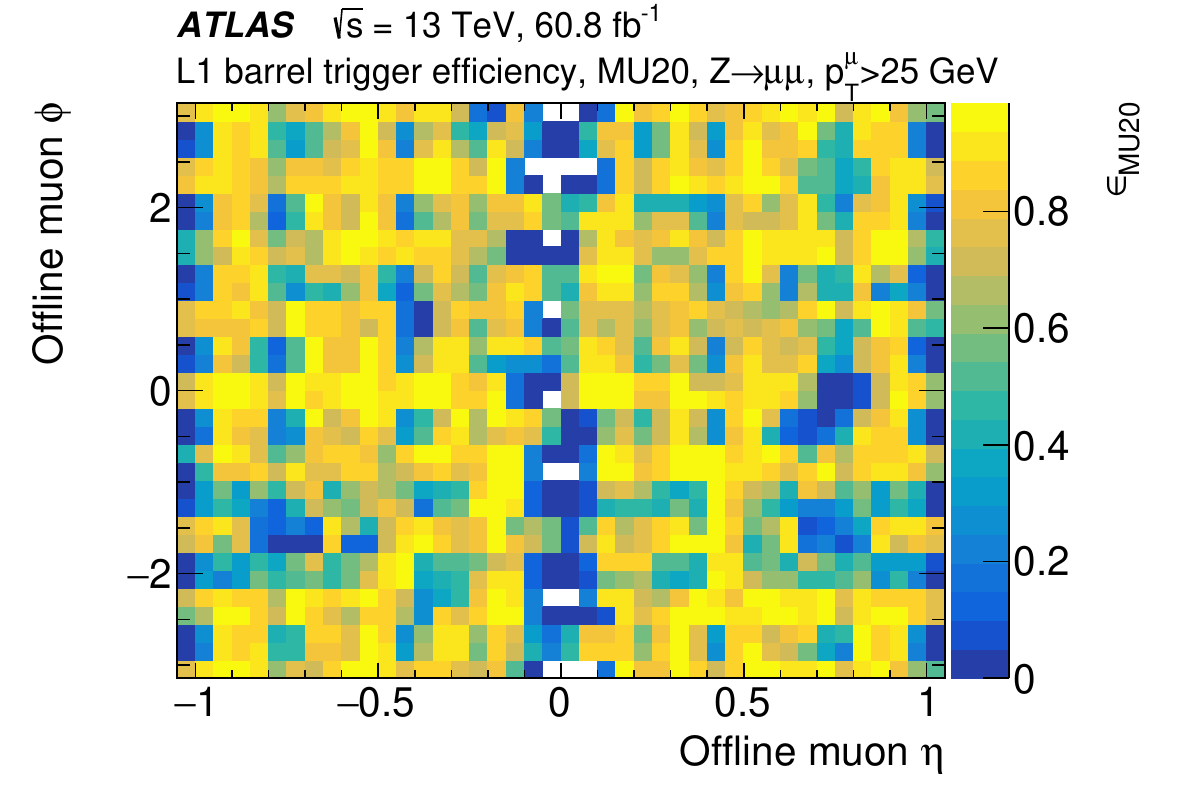}\label{pic:triggerEff2dMU20}}
\caption{L1 muon barrel trigger efficiency as a function of the muon pseudorapidity and azimuthal angle for the (a) MU10 and (b) MU20 triggers.}
\label{pic:triggerEff2d}
\end{figure}
 
\begin{figure}[!h]
\centering
\subfloat[]{\includegraphics[width=0.49\textwidth]{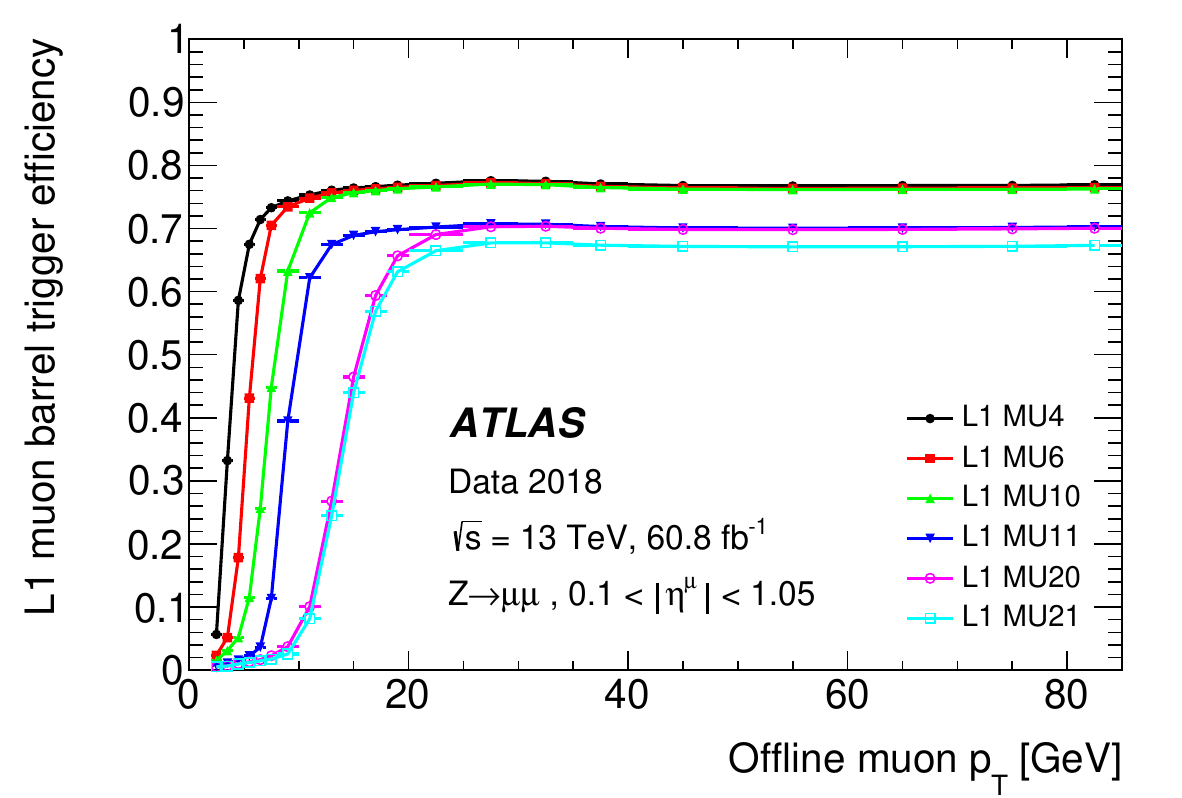} \label{pic:triggerEffPt}}
\subfloat[]{\includegraphics[width=0.49\textwidth]{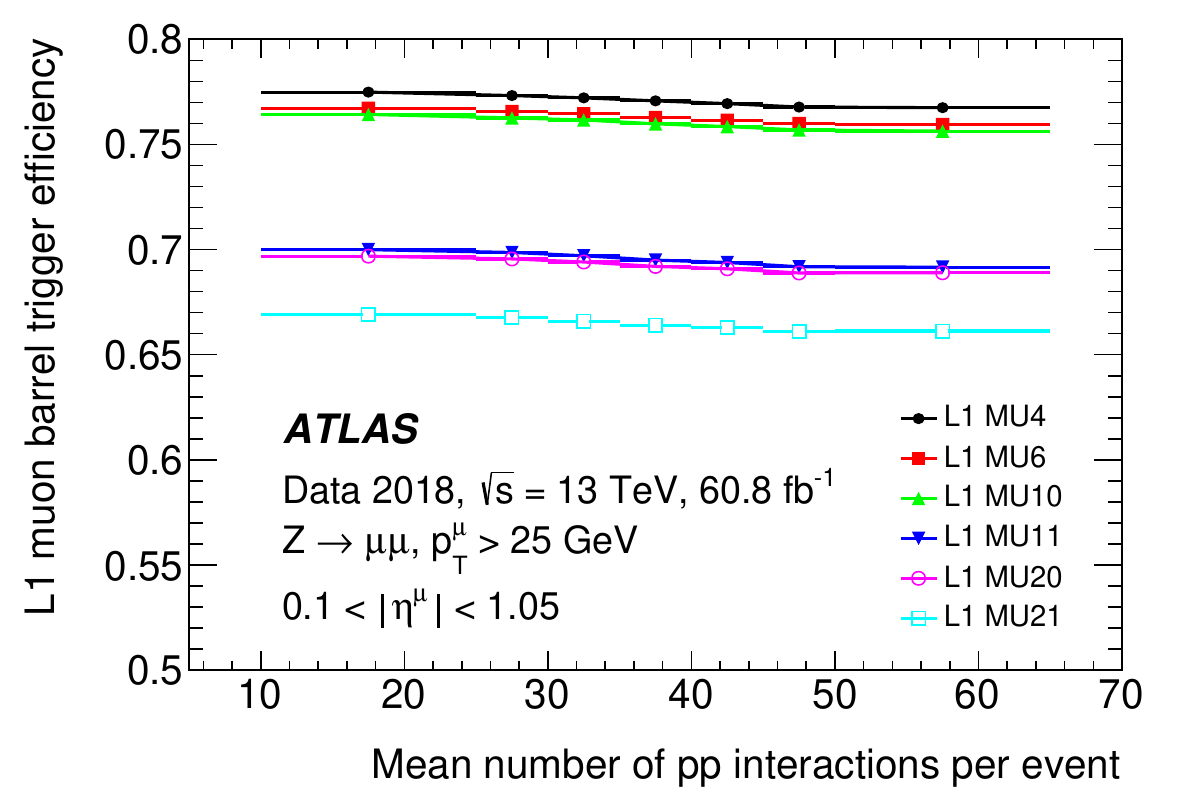} \label{pic:triggerEffMu1}}
\caption{(a) L1 muon barrel trigger efficiency plotted as a function of the probe muon $\pT$. (b) Overall L1 muon barrel trigger efficiency as a function of the mean number of proton--proton interactions per event.}
\label{pic:triggerEffs}
\end{figure}

The overall efficiency of the L1 muon barrel trigger system was measured as a function of the probe muon $\pT$ for six thresholds, as shown in Figure~\ref{pic:triggerEffPt} and quantified in Table~\ref{tab:trig_eff}. This overall efficiency includes the geometrical acceptance effects illustrated in Figure~\ref{pic:triggerEff2d}, the RPC efficiency to detect a muon signal, and the efficiency of trigger logic conditions. Probe muons falling in the region $|\eta|<0.1$ were removed from the analysis because this region is not instrumented with RPCs. The three low-$\pT$ triggers (MU4, MU6, MU10) require a geometrical and timing coincidence of at least three $\eta$ hits and three $\phi$ hits in the four innermost RPC layers. The three high-$\pT$ triggers (MU11, MU20, MU21) require that the highest threshold of the three low-$\pT$ triggers is satisfied by the muon candidate. They also require that the low-$\pT$ muon candidate is matched with at least one $\eta$ and one $\phi$ hit in the two outer RPC layers. This additional requirement reduces the $\sim 76\%$ efficiency of the MU10 trigger to approximately 70\% for the MU20 trigger.
 
The MU20 trigger was the primary L1 muon trigger for data-taking in 2015--2018~\cite{Aad:2020uyd}. The MU20 barrel trigger selects muon candidates using the entire RPC detector, including the new chambers in the ATLAS feet region which were commissioned during 2015 and 2016. The new feet trigger chambers were not included in the MU21 trigger logic, as demonstrated by the 3\% higher efficiency of the MU20 trigger in Figure~\ref{pic:triggerEffPt} compared to that of MU21. In the feet region, the trigger logic for the high-$\pT$ thresholds is similar to the low-$\pT$ trigger logic because only two RPC doublet layers are traversed by a muon produced at the collision point. This leads to a higher overall trigger rate and to a higher fraction of trigger candidates due to background events, as discussed in Sections~\ref{subsec:trigRate} and~\ref{subsec:trigComp}. The MU21 trigger was included as a backup in case the event acceptance rate by the MU20 trigger exceeded the allowed limit.
 
The overall efficiency of the L1 muon barrel trigger system was measured as a function of the mean number of proton--proton interactions per event using probe muons with $\pT>25$~\GeV, as shown in Figure~\ref{pic:triggerEffMu1}. The barrel trigger efficiency decreases by about 1\% between the lowest and highest instantaneous luminosity values recorded in 2018. This decrease corresponds to a similar decrease observed for the RPC detector, as discussed in Section~\ref{subsec:det_eff}.

\begin{table}[!h]
\centering
\caption{L1 muon barrel trigger efficiency in each $\pT$ bin of Figure~\ref{pic:triggerEffPt} for different L1 trigger thresholds. The statistical uncertainty for all table entries is smaller than 0.02\%.}
\vspace{0.2cm}
\resizebox{\textwidth}{!}{
\begin{tabular}{lccccccccccc}
\hline
Threshold &2--3\,\GeV &3--4\,\GeV &4--5\,\GeV &5--6\,\GeV &6--7\,\GeV &7--8\,\GeV &8--10\,\GeV &10--12\,\GeV &12--14\,\GeV &14--16\,\GeV &16--18\,\GeV \\
\hline
L1 MU4    & 5.7\%~~ & 33.2\%~~ & 58.6\%~~ & 67.5\%~~ & 71.4\%~~ & 73.3\%~~ & 74.5\%~~ & 75.3\%  & 76.1\% & 76.4\% & 76.6\% \\
\hline
L1 MU6    & 2.35\%   & 5.2\%       & 17.8\%~~ & 43.1\%~~ & 62.1\%~~ & 70.5\%~~ & 73.4\%~~ & 74.8\%  & 75.7\% & 76.1\% & 76.3\% \\
\hline
L1 MU10  & 1.54\%   & 3.1\%       & 5.2\%       & 11.6\%~~  & 25.7\%~~ & 44.9\%~~ & 63.3\%~~ & 72.5\%  & 74.9\% & 75.7\% & 76.0\% \\
\hline
L1 MU11  & 0.72\%   & ~~1.10\% & ~~1.58\% & ~~2.25\% & 3.6\%       & 11.4\%~~  & 39.5\%~~ & 62.2\%  & 67.5\% & 68.9\% & 69.5\% \\
\hline
L1 MU20  & 0.54\%   & ~~0.68\% & ~~1.04\% & ~~1.31\% & ~~1.66\% & ~~2.28\% & 3.7\%       & 10.1\%   & 26.8\% & 46.5\% & 59.4\% \\
\hline
L1 MU21  & 0.54\%   & ~~0.65\% & ~~0.98\% & ~~1.21\% & ~~1.47\% & ~~1.75\% & ~~2.56\% & ~~8.1\% & 24.5\% & 44.0\% & 56.9\% \\
\hline
\end{tabular}
}
 
\vspace{0.7cm}
 
\resizebox{\textwidth}{!}{
\begin{tabular}{lcccccccccc}
\hline
Threshold &18--20\,\GeV & 20--25\,\GeV & 25--30\,\GeV & 30--35\,\GeV & 35--40\,\GeV & 40--50\,\GeV & 50--60\,\GeV & 60--70\,\GeV & 70--80\,\GeV & 80--85\,\GeV \\
\hline
L1 MU4    & 76.9\% & 77.2\% & 77.6\% & 77.5\% & 77.1\% & 76.8\% & 76.8\% & 76.8\% & 76.8\% & 76.9\% \\
\hline
L1 MU6    & 76.5\% & 76.8\% & 77.2\% & 77.1\% & 76.7\% & 76.4\% & 76.4\% & 76.4\% & 76.4\% & 76.5\% \\
\hline
L1 MU10   & 76.3\% & 76.6\% & 77.1\% & 77.0\% & 76.5\% & 76.3\% & 76.2\% & 76.2\% & 76.2\% & 76.3\% \\
\hline
L1 MU11   & 69.8\% & 70.2\% & 70.7\% & 70.7\% & 70.3\% & 70.1\% & 70.0\% & 70.1\% & 70.2\% & 70.3\% \\
\hline
L1 MU20   & 65.7\% & 69.0\% & 70.3\% & 70.4\% & 70.1\% & 69.9\% & 69.8\% & 69.9\% & 69.9\% & 70.1\%	 \\
\hline
L1 MU21   & 63.1\% & 66.5\% & 67.7\% & 67.7\% & 67.4\% & 67.2\% & 67.1\% & 67.1\% & 67.2\% & 67.3\% \\
\hline
\end{tabular}
}
\label{tab:trig_eff}
\end{table}
 
\begin{figure}[!h]
\centering
\subfloat[]{\includegraphics[width=0.49\textwidth]{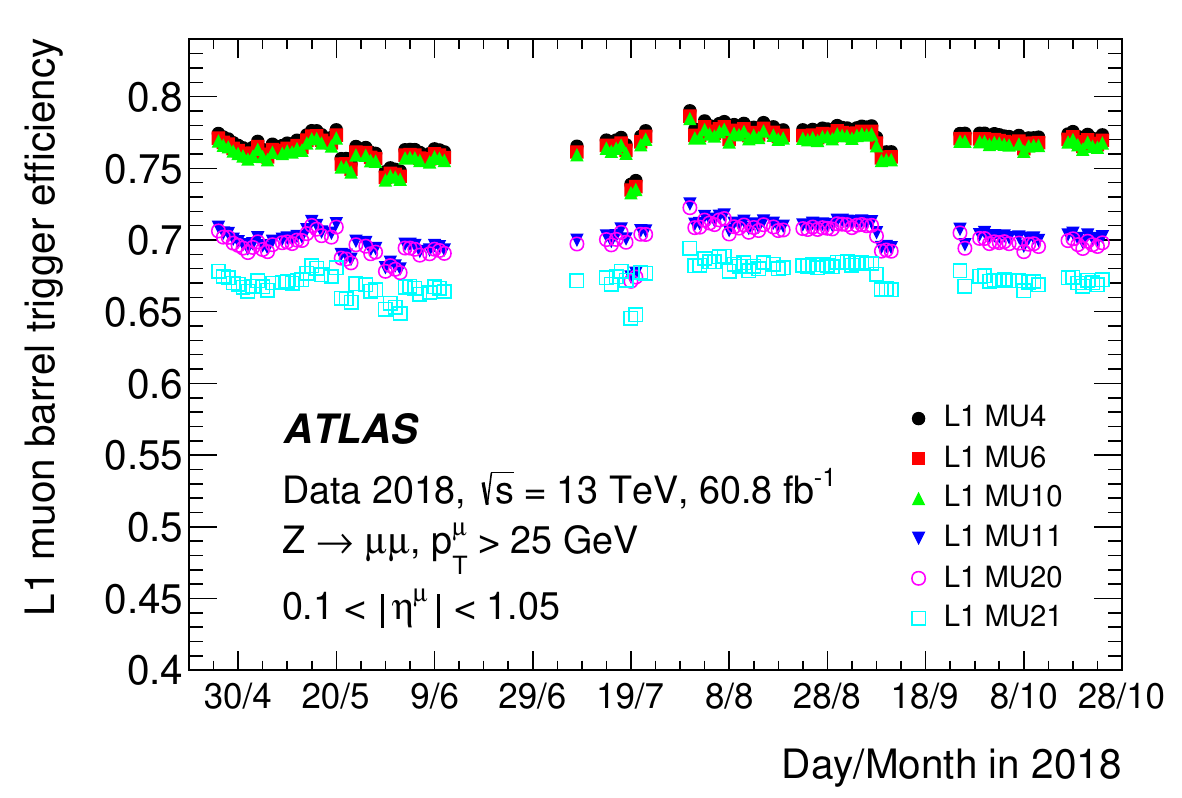} \label{pic:triggerEffTime}}
\subfloat[]{\includegraphics[width=0.49\textwidth]{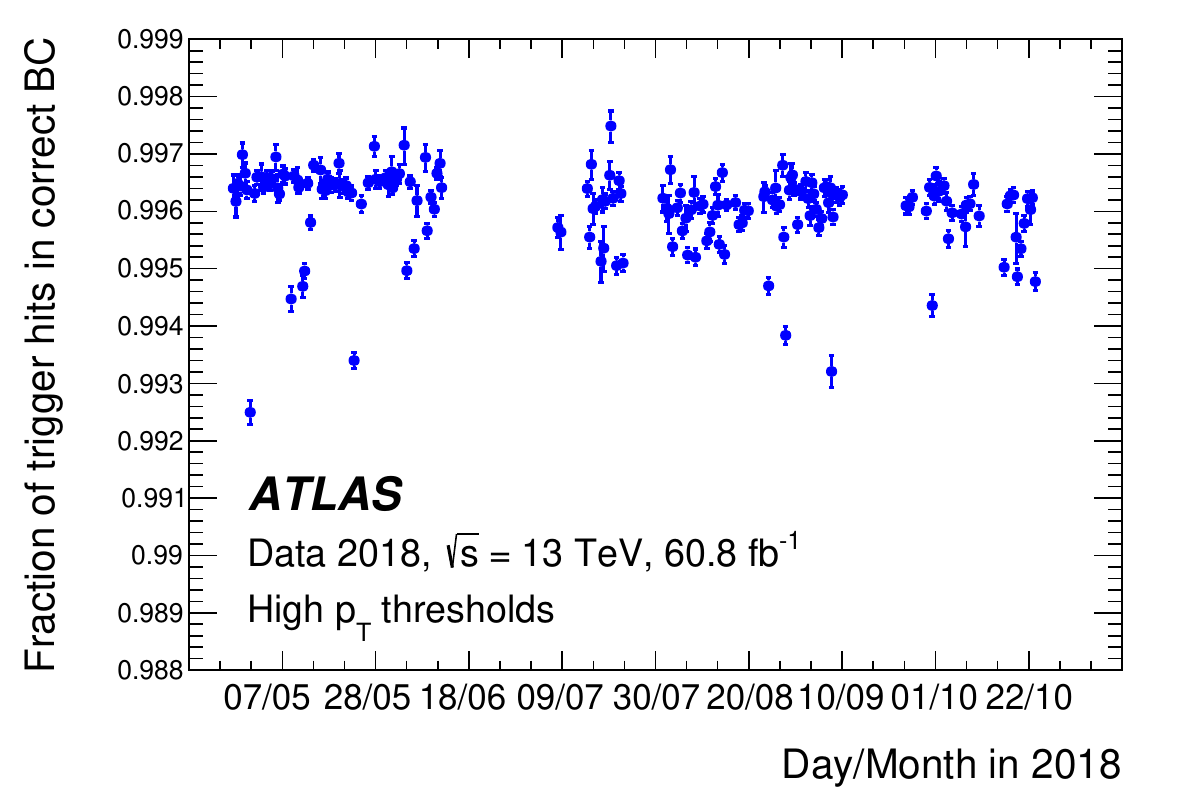} \label{pic:fbc0vstime}}
\caption{(a) The overall measured L1 muon barrel trigger efficiency as a function of time. (b) The fraction of high-$\pT$ trigger hits associated with the correct bunch crossing as a function of time. Each point corresponds to a separate run recorded by the ATLAS experiment in 2018.}
\end{figure}

The overall efficiency of the L1 muon barrel trigger was measured in individual runs recorded by the ATLAS experiment in 2018. The efficiency to detect muons with $\pT>25$~GeV is plotted as a function of time in Figure~\ref{pic:triggerEffTime} for all six trigger thresholds. Each point corresponds to a separate ATLAS run, where only runs with an integrated luminosity greater than 50~pb$^{-1}$ were included. The small variations in the trigger efficiency between individual runs were due to changes in detector conditions. For example, a small fraction of the detector elements may have been disabled in a given run due to transient problems with the RPC modules or with the data acquisition system of the RPC detector. The removed elements, which impact the trigger efficiency by at most 4\%, were usually recovered and included in the next run.
 
The assignment of an L1 muon trigger candidate to the correct bunch crossing requires accurate timing calibration of the RPC readout system. These timing calibrations~\cite{Anulli:2009zz} correct for the different signal propagation times of different RPC channels. These calibrations were initially performed in 2010--2012 using timing corrections in steps of 3.125~ns. The stability of the timing calibrations was monitored by measuring the fraction of hits belonging to trigger candidates that were associated with the correct bunch crossing. This fraction, averaged over the entire detector, is plotted in Figure~\ref{pic:fbc0vstime} as a function of time for hits associated with high-$\pT$ trigger candidates. This fraction was around 99.6\% and remained stable during the data-taking period. In some runs, this fraction was slightly lower due to the absence of a few trigger towers that lost synchronisation with the readout system. The impact of removing one tower is of the order of $1/432 \approx 0.23\%$, where 432 is the total number of the RPC trigger towers.  Trigger towers removed during a run were usually recovered in the next run.
\FloatBarrier
 
\subsection{Trigger rates}
\label{subsec:trigRate}
The ATLAS data acquisition and trigger system~\cite{ATLASTDAQ:2016pov} has an upper limit of approximately 100~kHz for the total rate of standard physics events accepted by the L1 trigger system~\cite{Aaboud:2016leb}. Muon signatures play an important role in the ATLAS physics programme and therefore a significant fraction of this L1 bandwidth is allocated to muon triggers~\cite{Aad:2020uyd}. The L1 muon trigger was designed to be efficient in detecting muon candidates while keeping L1 muon rates within acceptable limits. In this section, rates of the L1 muon trigger as a function of the instantaneous luminosity are presented for the full L1 muon trigger and separately for the barrel and endcap triggers.
 
Several representative runs recorded in 2018 were used for measuring L1 muon trigger rates. Small random fractions of MU10 and MU20 candidates were recorded at the constant rate of about 1~Hz each, without any additional selections applied by the HLT. The constant rate was maintained by varying the sampling fraction as a function of the instantaneous luminosity. Trigger rates were computed by counting the number of MU10 or MU20 candidates in fixed time intervals of about 60~s and using the known sampling fraction to obtain the full rate. The instantaneous luminosity was assumed to be constant during each 60~s time interval. The trigger rates shown here were averaged over several time intervals that belong to a single instantaneous luminosity bin with a width of $\silumi{32}$ and plotted as a function of the instantaneous luminosity.
 
\begin{figure}[h!]
\centering
\subfloat[]{\includegraphics[width=0.50\textwidth]{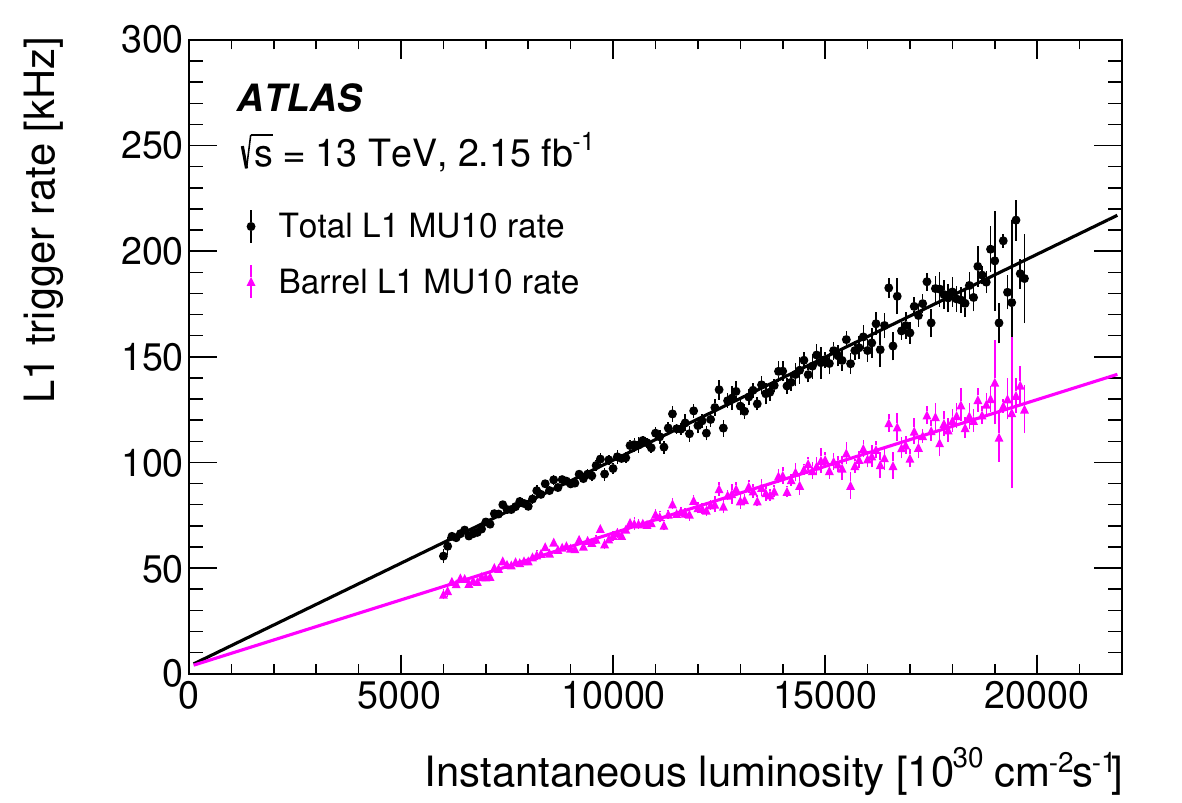}\label{pic:MU10}}
\subfloat[]{\includegraphics[width=0.50\textwidth]{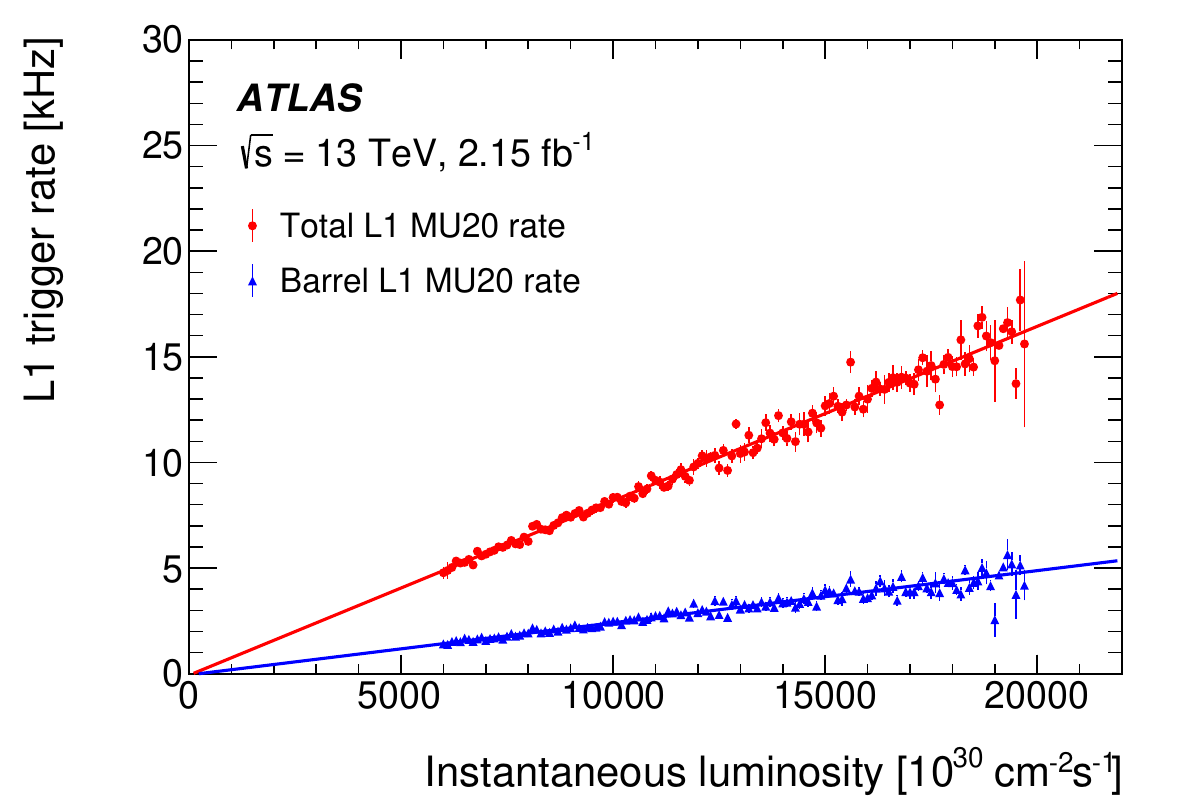}\label{pic:MU20}}
\caption{Rates of (a) MU10 and (b) MU20 L1 triggers, plotted as a function of the instantaneous luminosity. The fits demonstrate the linearity of the trigger rates as a function of the instantaneous luminosity.}
\label{pic:trigger_rates}
\end{figure}
 
The rates of events containing one or more MU10 or MU20 trigger candidates are plotted as a function of the instantaneous luminosity in Figure~\ref{pic:trigger_rates}. The total L1 muon trigger rate and the L1 barrel muon trigger rate are shown separately. The rate of MU10 triggers was well beyond the L1 bandwidth allocated to muon triggers. Therefore, only a small, random fraction of the events selected by the MU10 trigger was accepted by the L1 trigger system. On average, one MU10 event out of approximately 200\,000 (50\,000) was selected at the start (end) of a typical ATLAS run. Moreover, MU10 and other low-$\pT$ L1 muon triggers were used in combination with other L1 trigger objects in order to reduce L1 rates to acceptable levels, for example by requiring presence of two or more MU10 trigger candidates in a single event~\cite{Aad:2020uyd}.
 
Since the MU20 trigger was a primary L1 muon trigger, all MU20 candidates were accepted by the L1 system for further processing by the HLT. This corresponds to a rate of approximately 16~kHz and 5~kHz for the full and barrel-only MU20 trigger, respectively, when evaluated at the instantaneous luminosity of $2 \times \silumi{34}$. The MU20 barrel trigger rate is substantially lower than the MU10 barrel trigger rate due to the higher $\pT$ threshold value, implemented as the additional hit coincidence requirement in the outermost doublet layer (RPC3). This requirement leads to a more precise $\pT$ measurement due to the longer muon trajectory in the barrel magnetic field between the RPC2 and RPC3 layers. The different relative contributions of the barrel triggers to the total MU10 and MU20 trigger rates are due to different definitions of trigger roads for the RPC and TGC detectors. For the low-$\pT$ MU10 threshold, the trigger rate in the barrel region is higher than the endcap rate. This is because the barrel trigger uses a looser requirement, based on the coincidence of signals between the two innermost RPC doublet layers, than is used in the endcap trigger, where the coincidence of three layers of TGC chambers~\cite{Aad:2012xs,Aaboud:2016leb} is required.
 
\begin{figure}[h!]
\centering
\includegraphics[width=0.50\textwidth]{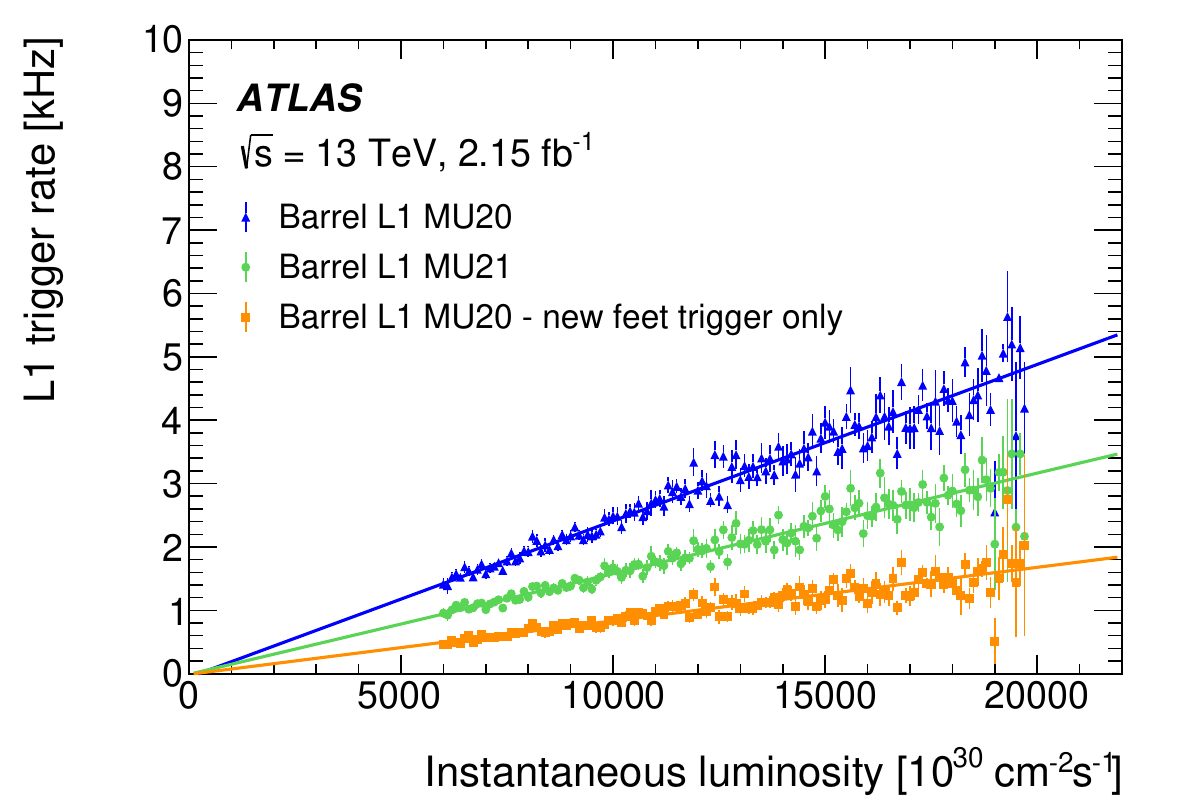}\label{pic:MU2021}
\caption{Rates of MU20 and MU21 triggers, and the contribution due to the new feet trigger component of MU20, plotted as a function of the instantaneous luminosity. The fits demonstrate the linearity of the trigger rates as a function of the instantaneous luminosity.}
\label{pic:trigger_rates_2}
\end{figure}

The muon barrel trigger rates were also measured separately for the MU20 and MU21 triggers, and also for the new feet trigger component of MU20, as shown in Figure~\ref{pic:trigger_rates_2}. New feet trigger candidates were selected by requiring that an L1 muon barrel candidate satisfies the MU20 trigger logic and fails the MU21 trigger logic. The contribution from the new feet chambers is approximately 30\% of the MU20 barrel trigger rate, while these chambers account for only $\sim$3\% of the L1 muon barrel trigger coverage. This higher relative rate was due to using only two out of four coincidences for high-$\pT$ L1 muon triggers in the feet chambers, as discussed in Section~\ref{subsec:trigEff}.
 
\subsection{Trigger composition}
\label{subsec:trigComp}
This section presents the results of studying the composition of events accepted by the L1 muon barrel trigger system. High rates of proton--proton collisions produce high levels of ionising radiation in the ATLAS cavern, resulting in background (non-muon) candidates being selected by the L1 muon trigger system. These background events originate from high-momentum charged hadrons that punch through the calorimeter system or from secondary particles produced in interactions of high-momentum particles with the detector material. This study estimates the fraction of muon candidates accepted by the L1 muon barrel trigger system that are due to background events. This fraction was measured for an inclusive sample of selected L1 muon trigger candidates and for candidates selected in different detector regions. This fraction was also estimated as a function of the mean number of interactions per event.
 
This study uses MU10 and MU20 candidates that were recorded using only the criteria imposed in the L1 trigger system, as detailed in Section~\ref{subsec:trigRate}, without any additional selections in the HLT. One representative run recorded in 2018 with an integrated luminosity of 0.51~$\text{fb}^{-1}$ was analysed for this study. Inclusive reconstructed muons with $\pT>3$~\GeV\ and $|\eta| < 1.05$ were used for this study (offline muons), without requiring the selection criteria described in Section~\ref{sec:analysis}. RoIs selected by the L1 muon barrel trigger were matched to the closest offline muon candidate using a geometrical requirement of $\Delta R < 0.3$ between the muon track and centre of the RoI. The muon reconstruction is highly efficient for selecting muons produced in collisions and has a small rate for misidentifying other particles as muons. Therefore, it is assumed that any RoI not matched with an offline muon candidate was associated with a background candidate traversing the RPC system, as described above.

\begin{table}[!h]
\caption{Fractions of RoIs that were not matched to offline muon candidates with $\pT>3$~\GeV\ or with $\pT>20$~\GeV. Fractions are shown for RoIs selected in the barrel region by MU10, MU20, MU21, and the new feet trigger component of MU20. The statistical uncertainty for all table entries is smaller than 0.7\%.}
\vspace{0.2cm}
\begin{center}
\resizebox{\textwidth}{!}
{
\begin{tabular}{lcc}
\hline
L1 trigger & Unmatched RoI fraction for $\pT>3$~\GeV & Unmatched RoI fraction for $\pT>20$~\GeV \\
\hline
MU10              & 71.1\% & ---\\
MU20              & 27.8\% & 88.3\% \\
MU21              &  ~~6.7\% & 82.7\% \\
MU20 new feet trigger  & 66.8\% & 98.6\% \\
\hline
\end{tabular}
}
\end{center}
\label{table:fakeTrigger}
\end{table}
 
\begin{figure}[!h]
\centering
\subfloat[]{\includegraphics[width=0.49\textwidth]{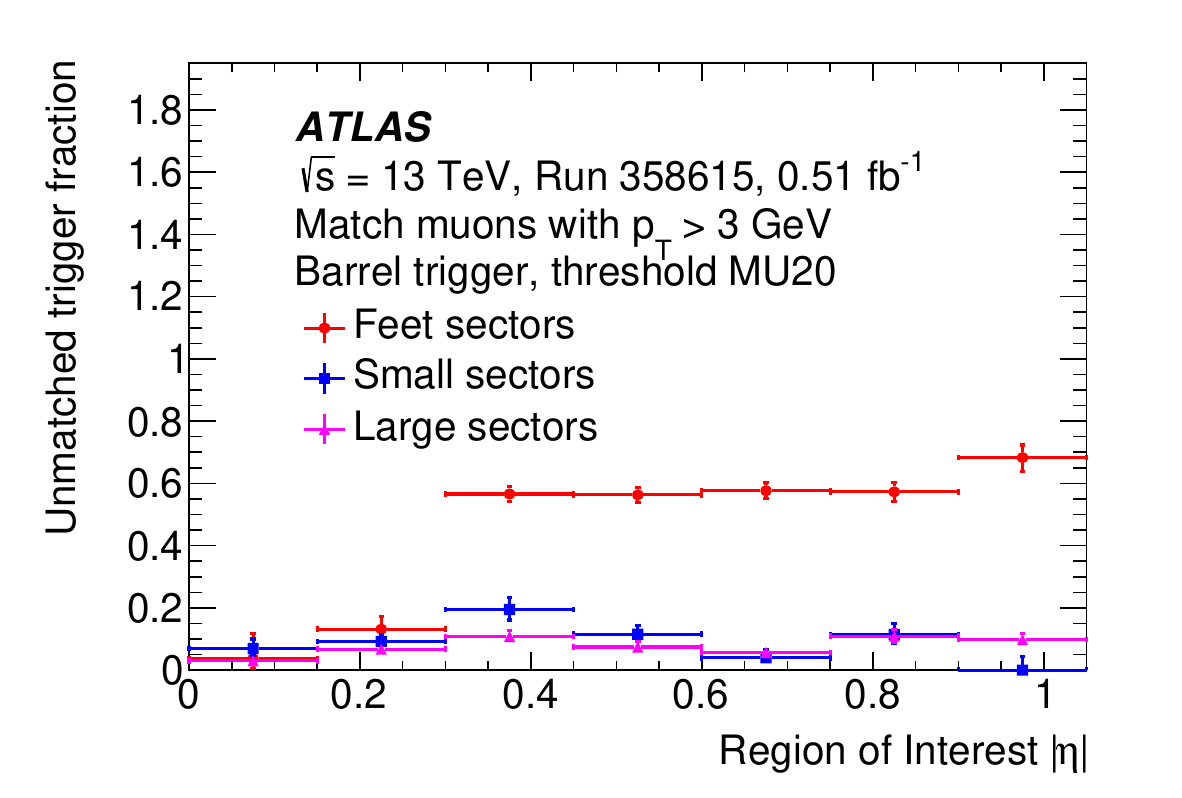}\label{pic:fakeTrigger2}}
\subfloat[]{\includegraphics[width=0.49\textwidth]{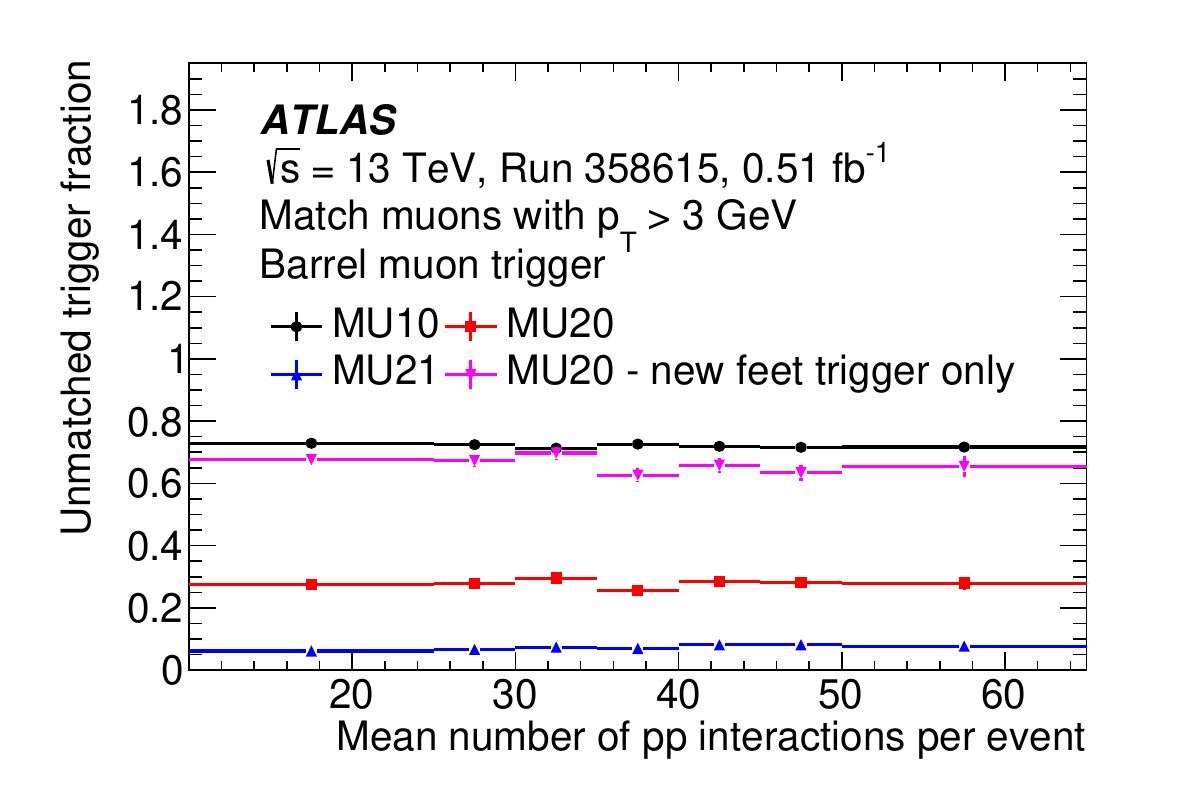}\label{pic:fakeTrigger1}}
\caption{(a) Fractions of MU20 RoIs not matched to offline muon candidates with $\pT>3$~\GeV\ plotted as a function of the RoI pseudorapidity and shown separately for the small, large and feet sectors of the RPC system. (b) Fractions of RoIs not matched to offline muon candidates with $\pT>3$~\GeV\ plotted as a function of the mean number of interactions per event for different L1 muon barrel trigger thresholds.}
\label{pic:fakeTrigger12}
\end{figure}

Table~\ref{table:fakeTrigger} shows fractions of RoIs selected by the L1 muon barrel trigger system that were not matched to offline muon candidates with $\pT>3$~\GeV. These unmatched RoIs are assumed to be due to background events. The unmatched RoI fraction of the MU10 trigger is higher than that of the MU20 trigger due to its looser trigger coincidence criteria, in which only four out of six RPC layers are used. For muon candidates with $\pT>3$~\GeV, the unmatched RoI fractions for the MU20 and MU21 triggers are 27.8\% and 6.7\%, respectively. The new feet trigger is excluded from the MU21 trigger, which is otherwise identical to the MU20 trigger. The unmatched RoI fraction for the new feet trigger is high because only four RPC layers are used by the new feet trigger component of the MU20 trigger. This effect is illustrated in Figure~\ref{pic:fakeTrigger2} where the unmatched RoI fraction is plotted  as a function of the RoI pseudorapidity separately for large, small and feet RPC sectors. The new feet trigger is installed in the region $|\eta| \gtrsim 0.35$, resulting in the large increase of unmatched RoIs in this region.
 
The unmatched RoI fraction is also plotted in Figure~\ref{pic:fakeTrigger1} as a function of the mean number of interactions per event for several L1 muon barrel triggers. The stability of the unmatched trigger fraction over the full range of the mean number of interactions per event implies that the relative contribution of background events to the trigger rate is independent of instantaneous luminosity.
 
Table~\ref{table:fakeTrigger} also shows fractions of RoIs selected by the L1 muon barrel trigger system that were not matched to offline muon candidates with $\pT>20$~\GeV. The unmatched RoI fraction for the MU20 trigger increases from 27.8\% to 88.3\% when the minimum offline muon $\pT$ requirement is raised from 3~\GeV\ to 20~\GeV.
 
\begin{figure}[!h]
\centering
\subfloat[]{\includegraphics[width=0.49\textwidth]{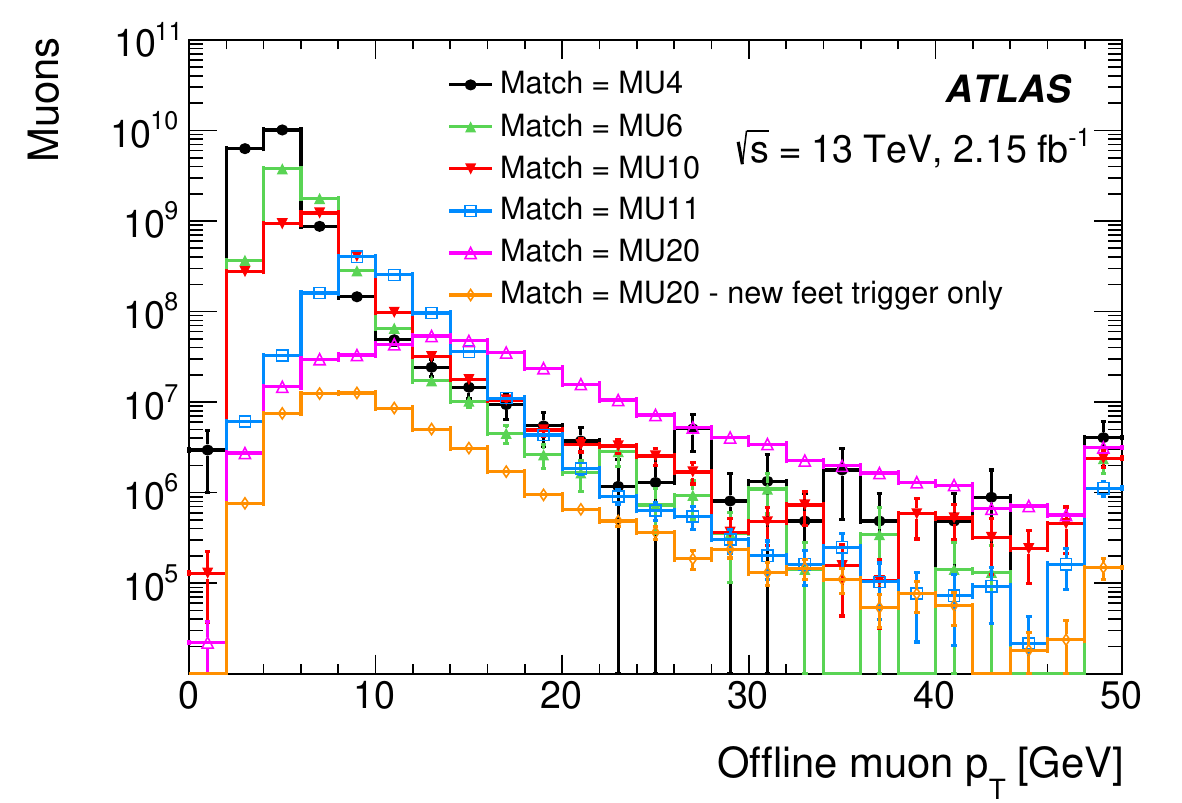}\label{pic:muonPtExc}}
\subfloat[]{\includegraphics[width=0.49\textwidth]{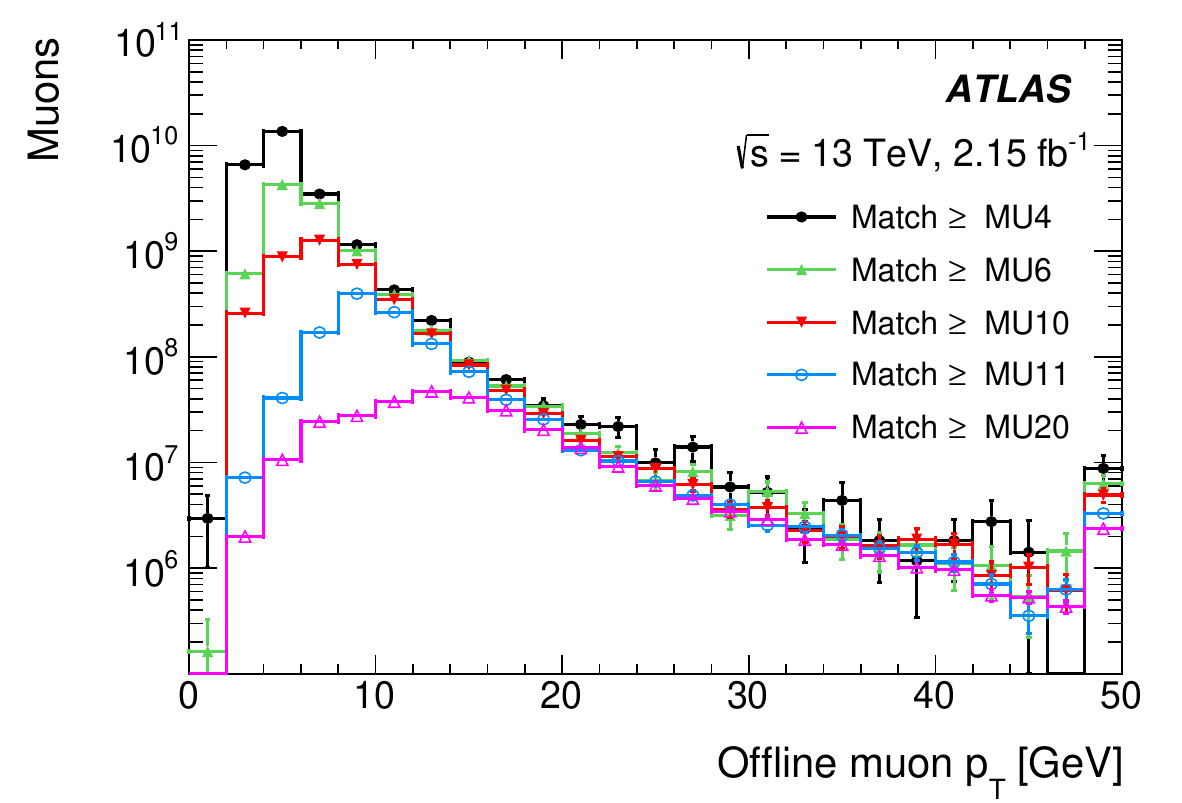}\label{pic:muonPtInc}}
\caption{The $\pT$ distribution of offline muon candidates matching the L1 muon barrel RoIs for the (a) exclusive and (b) inclusive cases. The L1 muon barrel trigger system reports the highest $\pT$ threshold passed by a muon candidate. For high muon $\pT$ values, muon candidates are therefore more likely to pass a corresponding high-$\pT$ threshold, thus leading to a smaller fraction of muons selected by the low-$\pT$ triggers. The rightmost bin includes the overflow entries.}
\label{pic:muonPt}
\end{figure}

Figure~\ref{pic:muonPt} shows the $\pT$ spectra of the offline muons matching the L1 muon barrel RoIs for two different cases. Figure~\ref{pic:muonPtExc} shows the exclusive case where muons are matched only to those RoIs that satisfy the specific trigger threshold and fail higher thresholds. Figure~\ref{pic:muonPtInc} shows the inclusive case where muons are matched to the RoIs that satisfy either the specific trigger threshold or any higher threshold. For the inclusive muon $\pT$ distributions, only events with one muon RoI are considered in order to simplify analysis procedures. The selected RoI is required to be generated in the muon barrel region. For the exclusive muon $\pT$ distributions, this requirement is not applied. The majority of muon candidates selected by the MU20 trigger are muons with $\pT<20$~\GeV. This mismeasurement of the muon $\pT$ by the RPC detector is due to its position resolution being less precise than that of the MDT detector, which is used for offline muon reconstruction in the barrel region. This is expected because the RPC detector was optimised for fast triggering with sufficient $\pT$ resolution to keep trigger rates within allowed limits, as discussed in Section~\ref{subsec:trigRate}.
 
\begin{figure}[!h]
\centering
\subfloat[]{\includegraphics[width=0.49\textwidth]{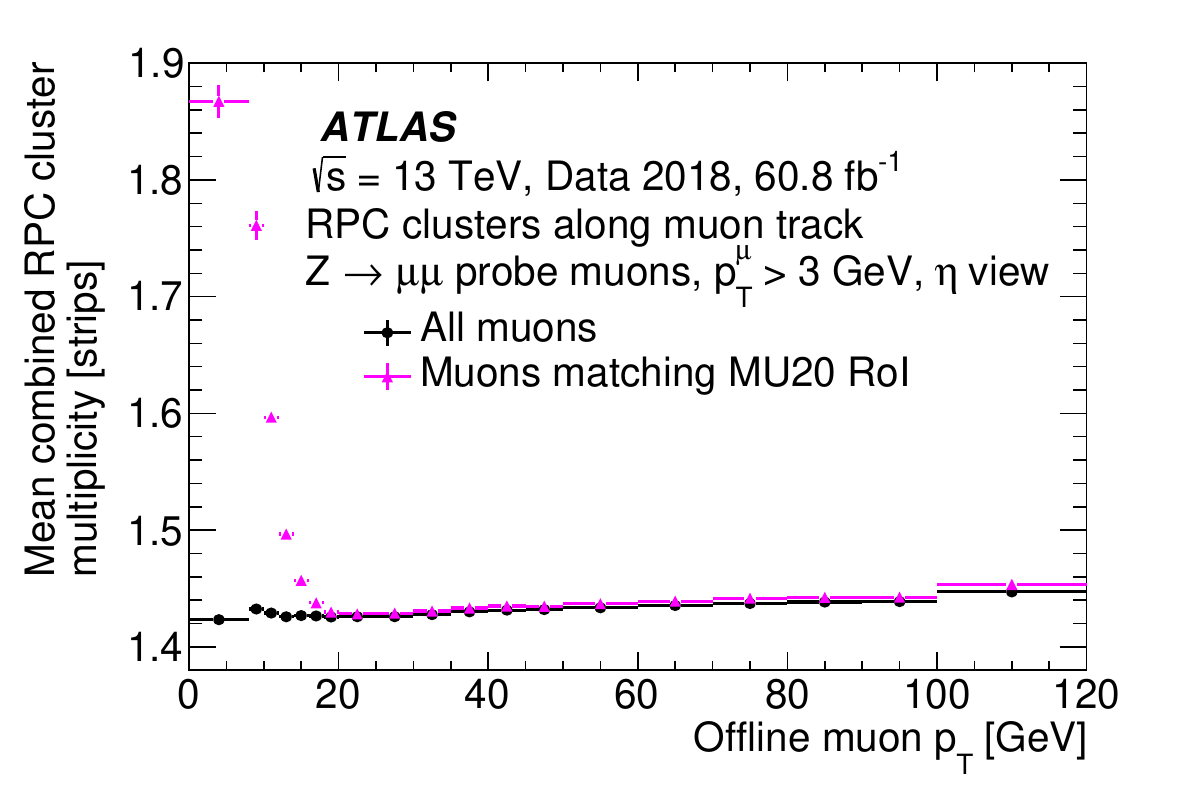} \label{pic:cluster_vs_pt_eta}}
\subfloat[]{\includegraphics[width=0.49\textwidth]{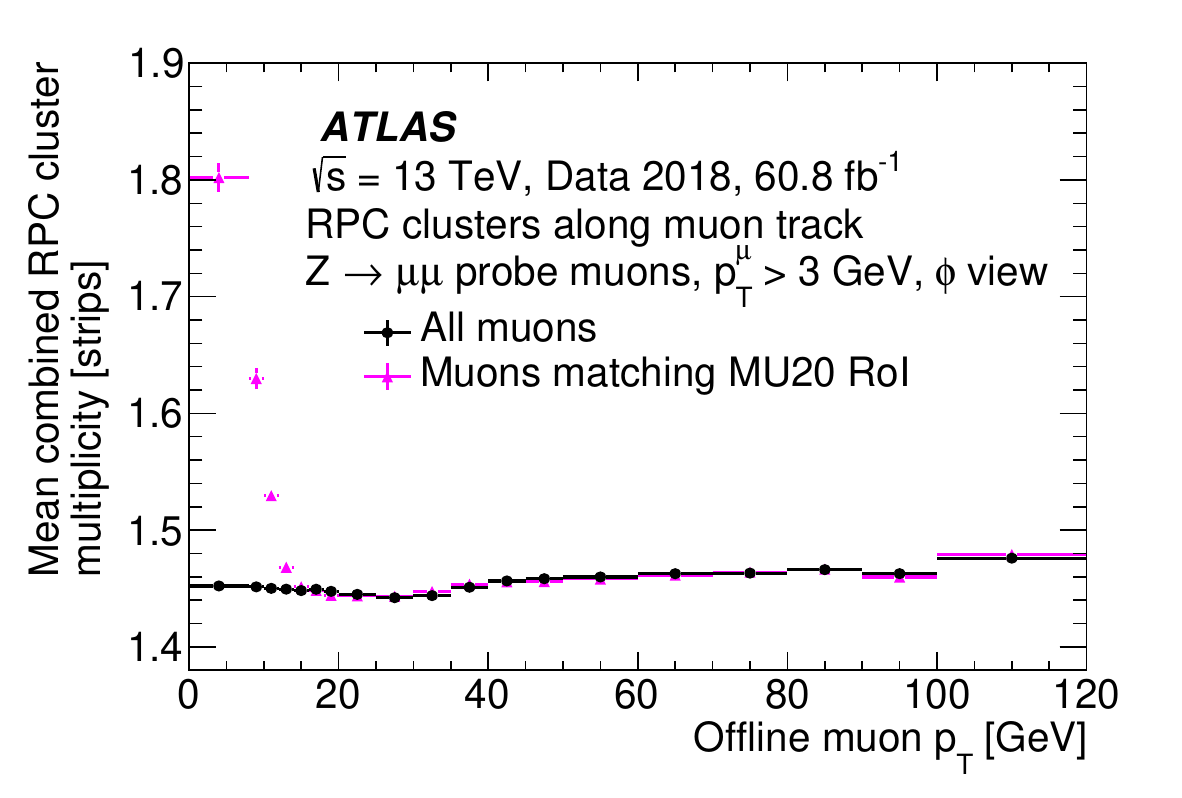} \label{pic:cluster_vs_pt_phi}}
\caption{Mean cluster hit multiplicity for (a) $\eta$ panels and (b) $\phi$ panels averaged over all RPC modules plotted as a function of muon $\pT$ using $Z \rightarrow \mu\mu$ probe muons. The mean cluster hit multiplicity is shown for all probe muons and separately for probe muons matched to the MU20 trigger RoI. The rightmost bin includes the overflow entries.}
\label{pic:cluster_vs_pt}
\end{figure}
 
A muon with $\pT < 20$~\GeV\ would bend outside a geometrical matching window of the MU20 trigger and therefore normally would be expected to fail the MU20 trigger logic. In addition to intrinsic limitations of the RPC position resolution, low-$\pT$ muons can also satisfy the MU20 trigger requirement by generating larger than typical clusters because such muons are more likely to produce hits within the trigger coincidence window. To illustrate this effect, the mean cluster hit multiplicity is plotted in Figure~\ref{pic:cluster_vs_pt} as a function of muon $\pT$ for all probe muons and only for probe muons matching the MU20 RoIs. The mean cluster hit multiplicity is approximately constant within the considered $\pT$ range for the inclusive muon sample. A small increase is observed at high-$\pT$ values, and is due to muon bremsstrahlung radiation. For probe muons matching the MU20 RoIs, the mean cluster hit multiplicity increases by up to 30\% at low muon $\pT$ values. This observation confirms that the low-$\pT$ muons that produce larger-than-typical clusters are more likely to satisfy the MU20 trigger requirement.

 
\FloatBarrier
 
\section{Measurements of RPC currents and counting rates}
\label{sec:current_rate}
Nearly all of the RPC counting rates and ionisation currents are induced by secondary particles, mostly photons and neutrons, that are produced in interactions of primary particles, i.e., those that originate in proton--proton collisions, with the detector and beam-pipe material. The flux of the secondary particles in the MS was computed using simulation and presented in Appendix~A of Ref.~\cite{CERN-LHCC-2017-017}. These secondary particles induce a dominant fraction of RPC avalanche events, which causes the resulting RPC ionisation currents and counting rates to be proportional to the instantaneous luminosity.
 
In the previous two sections, muons were used to study the detector and trigger performance. In this section, the RPC detector response is evaluated using zero-bias proton--proton collisions. Sections~\ref{subsec:current} and~\ref{subsec:hitRate} present measurements of the RPC counting rates and RPC ionisation currents, respectively. These measurements are performed as a function of the instantaneous luminosity for RPCs located in different detector regions in order to study their dependence on the distance to the collision point. In Section~\ref{subsec:charge}, these measurements are combined in order to measure the average avalanche charge of the ATLAS RPC detector.
 
\subsection{RPC current measurements}
\label{subsec:current}
The total ionisation current for each RPC gas volume is inferred by measuring the voltage across a 100~k$\si{\ohm}$ resistor placed between the RPC readout panel and ground~\cite{Aielli:2013nea}. Current measurements are recorded and time-stamped by the DCS~\cite{Polini:2012zz}. The current measurements used in this analysis were recorded during periods of proton--proton collisions with stable LHC and detector conditions. The baseline current measured for each chamber using the nominal voltage settings, but without beams in the LHC, is subtracted in order to study the net effect of collisions. The current measurements are normalised to the active surface area of each module (current density) to allow comparisons between modules of different size. About 90\% of the RPC modules without known problems were selected for this analysis. The RPCs that were excluded comprise modules disconnected due to gas leaks, modules where the DCS current readout had problems, and modules that had low efficiency for detecting muons.
 
\begin{figure}[!h]
\begin{center}
\subfloat[]{\includegraphics[width=0.495\textwidth]{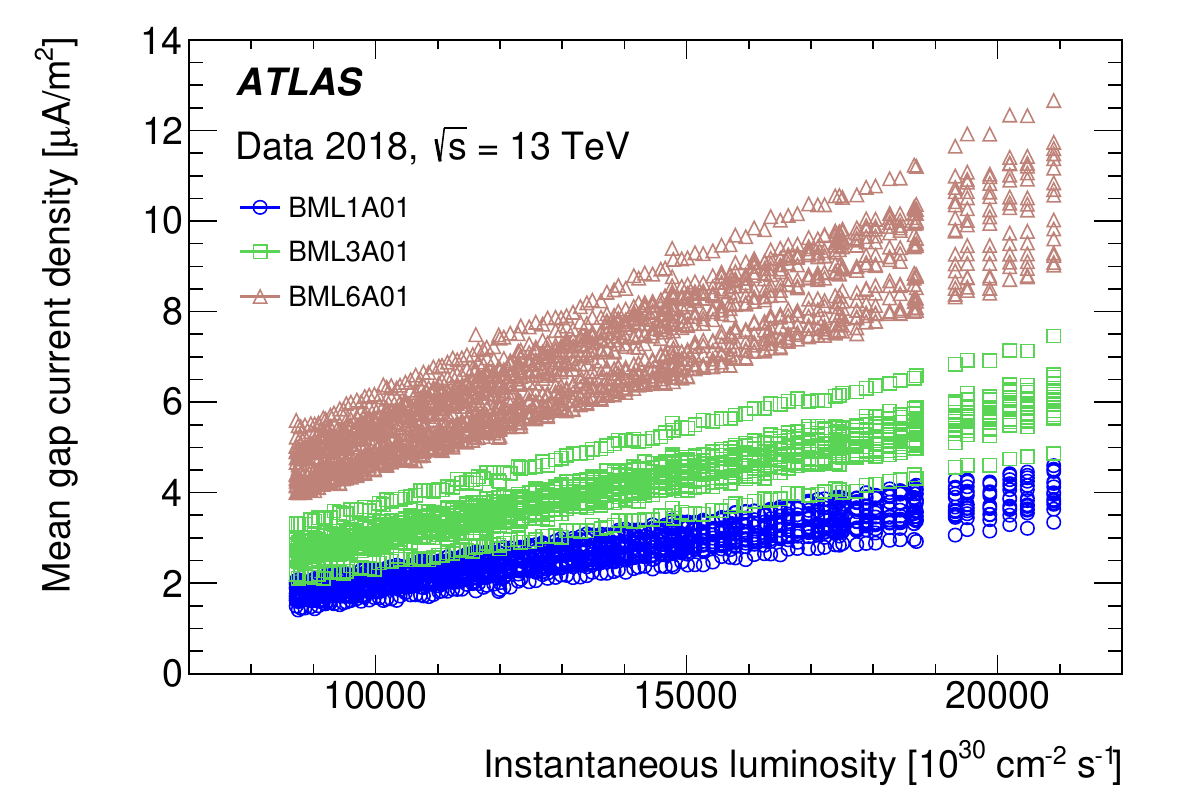}\label{pic:lumi3}}
\subfloat[]{\includegraphics[width=0.495\textwidth]{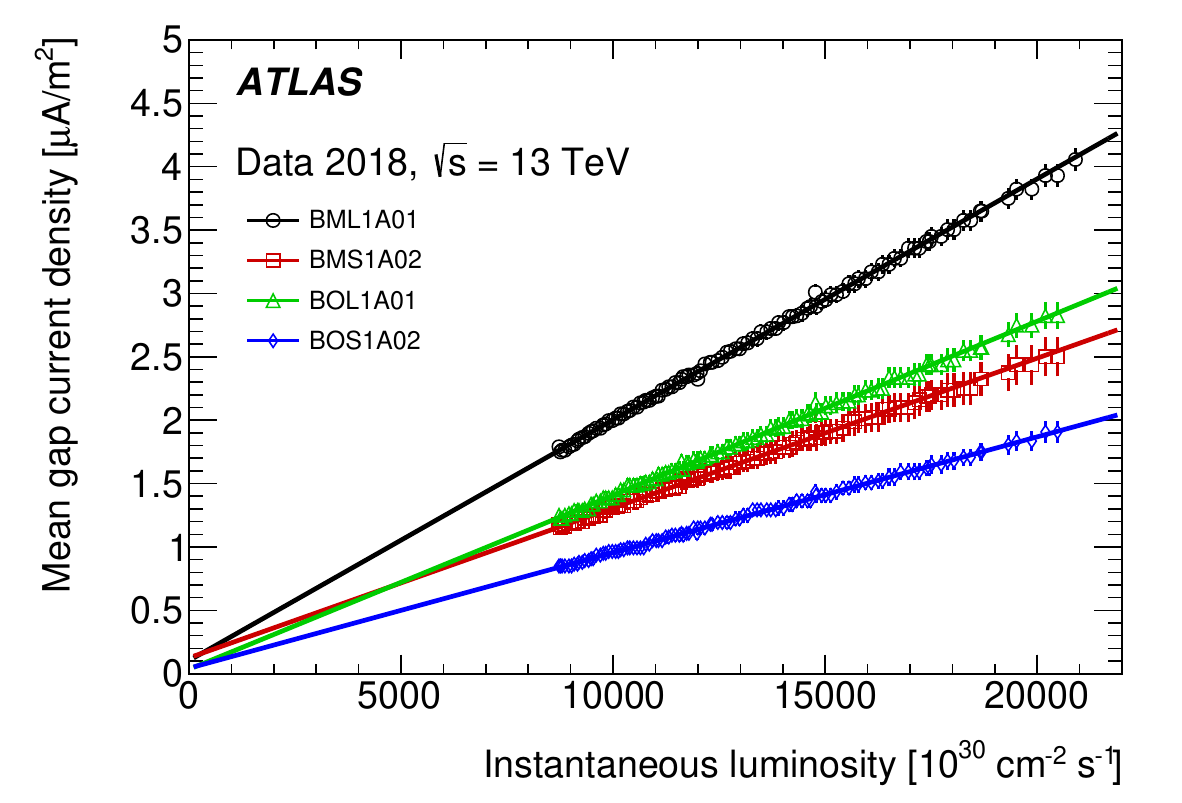}\label{pic:lumi1}}
\caption{RPC current density shown as a function of instantaneous luminosity for several representative modules. The nominal voltage setting of 9.6~kV was used for these measurements. (a) Each set of points corresponds to one individual module belonging to one of the three stations located at different positions along the \textit{z}-direction. (b) Each line corresponds to the current averaged over all the modules in each RPC station. The four stations belong to small and large sectors in the RPC1 and RPC3 layers.}
\label{pic:current_lumi}
\end{center}
\end{figure}

The measured current density is shown in Figure~\ref{pic:current_lumi} as a function of the instantaneous luminosity for several representative modules in order to illustrate the typical RPC response in different detector regions. The nominal voltage setting of 9.6~kV was used for these measurements. Figure~\ref{pic:lumi3} shows the current density for the individual RPC modules located in the stations with $\eta$ indices 1, 3 and 6; only the BML RPCs belonging to the large $\phi$ sector number 1 of ATLAS side A were used for these plots. The current density increases at higher $\eta$ values due to the higher flux of hadrons in the regions outside the endcap calorimeters, as shown in Appendix~A of Ref.~\cite{CERN-LHCC-2017-017}.
 
Figure~\ref{pic:lumi1} shows the current density measurements averaged over the modules belonging to the stations with $\eta$ index 1, plotted as a function of the instantaneous luminosity for the BML, BMS, BOL and BOS sectors. Only the RPCs located in $\phi$ sectors 1 and 2 of side A were used for these measurements. The current density decreases at larger radii due to the reduced particle flux density in the BOL and BOS layers. The differences between small and large chambers are due to the presence of the barrel magnet coils which shadow the small sectors from the collision point. In addition, the small chambers are located at slightly larger radii than the large chambers in the same layer and therefore detect smaller particles fluxes. These results confirm the expected linear dependence of the RPC current density as a function of the instantaneous luminosity within the currently accessible luminosity range. Therefore, it is assumed that the RPC current density can be extrapolated linearly to the HL-LHC instantaneous luminosity of $7.5 \times \silumi{34}$.
 
\begin{figure}[!h]
\centering
\includegraphics[width=0.495\textwidth]{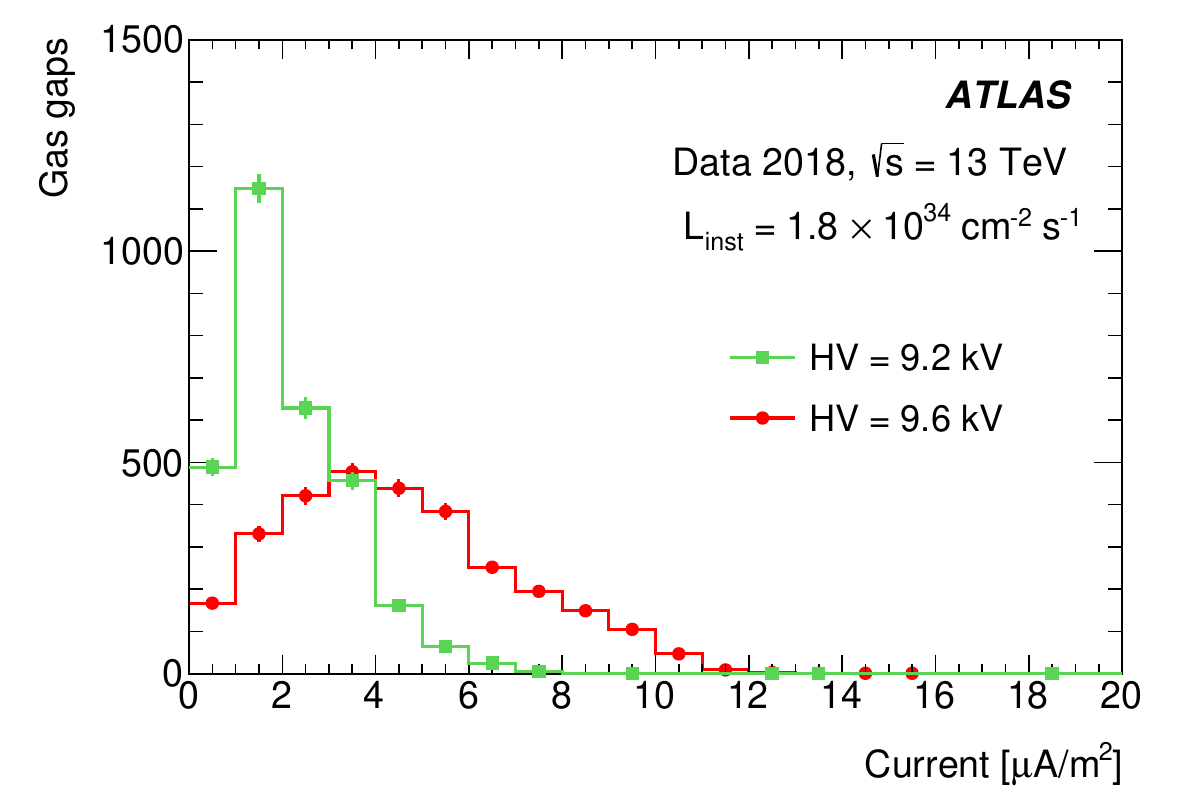}
\caption{Distributions of the measured current density for the selected RPCs obtained at an instantaneous luminosity of $1.8 \times \silumi{34}$. The measurements were taken during one representative run in 2018.}
\label{pic:Igap_vs_HV_2018lumi}
\end{figure}

Figure~\ref{pic:Igap_vs_HV_2018lumi} shows the current density distributions for all selected RPCs at an instantaneous luminosity of $1.8 \times \silumi{34}$ for voltages of 9.2~kV and 9.6~kV. Since the voltages applied to the detector modules can be corrected for temperature and pressure effects, the actual applied voltage could differ from the set value. These corrections were taken into account to allow comparisons between different modules. The current density for each module was scaled to obtain current density values at 9.2~kV and 9.6~kV, using the linear relationship between current and applied voltage that was measured for that particular module.
 
\begin{figure}[!h]
\begin{center}
\subfloat[]{\includegraphics[width=0.495\textwidth]{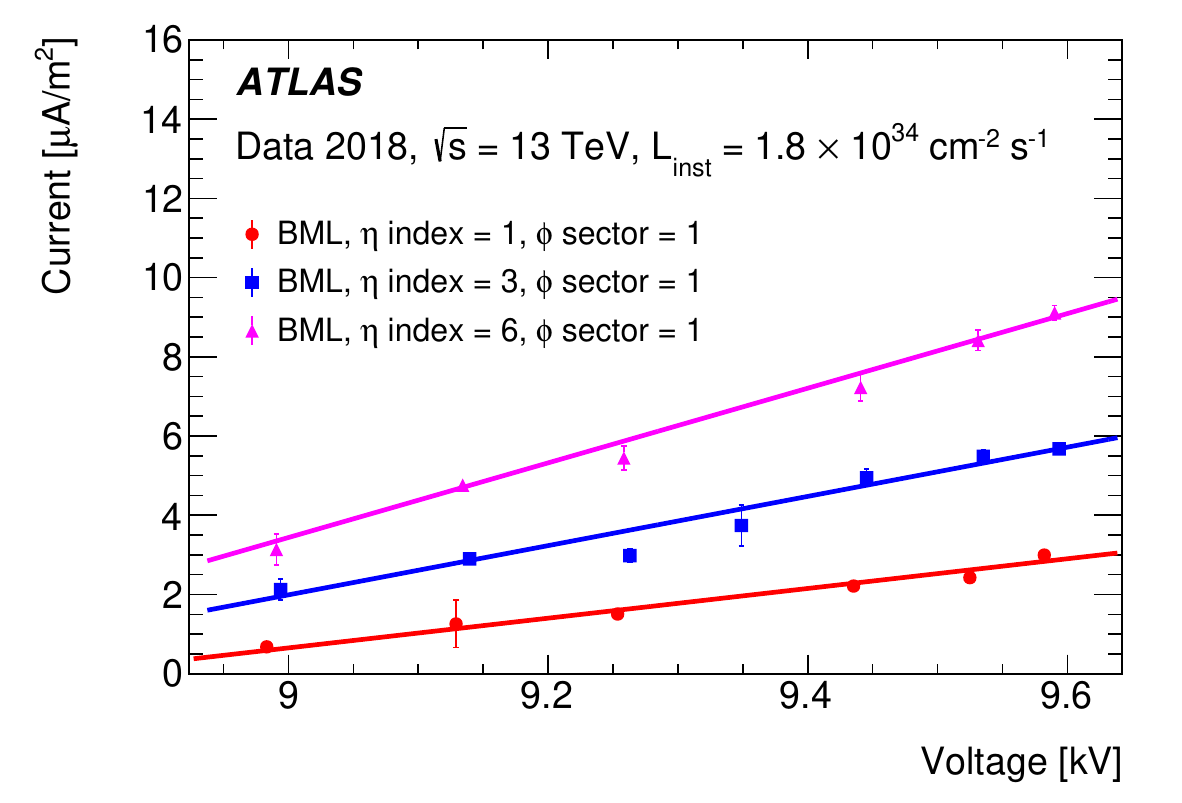}\label{pic:Igap_vs_HV_wb_public}}
\subfloat[]{\includegraphics[width=0.495\textwidth]{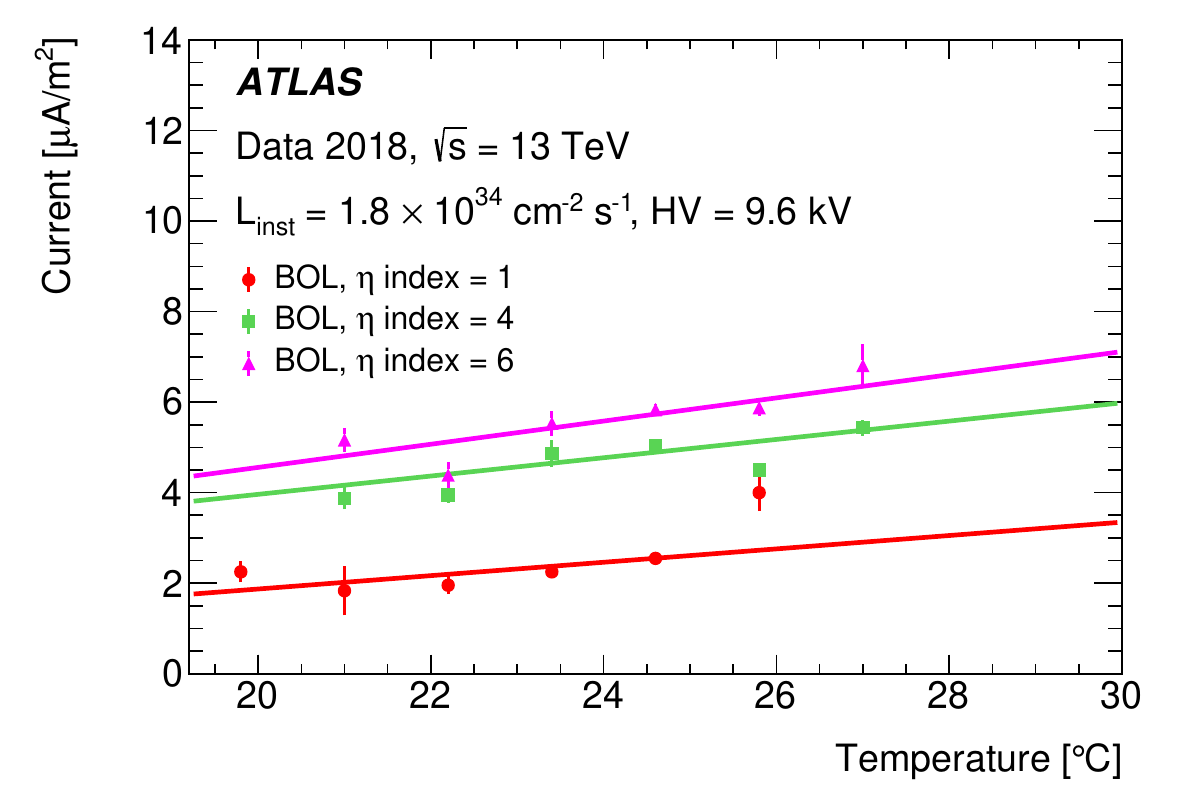}\label{pic:Igap_vs_T}}
\caption{(a) RPC current density shown as a function of the applied voltage for representative BML modules located in $\eta$ stations 1, 3, and 6, and belonging to the large $\phi$ sector 1. (b) RPC current density shown as a function of temperature for BOL modules located in $\eta$ stations 1, 4, and 6. The current density is averaged over $\phi$ sector modules included for each temperature bin.}
\label{pic:Igap_vs_HV_2018}
\end{center}
\end{figure}

The current density in each selected RPC was measured as a function of the operating voltage at the beginning of several proton--proton collision runs recorded in 2018. These measurements were performed while the RPC operational voltage was increasing from the standby setting of 9.0~kV to the nominal setting of 9.6~kV. The standby voltage setting was used in order to protect the RPCs during periods with unstable LHC beam conditions. The current density measurements for each individual module were linearly extrapolated to an instantaneous luminosity of $1.8 \times \silumi{34}$. Figure~\ref{pic:Igap_vs_HV_wb_public} shows the current density as a function of the applied voltage for representative BML modules located in $\eta$ stations 1, 3, and 6, and belonging to $\phi$ sector 1. The current density has an approximately linear dependence on voltage within the measurement range of 9.0~kV to 9.6~kV. Each current measurement was taken twice in order to estimate the uncertainty arising from variations in the instantaneous luminosity within
the approximately one minute interval during which the instantaneous luminosity measurement was performed; the analysis assumes the instantaneous luminosity was constant within this interval. This uncertainty was estimated as half of the difference between the two measurements taken at the same voltage and separated by approximately 60~s.
 
The temperature gradient in the ATLAS cavern is about 10$^\circ$C, as illustrated in Figure~8 of Ref.~\cite{Amram:2011ue}. This gradient is due to the air circulation in the cavern, with lower temperatures at the bottom of the detector. Figure~\ref{pic:Igap_vs_T} shows the extrapolated current density measured as a function of temperature for BML modules located in $\eta$ stations 1, 4, and 6. The measured current density in each module is extrapolated to the value expected at the nominal voltage of 9.6~kV using the procedure described earlier. The current density measurements were averaged over modules from all $\phi$ sectors that belong to the same temperature bin. An approximately linear dependence of the current density as a function of temperature is observed. The small deviation from linearity for the BOL temperature at around 26$^\circ$C is likely to be due to the
local module temperatures differing from the temperature recorded by the nearest temperature sensor, which is spatially separated from the modules.
 
\subsection{RPC counting rate measurements}
\label{subsec:hitRate}
Counting rates were measured using collision events selected online with a primary single-muon trigger~\cite{Aad:2020uyd} that requires the presence of an isolated muon with $\pT > 26$~\GeV. This single-muon trigger selects energetic collisions that typically contain a $W$ or $Z$ boson, or a heavy-flavour jet with high $\pT$. Because such energetic events include a hard-scattering process that produces many secondary hadrons and photons, they are not representative of an average LHC bunch crossing. Since the production rates of $W$ and $Z$ bosons are directly proportional to the instantaneous luminosity, sampling the three bunch crossings preceding the trigger provides an unbiased sample of LHC collisions at a given instantaneous luminosity.
 
RPC hits were required to belong to the 75~ns time window preceding the bunch crossing selected by the trigger. Selected hits have times between $-$87.5~ns and $-$12.5~ns, where the zero time corresponds to the bunch crossing selected by the muon trigger. This criterion selects the three bunch crossings immediately preceding the triggered bunch crossing. Counting rates were computed over a time period of approximately one minute during which the instantaneous luminosity was assumed to be constant. A statistically weighted average of the counting rates from several such time intervals was used to compute the counting rate as a function of the instantaneous luminosity. This averaging was performed using an instantaneous luminosity binning of $\silumi{32}$. The resulting counting rates were then normalised to the active module surface area. Only modules with detector efficiency larger than 50\% and with a nominal voltage of 9.6~kV were used for the measurements presented in this section.
 
\begin{figure}[!h]
\begin{center}
\centering
\subfloat[]{\includegraphics[width=0.495\textwidth]{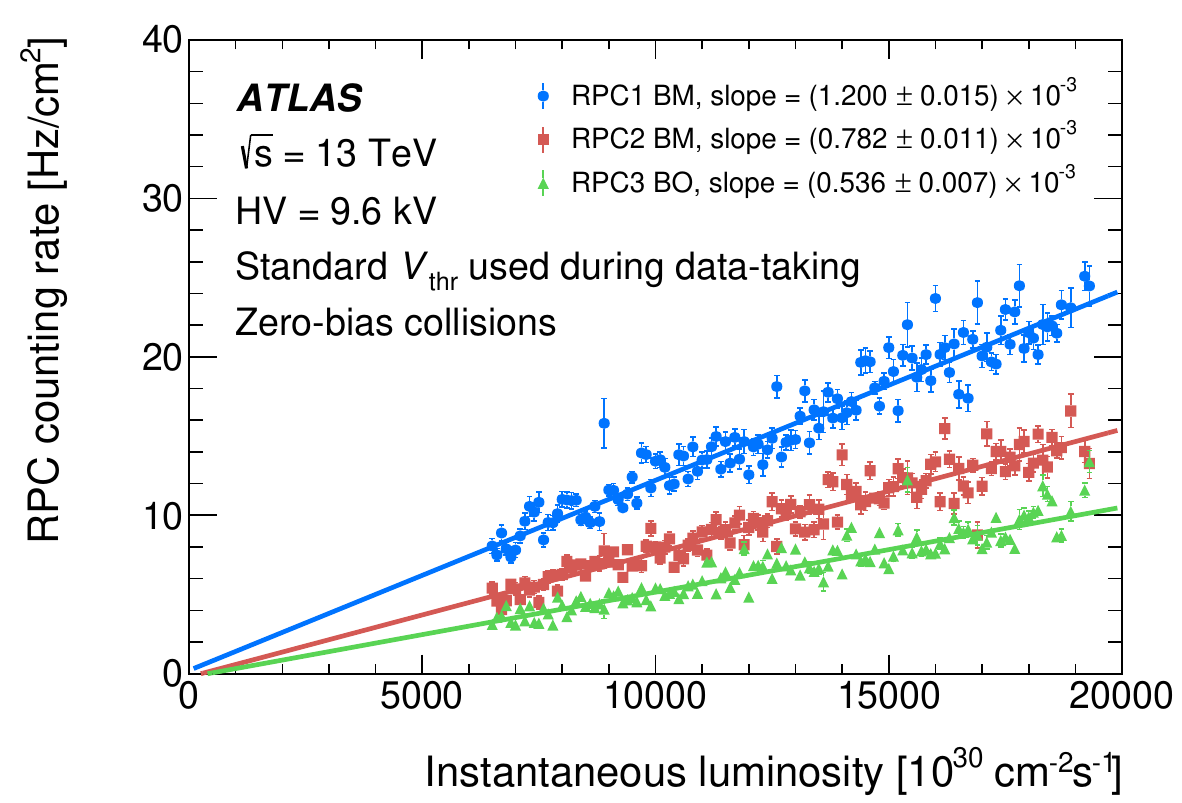}\label{fig:rateVsRadius}}
\subfloat[]{\includegraphics[width=0.495\textwidth]{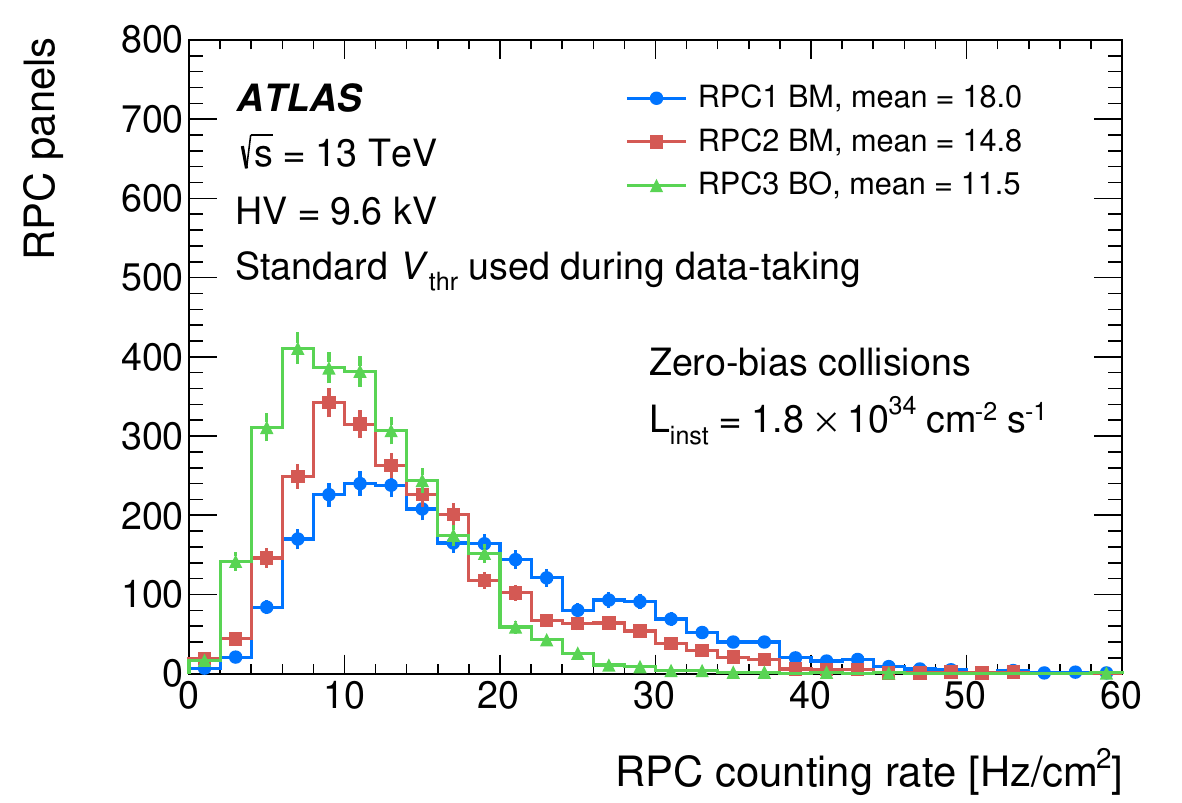}\label{fig:rate18Layer}}
\caption{(a) RPC counting rates measured per unit surface area plotted as a function of the instantaneous luminosity for three RPC $\phi$ panels located at different distances from the collision point, corresponding to RPC1, RPC2 and RPC3 layers. All $\phi$ panels used for this study belong to the large $\phi$ sector 1 and are located at $\eta$ station 2. The slope parameters were extracted from a linear fit to the data points. (b) Distribution of the counting rates for all RPC panels, separated into three sets of chambers in different detector doublet layers, at an instantaneous luminosity of $1.8 \times \silumi{34}$. Nominal voltage and FE threshold settings were used in this study.}
\label{fig:rates1}
\end{center}
\end{figure}
 
Figure~\ref{fig:rateVsRadius} shows the RPC counting rates as a function of instantaneous luminosity for three representative RPC panels located at radii of approximately 6.8~m, 7.4~m and 9.8~m from the beams. As expected, the counting rates depend linearly on the instantaneous luminosity in all detector regions. This observation confirms that secondary particles, produced in interactions of the primary particles with the detector and beam-pipe material, are the main source of RPC hits. Figure~\ref{fig:rate18Layer} shows the distribution of counting rates measured at an instantaneous luminosity of $1.8 \times \silumi{34}$ for the RPC panels located in three different detector layers. Panels located further from the collision point see smaller counting rates because the flux of secondary particles decreases with increasing distance from the centre of the detector.
 
\begin{figure}[!h]
\centering
\subfloat[]{\includegraphics[width=0.495\textwidth]{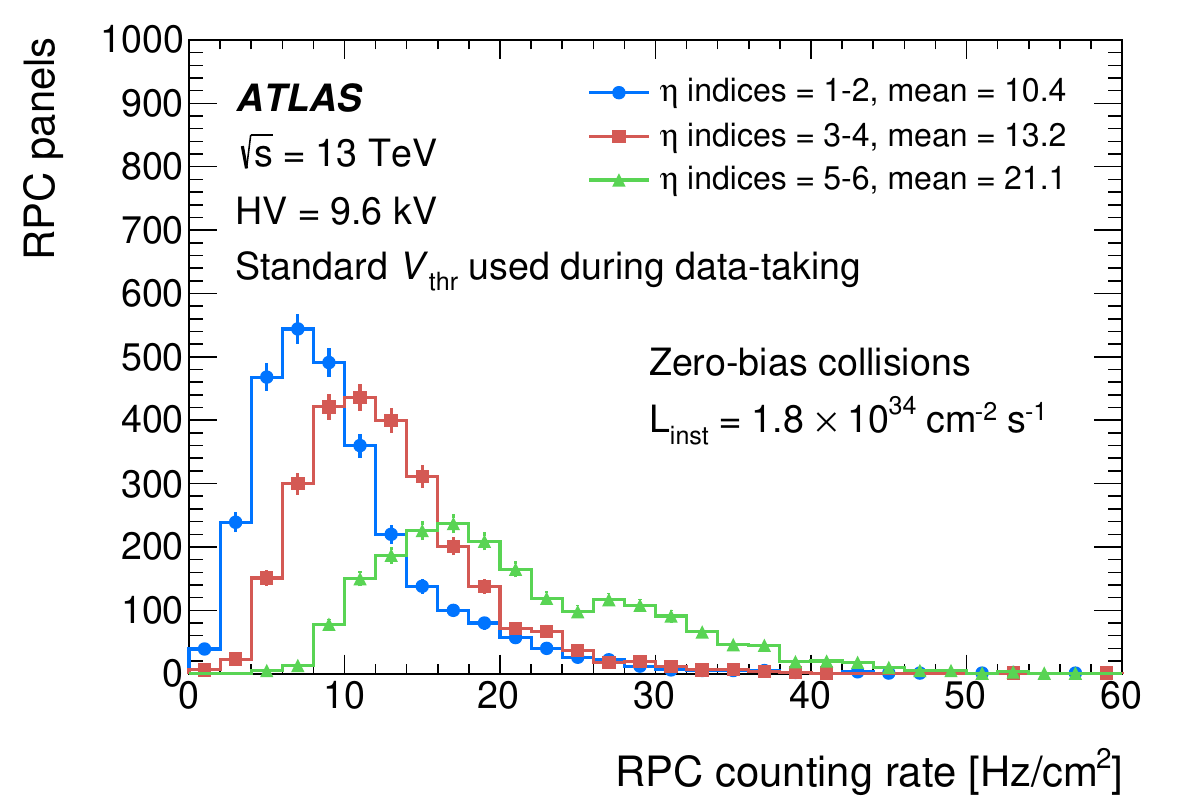}\label{fig:rate18Eta}}
\subfloat[]{\includegraphics[width=0.49\textwidth]{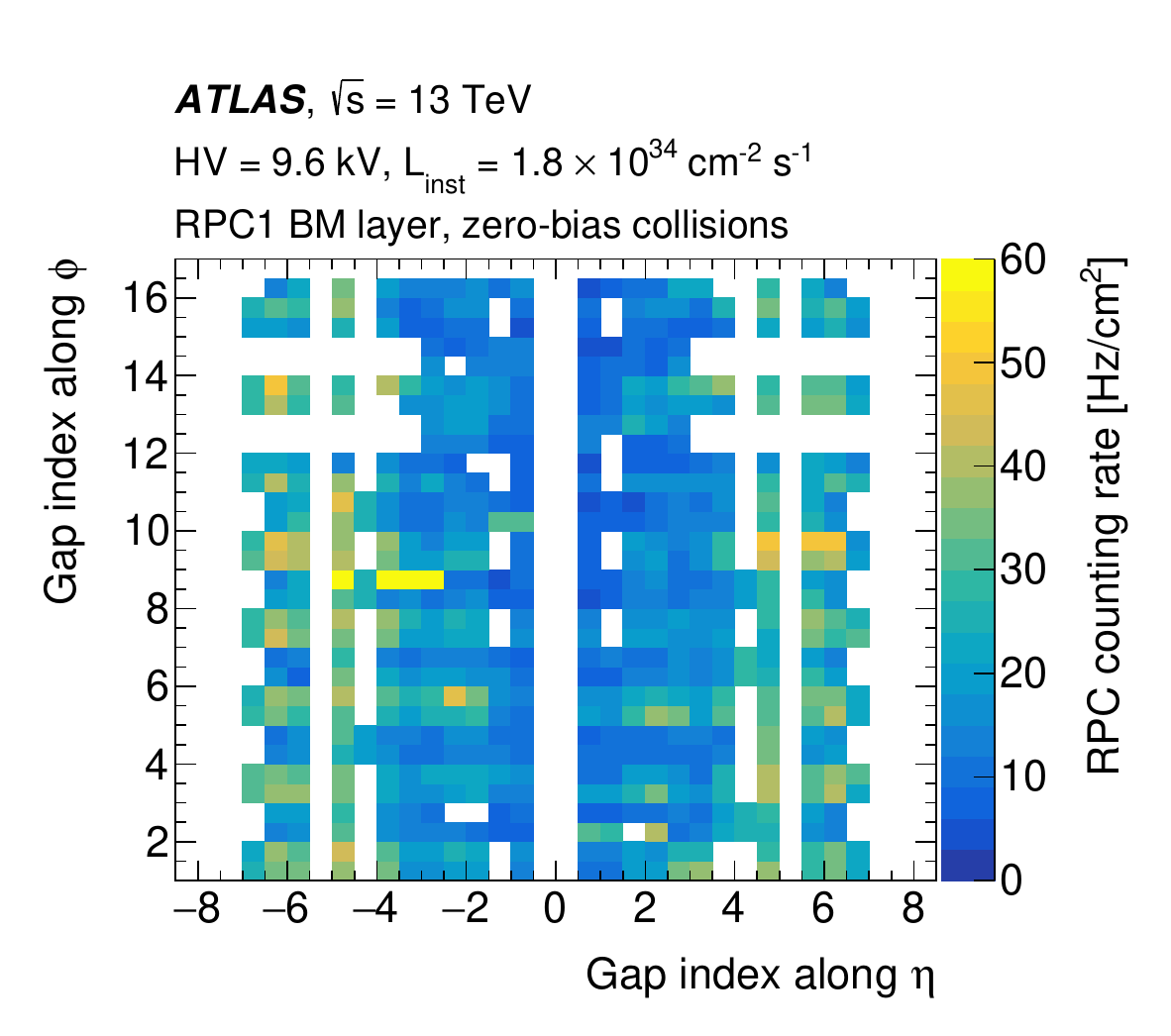}\label{fig:rate2D}}
\caption{RPC counting rates per unit surface area at an instantaneous luminosity of $1.8 \times \silumi{34}$ for nominal voltage and FE threshold settings. (a) Distribution of the counting rates for all RPC panels, separated into three sets of chambers with different $\eta$ indices. (b) Counting rates for the RPC1 layer shown as a function of RPC $\eta$ and $\phi$ coordinates. The values were averaged over the two individual sub-layers. Empty bins correspond to logical combinations of indices that do not represent installed RPCs or correspond to modules with efficiencies less than 50\%.}
\label{fig:rates2}
\end{figure}
 
The distributions of RPC counting rates at an instantaneous luminosity of $1.8 \times \silumi{34}$ are presented in Figure~\ref{fig:rate18Eta}, where RPC panels in each station are grouped by their $\eta$~indices. Figure~\ref{fig:rate2D} shows the counting rates at an instantaneous luminosity of $1.8 \times \silumi{34}$ for the RPC1 doublet layer as a function of RPC $\eta$ and $\phi$ coordinates. Because the flux of ionising particles increases at higher $|\eta|$ values, panels located in the more forward regions of the MS measure higher counting rates.
 
\subsection{RPC avalanche charge measurements}
\label{subsec:charge}
The mean avalanche charge is a key performance parameter of the RPC detector~\cite{Riegler:2002vg}. Its value depends on the properties of the ionising particle and on the gain of the electron amplification process in the gas. The gain of the amplification process is determined by the electrical field across the gas gap and by the gas mixture. The readout system of the ATLAS RPC detector only measures pulse time when the leading edge of the pulse is above a predetermined threshold. Therefore, the avalanche charge cannot be measured directly by the ATLAS RPC detector. Instead, the mean avalanche charge is estimated for each RPC module by combining simultaneous measurements of the current and counting rate for that module. Since both of these measurements integrate over many ionisation events, only the mean avalanche charge can be assessed by this procedure.
 
The mean avalanche charge, $Q$, was measured using the following expression:
\begin{equation*}
Q = \frac{\textrm{Current}}{\textrm{Counting rate}}
\label{equation_charge}
\end{equation*}
where the current and counting rate were measured simultaneously for each RPC module following the procedures described in Sections~\ref{subsec:current} and~\ref{subsec:hitRate}, respectively. These measurements were performed on the same ATLAS run recorded in October 2018, using standard operating voltage and FE threshold values. In this run, about 90\% of the FE channels had thresholds set to the nominal value of 1.0~V, while the majority of the remaining channels used higher-than-nominal thresholds.
 
Counting rate measurements include only those ionisation events that produce signals above the FE thresholds in a given module, while current measurements account for the total ionisation charge in that module. Therefore, the present study overestimates the mean avalanche charge since the ionisation signals below the FE thresholds contribute to the measured current but not to the counting rate. The fraction of ionisation events below the FE thresholds is expected to remain constant as a function of the instantaneous luminosity. This hypothesis was verified by checking that the measurements of the mean avalanche charge do not depend on the specific instantaneous luminosity used to obtain the measurements. This is expected since both the currents and counting rates depend linearly on the instantaneous luminosity as discussed in Sections~\ref{subsec:current} and~\ref{subsec:hitRate}, respectively.
 
The fraction of ionisation events produced by muons is underestimated by approximately 3\%, due to the inefficiency of the readout chain, which is discussed in Section~\ref{subsec:det_eff}. This implies that the mean avalanche charge produced by minimum-ionising particles is overestimated by a similar amount. Most RPC hits are due to photons and neutrons and their avalanche spectra cannot be estimated in situ. Therefore, the precise fraction of the missing ionisation events due to these particles cannot be determined in this analysis.  This implies that the measured mean avalanche charge is probably overestimated. In addition, the mean avalanche charge measurements depend on environmental parameters and on the specific FE threshold values used to obtain the counting rates. Despite these limitations, the mean avalanche charge measurements provide useful benchmarks for studies of detector performance as a function of time and for studies of the expected detector lifetime for the HL-LHC programme.

The counting rate measurements were performed independently for each $\eta$ and $\phi$ panel. The average of the $\eta$ and $\phi$ counting rates for each module was used for avalanche charge measurements. The current and rate measurements were also averaged over the two parallel modules of the doublet layer. The distance between the parallel modules of the RPC doublet layer is approximately 2~cm, so the counting rates are expected to be the same in these two modules.
 
\begin{figure}[!h]
\centering
\subfloat[]{\includegraphics[width=0.49\textwidth]{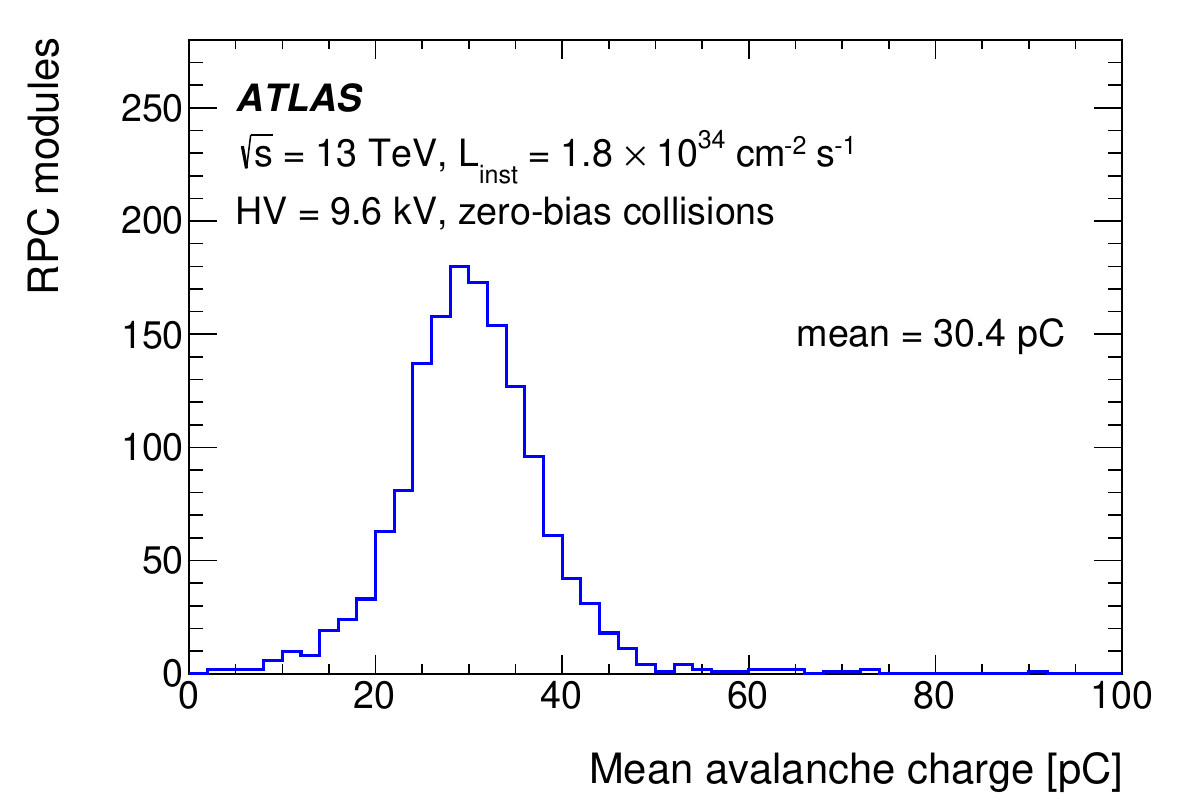}\label{pic:charge1D}}
\subfloat[]{\includegraphics[width=0.45\textwidth]{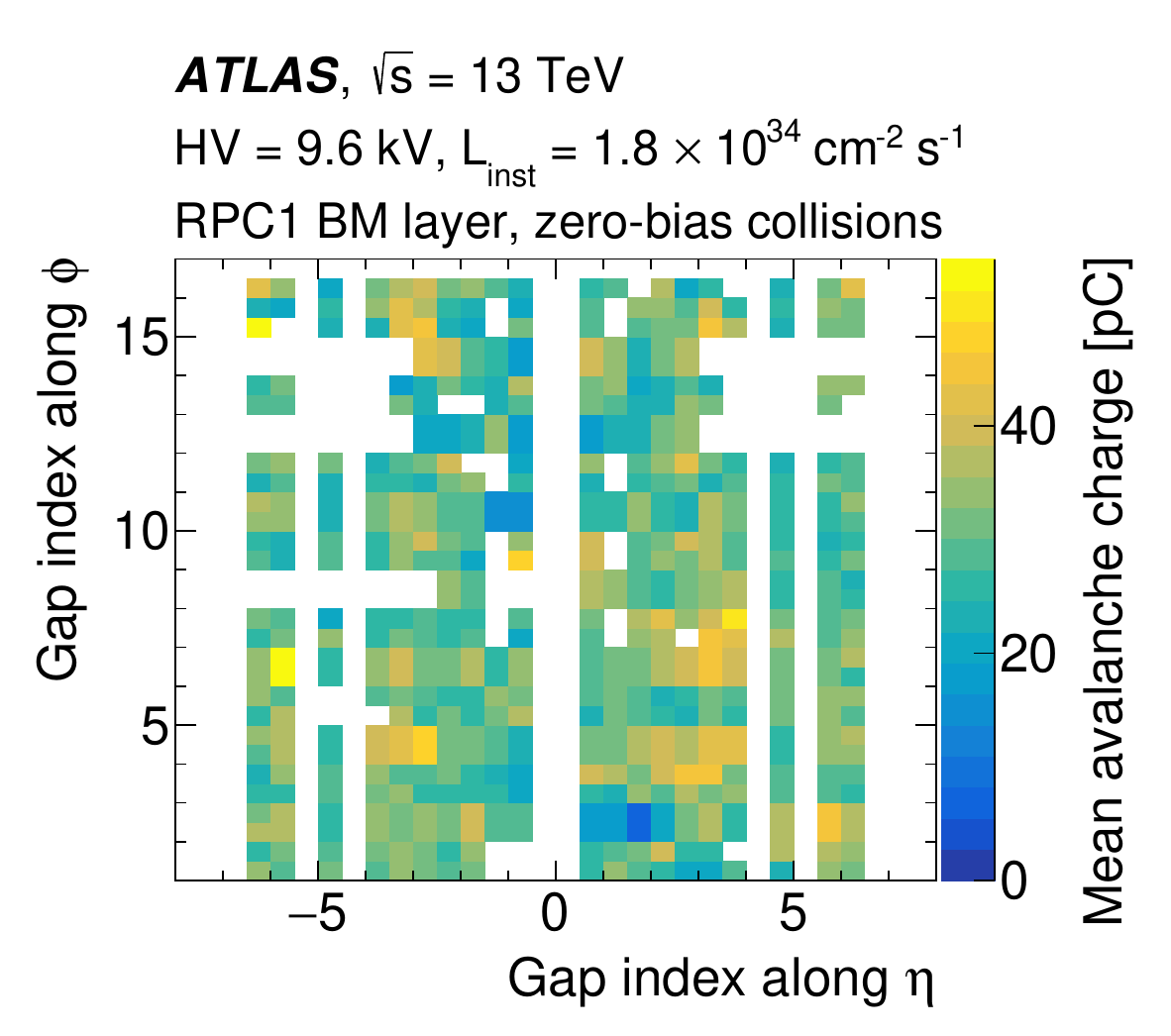}\label{pic:charge2DRPC1}}
\caption{(a) Distribution of the mean avalanche charge of the ATLAS RPC detector. (b) Mean avalanche charge measured as a function of the $\eta$ and $\phi$ positions of the RPCs belonging to the RPC1 doublet layer. Empty bins correspond to logical combinations of indices that do not represent installed RPCs or to modules with efficiency less than 50\%.}
\end{figure}
 
The distribution of the mean avalanche charge is shown in Figure~\ref{pic:charge1D} for all active RPC modules. The same distribution shape was observed when it was computed using only the modules with the nominal FE thresholds. Therefore, the width and shape of this distribution is likely due to natural differences among the modules. The average of the mean avalanche charge distribution is 30.4~pC and is in good agreement with previous measurements performed using a gamma irradiation source, when the same FE thresholds are used, as shown in Figure~7 of Ref.~\cite{Aielli:2000sa}. Those measurements were obtained using RPCs of the same design as the ATLAS RPCs, with the same FE electronics, gas mixture and operating voltage.
 
Figure~\ref{pic:charge2DRPC1} shows the mean avalanche charge as a function of the $\eta$ and $\phi$ positions of the RPCs belonging to the RPC1 doublet layer. The observed variations between different chambers are due to intrinsic variations among modules and to variations in operating voltages and FE threshold settings. The operating voltages are expected to affect the measured current values, while the thresholds influence the measured counting rates.

 
\subsection{RPC efficiency as a function of counting rate}
\label{subsec:pileup}
As discussed in Sections~\ref{subsec:det_eff} and~\ref{subsec:trigEff}, the mean RPC detector efficiency and L1 muon barrel trigger efficiency decrease by approximately 1\% as the number of proton--proton collisions per event increases from 10 to 60 collisions. In this section, this effect is investigated further by correlating the RPC module efficiency with its counting rate. It is assumed that an avalanche multiplication process would deplete available charge on the RPC surface in the vicinity of the avalanche, resulting in a decreased efficiency in that localised area to detect a subsequent avalanche event shortly thereafter. The results presented in this section aim to correlate the muon detection efficiency with the mean counting rate for each individual RPC module and for the detector as the whole.
 
The full dataset of 60.8~fb$^{-1}$ was used for the efficiency measurements in order to maximise the number events with a $Z$ boson decaying into two muons. A smaller dataset of 3.6~fb$^{-1}$ was used for the counting rate measurements in order to simplify analysis procedures. Sufficient statistical precision was obtained using this smaller dataset. The mean RPC detector efficiency and counting rates were measured using only RPC panels with an efficiency greater than 50\%.
 
\begin{figure}[!h]
\begin{center}
\centering
\subfloat[]{\includegraphics[scale=0.4]{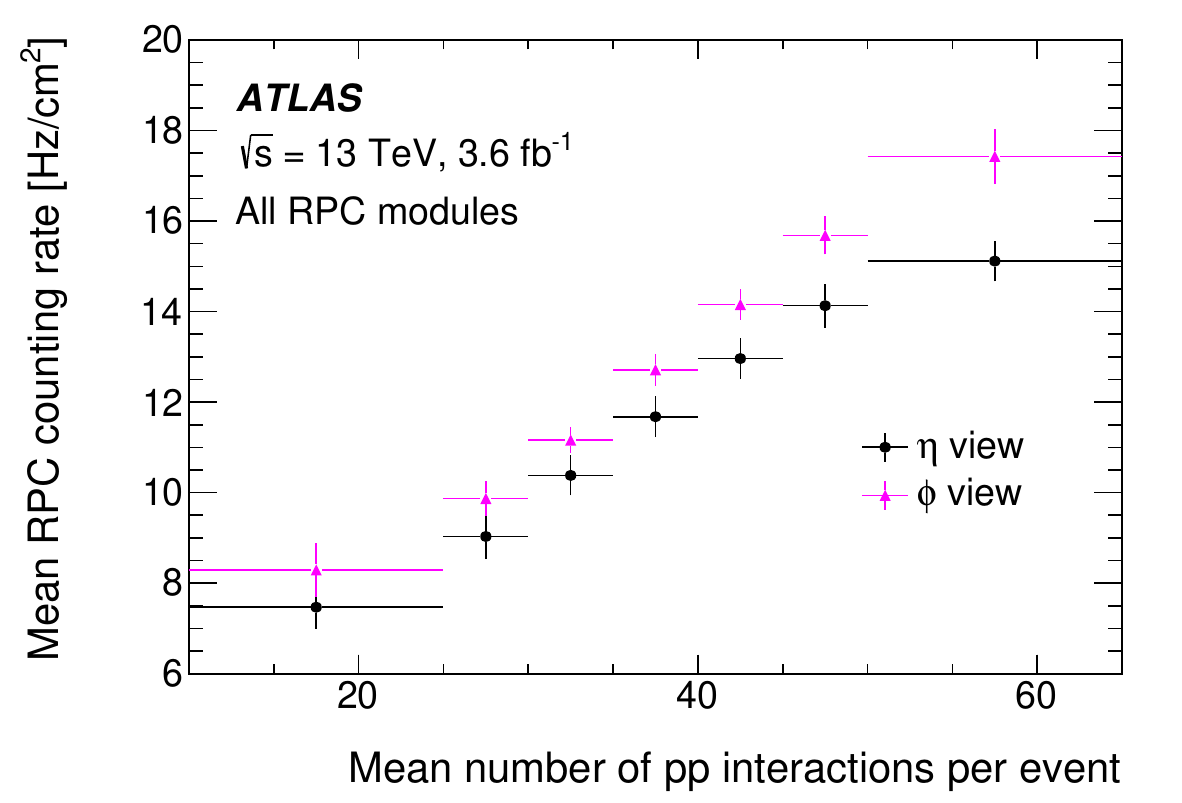}\label{fig:pu_perf_1}}
\subfloat[]{\includegraphics[scale=0.4]{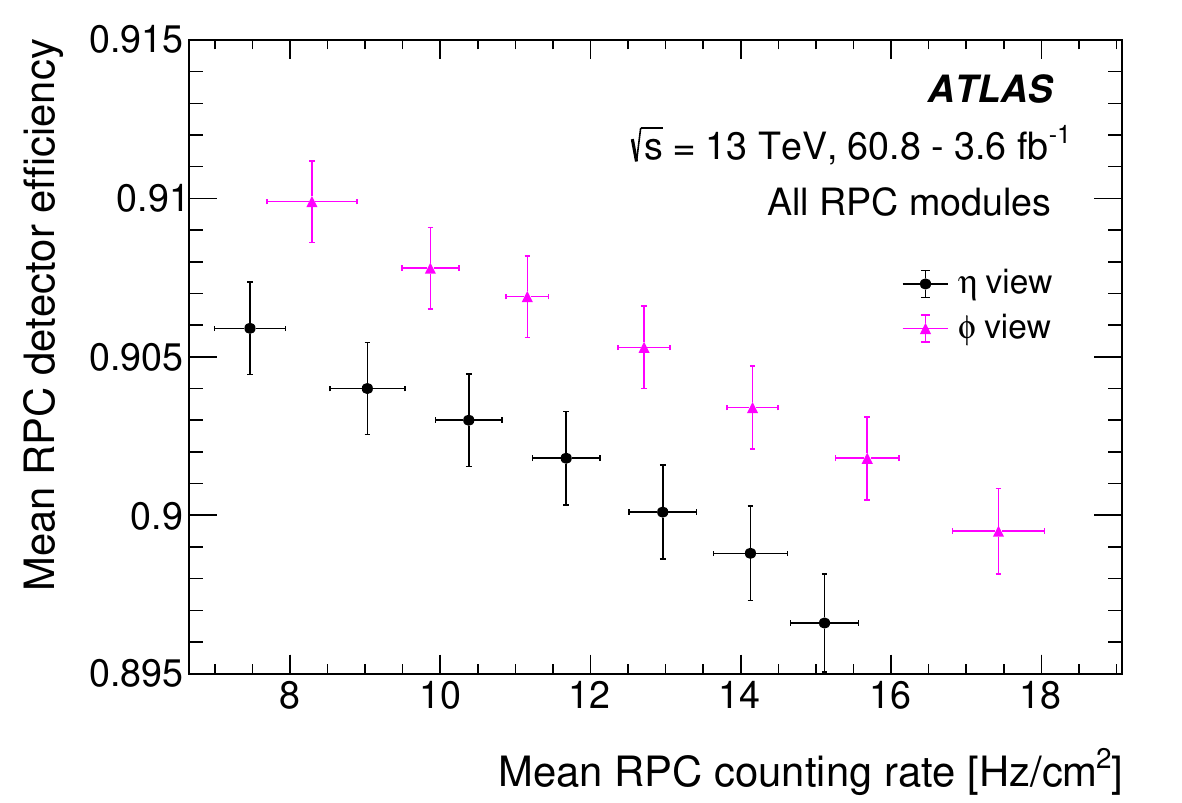}\label{fig:pu_perf_3}}
\caption{(a) Mean RPC counting rates measured in seven different intervals of the number of proton--proton collisions per event. The counting rate was averaged over all active RPC modules and is shown separately for $\eta$ and $\phi$ panels. (b) Correlation between the mean RPC detector efficiency and the mean RPC detector counting rate, measured in seven different intervals of the number of proton--proton collisions per event, shown separately for $\eta$ and $\phi$ panels. Two different datasets, corresponding to 60.8~fb$^{-1}$ and 3.6~fb$^{-1}$, are used for the efficiency measurements and the counting rate measurements, respectively. Both figures were obtained by averaging over all active $\eta$ and $\phi$ panels. Only RPC panels with an average efficiency larger than 50\% were used for these figures.}
\end{center}
\end{figure}

The mean RPC detector counting rate was measured in seven bins of the number of collisions per event, shown in Figure~\ref{fig:pu_perf_1}. A linear dependence of the counting rate was observed, in agreement with the measurements of the RPC counting rates as a function of the instantaneous luminosity presented in Section~\ref{subsec:hitRate}. The mean RPC detector efficiency in the same seven bins were presented in Section~\ref{subsec:det_eff}. Figure~\ref{fig:pu_perf_3} shows the mean RPC detector efficiency plotted as a function of the mean RPC detector counting rate computed separately for the seven bins of the number of collisions per event. A decrease of approximately 1\% is observed within the considered range.
 
\begin{figure}[!h]
\begin{center}
\centering
\subfloat[]{\includegraphics[width=0.49\textwidth]{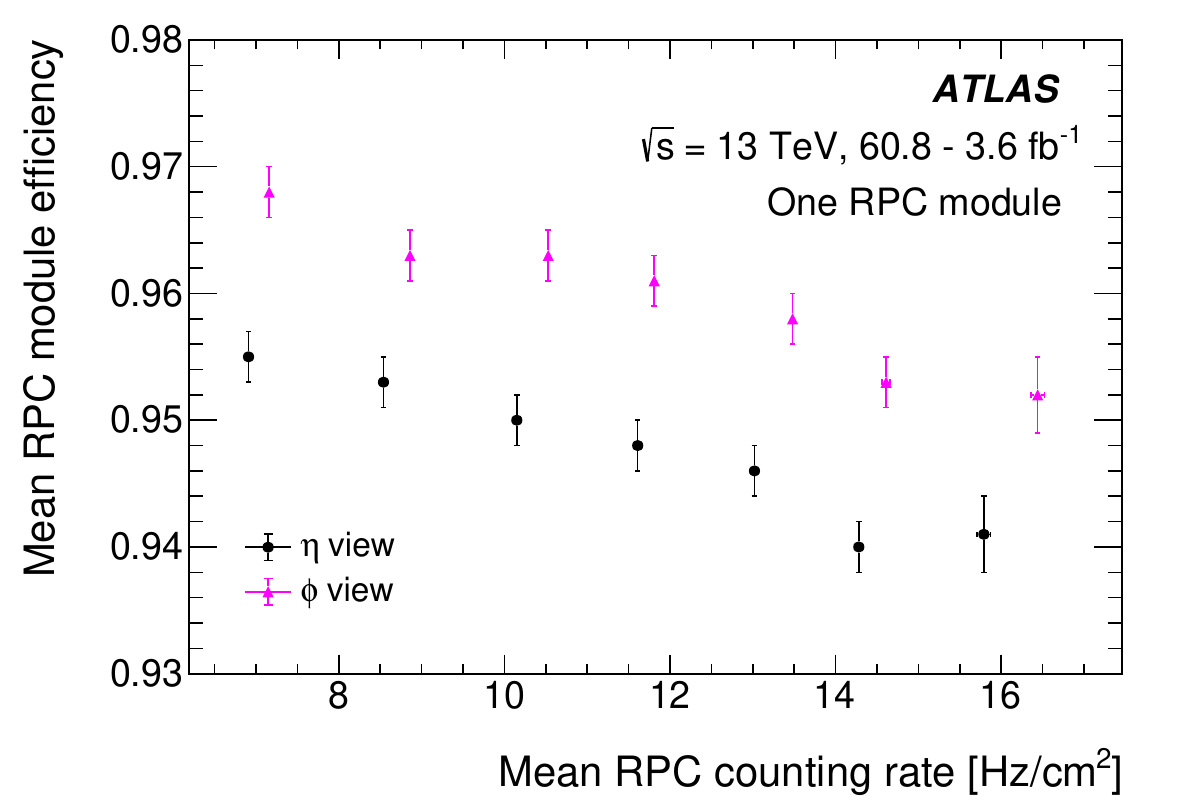}\label{fig:pileup_eff_rate_single}}
\subfloat[]{\includegraphics[width=0.49\textwidth]{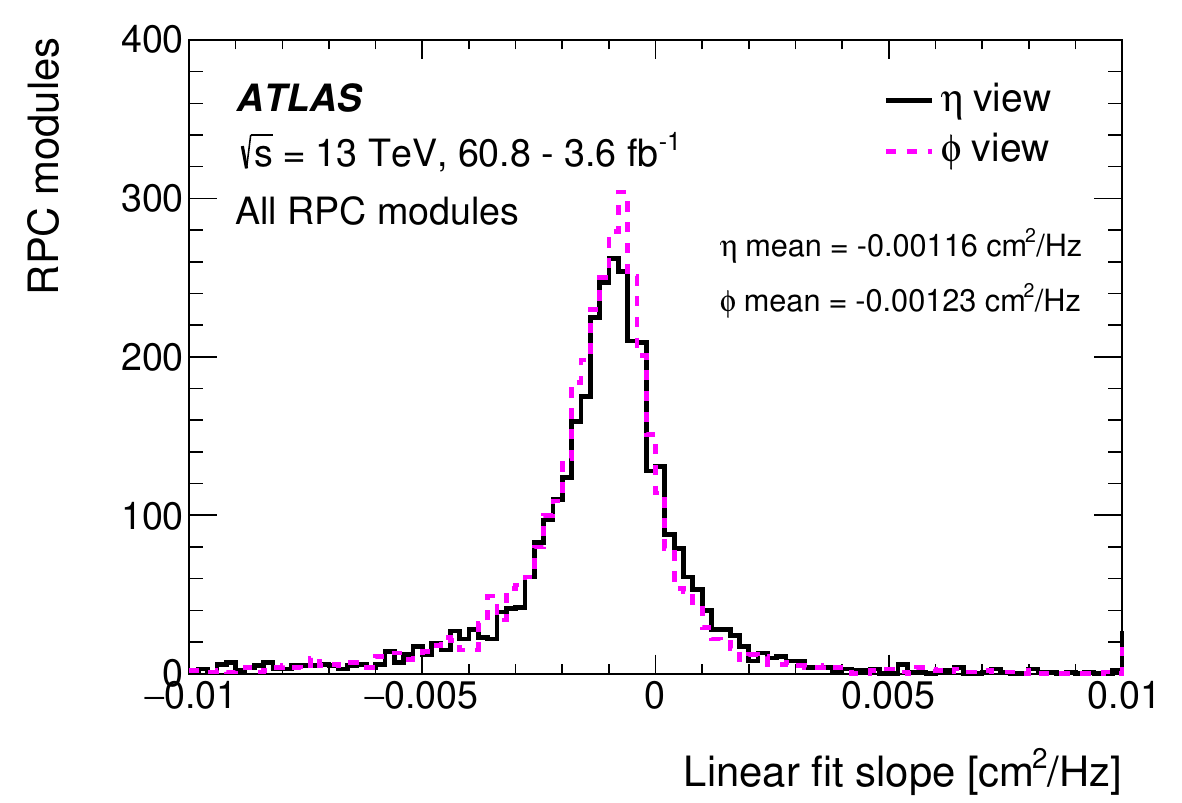}\label{fig:pileup_slope}}
\caption{(a) Correlation between the mean RPC module efficiency and the mean RPC counting rate, measured in seven different intervals of the number of proton--proton collisions per event, for one RPC module in both the $\eta$ and $\phi$ views. (b) Distribution of the slope parameters extracted from a linear fit of the efficiency versus counting rate dependence of each RPC panel. Two different datasets, corresponding to 60.8~fb$^{-1}$ and 3.6~fb$^{-1}$, are used for the efficiency measurements and the counting rate measurements, respectively. Only RPC panels with an average efficiency larger than 50\% were used for this plot.}
\label{fig:pu_perf_2}
\end{center}
\end{figure}

The muon detection efficiency and mean counting rate were also measured for each individual RPC panel in seven bins of the number of collisions per event. The mean module efficiency is plotted as a function of the mean counting rate in Figure~\ref{fig:pileup_eff_rate_single} for one representative RPC module. The slope of the efficiency versus the counting rate for each RPC panel was determined from a linear fit. The distribution of the linear slope parameter is shown in Figure~\ref{fig:pileup_slope} for all selected RPC panels. Only RPC panels with a fit $\chi^{2}$ per degree of freedom less than 5.0 were retained for the analysis. The negative mean value of the distribution indicates clearly that a majority of the RPC panels show a decrease in muon detection efficiency as a function of the counting rate. This result supports the hypothesis that the observed decrease in the RPC detector efficiency at higher instantaneous luminosity is due to the increase in RPC counting rates in individual modules.
 
\FloatBarrier
 
\section{Expected performance of the existing RPCs at HL-LHC}
\label{sec:HL-LHC}
The ATLAS detector is scheduled to operate with the HL-LHC until approximately 2040.  An extensive set of upgrades will be installed in 2025--2026.  These include upgrades of the MS and its electronics~\cite{CERN-LHCC-2017-017}, and of the trigger system~\cite{CERN-LHCC-2017-020}. In particular, three new layers of RPCs will be installed in the innermost region of the barrel MS in order to improve redundancy in the RPC system, to increase muon trigger efficiency and to reduce trigger rates.  More sophisticated, customisable trigger algorithms will be deployed utilising field-programmable gate array (FPGA) devices. The MDT detector will be also used by the hardware-level muon trigger~\cite{CERN-LHCC-2017-017}, allowing more effective rejection of muons with {\pt} below a target trigger threshold.
 
The ATLAS RPC system was originally designed to operate at an instantaneous luminosity of $\silumi{34}$ for a total integrated luminosity of $300~\text{fb}^{-1}$. The same RPCs and FE electronics will be used for HL-LHC operations at an instantaneous luminosity of $7.5 \times \silumi{34}$ for a total integrated luminosity of more than $3000~\text{fb}^{-1}$. The majority of RPCs are expected to stay within safe operating limits at the HL-LHC, as detailed in Ref.~\cite{CERN-LHCC-2017-017}. Nevertheless, the integrated charge collected by the chambers located at $|\eta| \approx 1$ will exceed the design specifications by more than a factor of three. In order to operate these RPCs safely during HL-LHC data-taking, the operating voltage will be lowered from the nominal value of 9.6~kV to approximately 9.2~kV, which will lead to a reduction of the muon detection efficiency.
 
This section presents several studies that extrapolate the performance of the existing RPCs to the conditions expected during HL-LHC operations. Section~\ref{subsec:HL_current} presents extrapolations of RPC currents to the expected HL-LHC instantaneous luminosity for both the default and reduced operating voltage settings. Section~\ref{subsec:HL_threshold} presents measurements of the RPC counting rates measured using zero-bias collisions, obtained at different operating voltage and FE threshold settings. This section also presents measurements of RPC detector efficiency and cluster size, obtained at different operating voltage and FE threshold settings, using muons produced in proton--proton collisions.
 
\subsection{Expected RPC currents and counting rates at the HL-LHC}
\label{subsec:HL_current}
\begin{figure}[!h]
\centering
\includegraphics[width=0.495\textwidth]{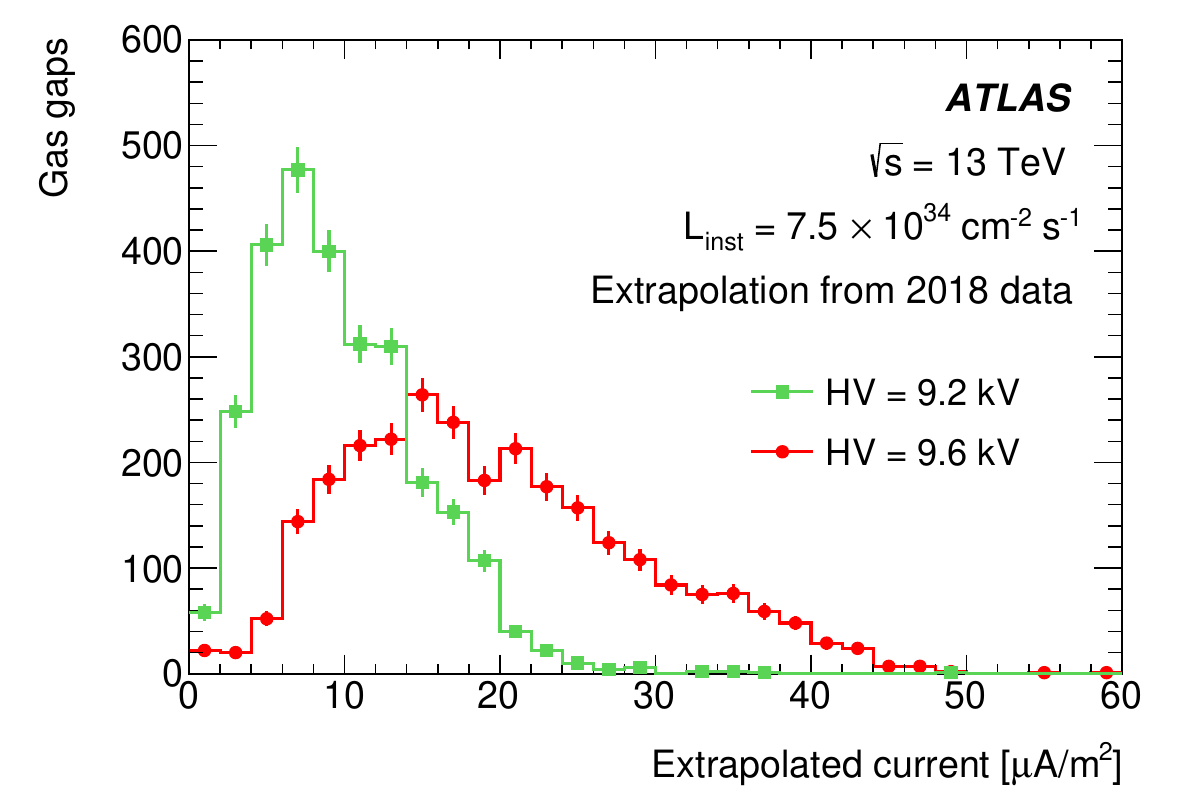}
\caption{Distributions of the expected current density for the selected RPCs extrapolated to an instantaneous luminosity of $7.5 \times \silumi{34}$.}
\label{pic:Igap_vs_HV_HL}
\end{figure}

In order to extrapolate RPC currents to the expected HL-LHC instantaneous luminosity of $7.5 \times \silumi{34}$, the measured current density of each module was scaled by a factor of 4.2. Figure~\ref{pic:Igap_vs_HV_HL} shows the distributions of the extrapolated current density for all selected RPCs at the instantaneous luminosity of $7.5 \times \silumi{34}$. For a fraction of the RPCs operating at 9.6~kV, the extrapolated current density exceeds the safety limit of about $\idens{30}$ required for stable RPC operations at the HL-LHC~\cite{CERN-LHCC-2017-017}. The expected current density in the affected RPCs can be reduced by lowering the operating voltage to 9.2~kV, thereby allowing safe operation of nearly all RPCs. The lower voltage also leads to a decrease of the muon detection efficiency that can be partially recovered by adjusting the thresholds of the FE electronics, as discussed in Section~\ref{subsec:HL_threshold}.
 
\begin{figure}[!h]
\begin{center}
\subfloat[]{\includegraphics[width=0.495\textwidth]{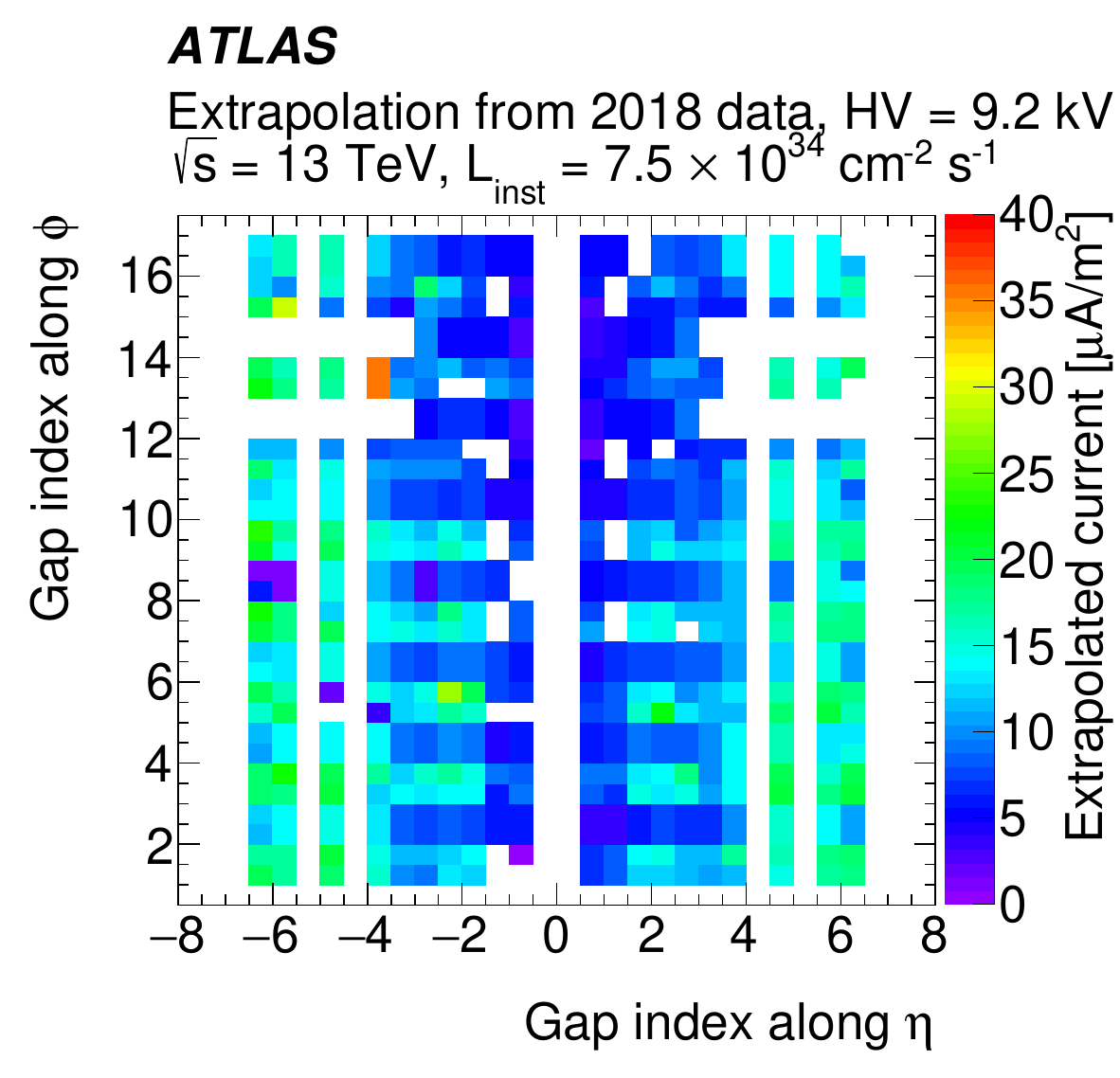}\label{pic:BM_CO_9_2}}
\subfloat[]{\includegraphics[width=0.495\textwidth]{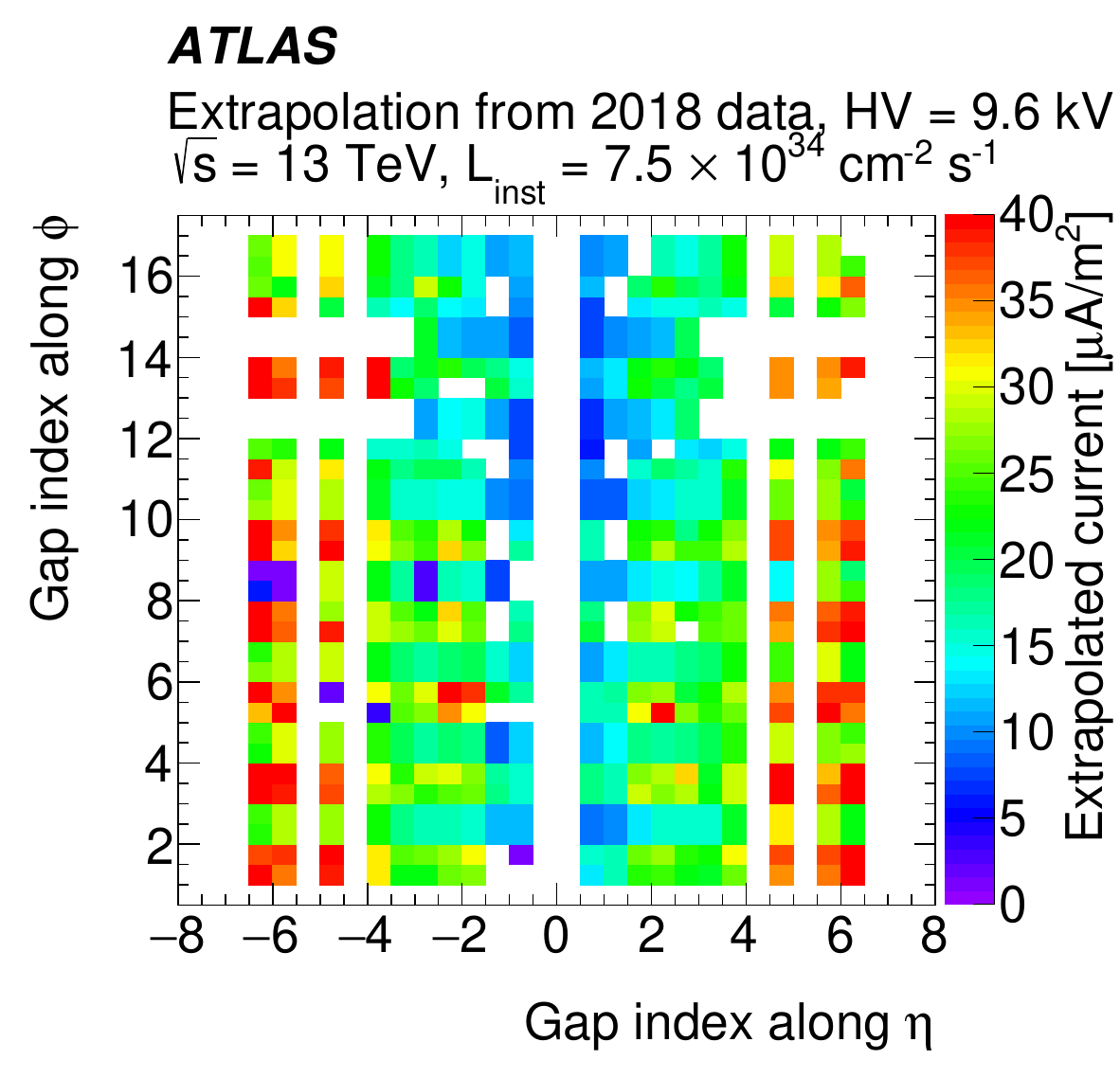}\label{pic:BM_CO_9_6}}
\caption{Expected current density at an instantaneous luminosity of $7.5 \times \silumi{34}$ shown as a function of the $\eta$ and $\phi$ position for the barrel RPCs belonging to the RPC middle doublet layer. The current density values were averaged over the two individual sub-layers and are shown for operating voltages of (a) 9.2~kV and (b) 9.6~kV. Empty bins correspond to logical combinations of indices that do not represent installed RPCs or to modules that failed the selection criteria.}
\label{pic:Igap_vs_HV_etaphi}
\end{center}
\end{figure}

Figures~\ref{pic:BM_CO_9_2} and \ref{pic:BM_CO_9_6} show the expected current density measurements at an instantaneous luminosity of $7.5 \times \silumi{34}$ for operating voltage values of 9.2~kV and 9.6~kV, respectively. The current density measurements are shown as a function of $\eta$ and $\phi$ position for the RPCs belonging to the RPC1 doublet layer. The values were averaged over the two individual layers of each RPC chamber. The expected current density for each module was obtained using the linear extrapolation method described previously. The RPCs with current density higher than the safety limit of $\idens{30}$ are located primarily at high values of $|\eta|$. The expected current density is slightly lower for the RPCs located at the bottom of the ATLAS detector, corresponding to the sectors with higher $\phi$ indices (from 10 up to 16). This effect is consistent with the temperature gradient in the ATLAS cavern, where lower temperatures are measured in the bottom part of the detector.
 
\subsection{RPC detector performance using different operating voltage and FE threshold settings}
\label{subsec:HL_threshold}
As discussed in Section~\ref{subsec:HL_current}, the operating voltage will be lowered to approximately 9.2~kV for some RPCs to permit safe operation at the HL-LHC. This reduction will be necessary to keep ionisation currents below the safety limit for those chambers. To check the resulting effects on RPC performance, three runs were recorded in 2018 with several modules operating at 9.2~kV and using three different FE threshold settings {\vfe}: 0.8~V, 0.9~V and 1.0~V. In order to minimise the impact on the muon trigger efficiency during ATLAS physics data-taking, these non-standard settings were applied to only one layer of the outermost RPC doublet layer (RPC3) of the large $\phi$ sector 13 in ATLAS side A.
 
\begin{figure}[!h]
\centering
\subfloat[]{\includegraphics[width=0.495\textwidth]{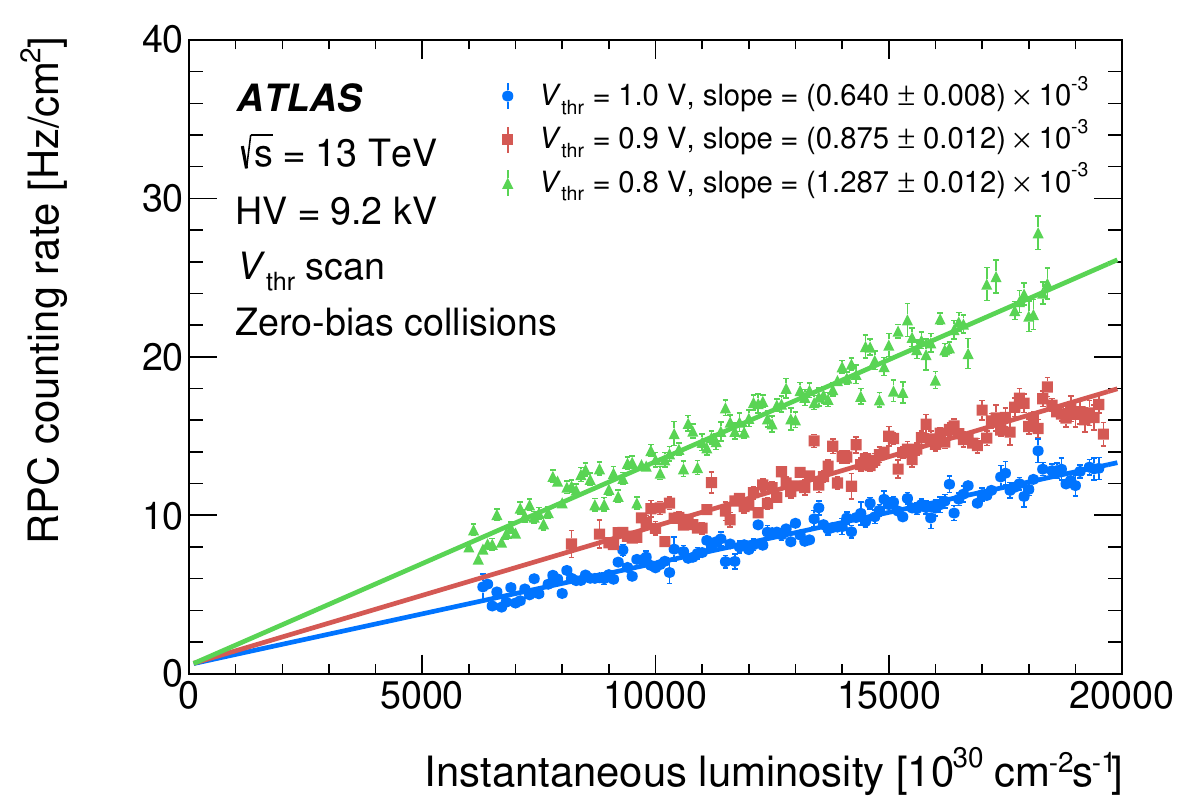}\label{fig:rate_9p2}}
\subfloat[]{\includegraphics[width=0.495\textwidth]{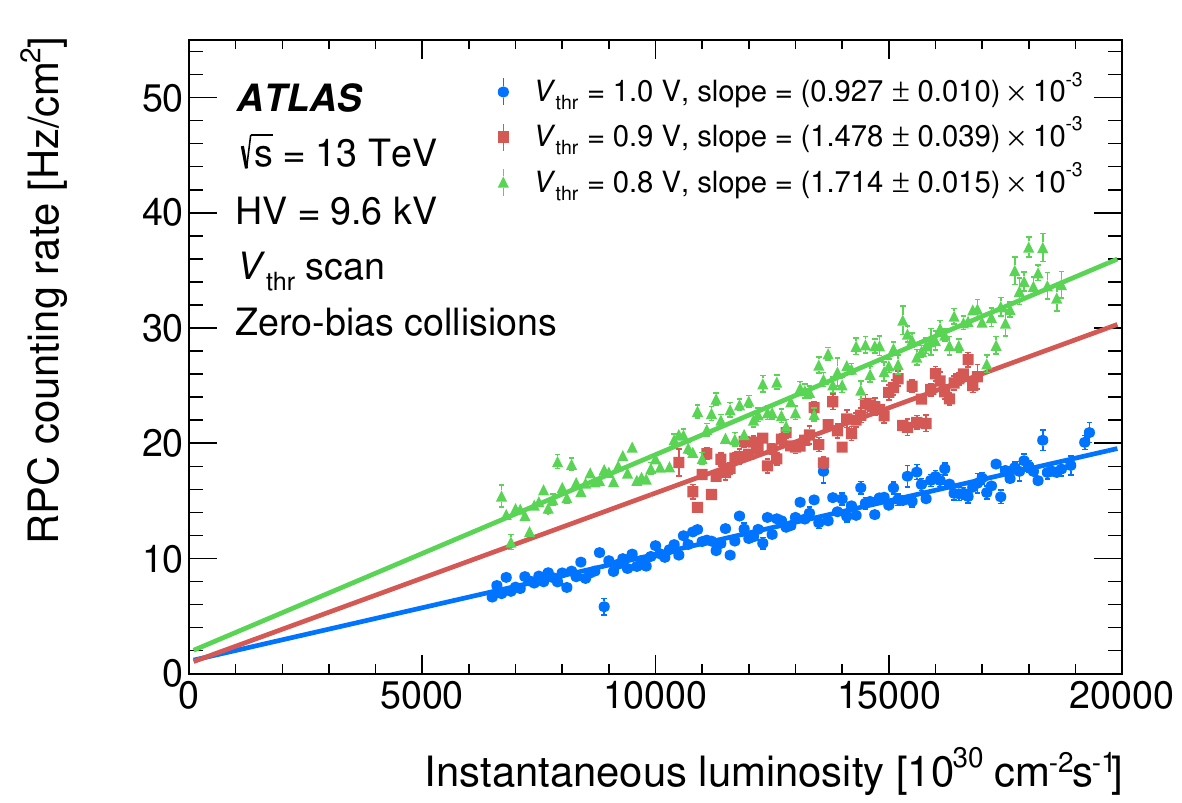}\label{fig:rate_9p6}}
\caption{RPC counting rates per unit surface area as a function of instantaneous luminosity for one representative $\eta$ panel operating at (a) 9.2~kV and (b) 9.6~kV, using different FE threshold settings. This module belongs to the BOL doublet layer and is located in large $\phi$ sector 1 and $\eta$ station 5. Slope parameters were extracted from a linear fit to data points.}
\label{fig:rate_vs_lumi_2}
\end{figure}
 
\begin{figure}[!h]
\centering
\includegraphics[width=0.495\textwidth]{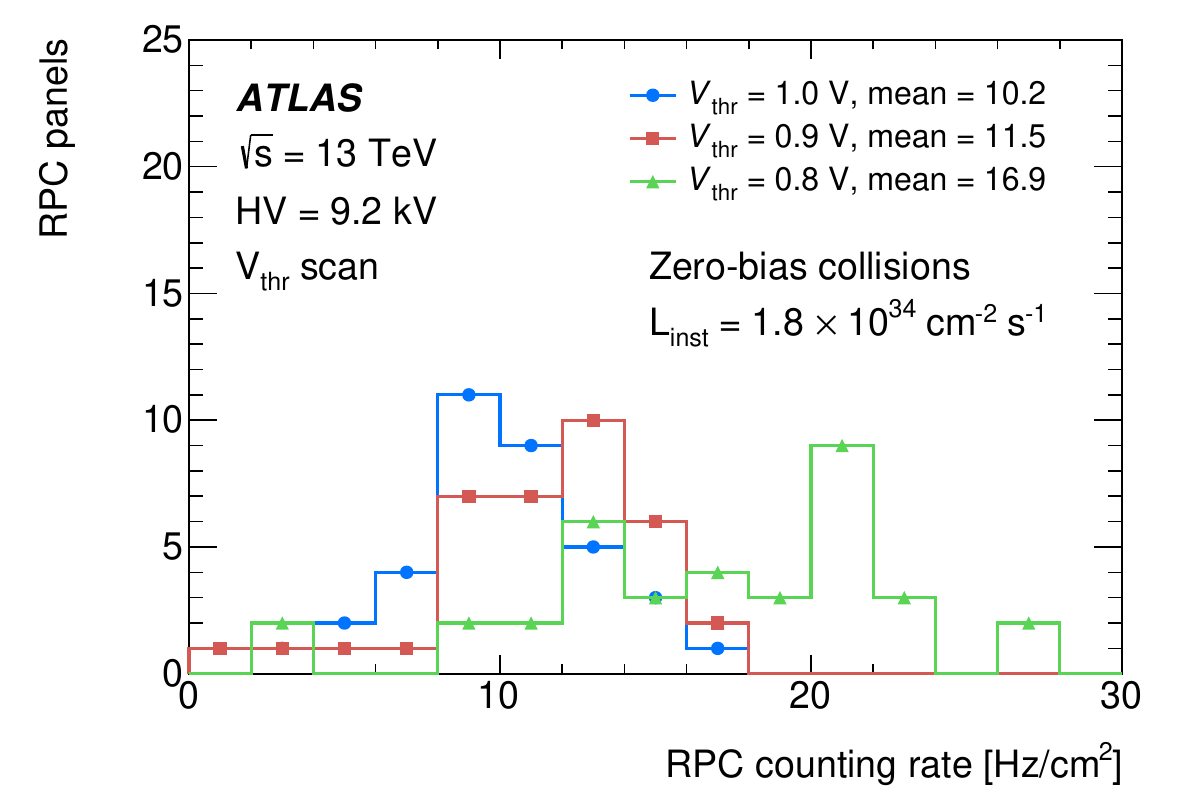}
\caption{Distribution of the measured RPC counting rates at an instantaneous luminosity of $1.8 \times \silumi{34}$ for panels operating at 9.2~kV using three different FE threshold settings.}
\label{fig:rateDistributions}
\end{figure}

Figure~\ref{fig:rate_vs_lumi_2} shows the area-normalised RPC counting rates as a function of instantaneous luminosity for one representative RPC $\eta$ panel operating at 9.2~kV and 9.6~kV, and using three different FE threshold settings. Decreasing the FE threshold from 1.0~V to 0.8~V results in an increase of the RPC counting rates by about a factor of~two. This effect is also demonstrated in Figure~\ref{fig:rateDistributions}, which shows the distributions of the counting rates measured at an instantaneous luminosity of $1.8 \times \silumi{34}$ for 36 tested panels operating at 9.2~kV and using three different FE thresholds.

\begin{figure}[!h]
\begin{center}
\centering
\subfloat[]{\includegraphics[width=0.495\textwidth]{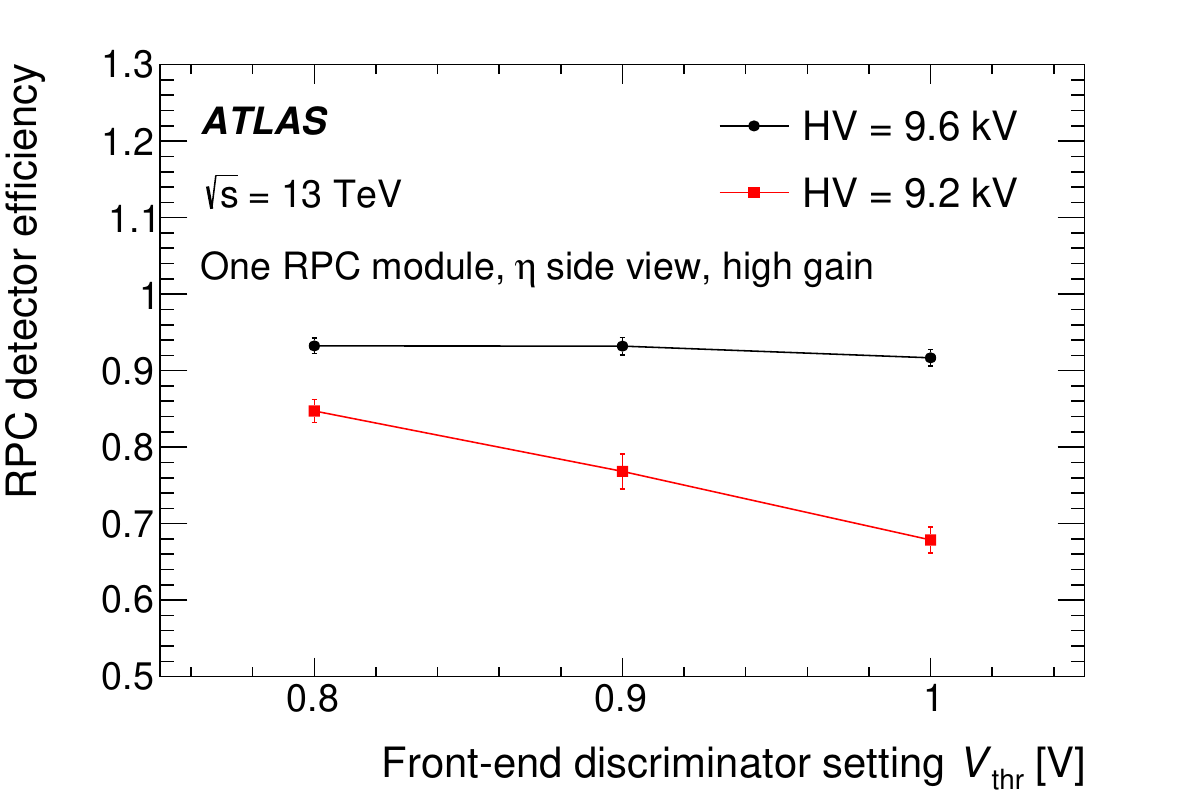}\label{fig:scanVthEta}}
\subfloat[]{\includegraphics[width=0.495\textwidth]{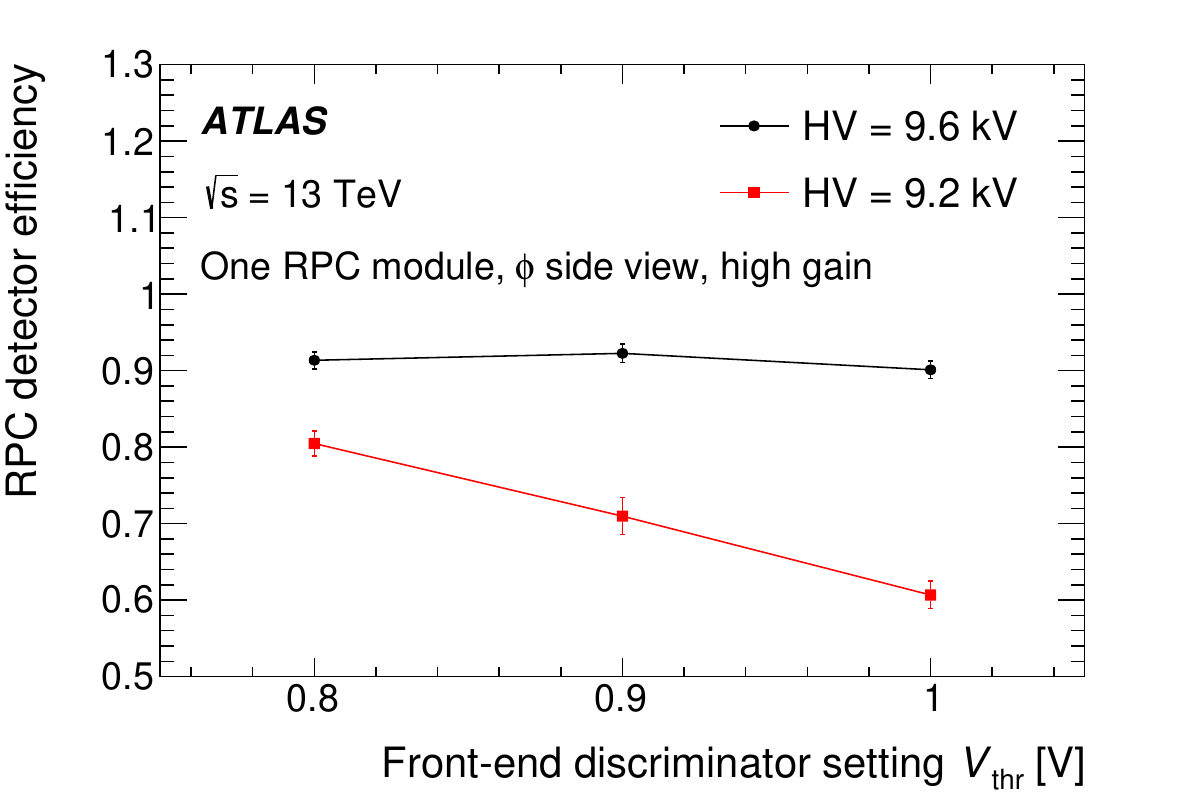}\label{fig:scanVthPhi}}
\caption{RPC detector efficiency plotted as a function of the FE threshold setting {\vfe} for the (a) $\eta$ panel and (b) $\phi$ panel of one representative RPC module, measured at the operating voltage values of 9.2~kV and 9.6~kV.}
\label{fig:scanVth}
\end{center}
\end{figure}

Figure~\ref{fig:scanVth} shows the muon detection efficiency as a function of {\vfe} values for the representative $\eta$ and $\phi$ panels belonging to the same module operating at 9.2~kV and 9.6~kV. When the voltage was lowered to 9.2~kV and the FE threshold was kept at the nominal value of 1.0~V, the absolute efficiency was reduced by about 24\% and 30\% for the $\eta$ and $\phi$ panel, respectively. Then decreasing the FE threshold from 1.0~V to 0.8~V leads to an absolute efficiency gain of about 17\% for the $\eta$ panel and 20\% for the $\phi$ panel. The efficiency gain is computed as the difference between the muon detection efficiencies computed at the {\vfe} of 1.0~V and 0.8~V, when operating at 9.2~kV. A constant detector efficiency is observed while varying the FE thresholds at 9.6~kV because the RPC is operating at the efficiency plateau and therefore it is not sensitive to changes of the FE threshold values.

\begin{figure}[!h]
\begin{center}
\centering
\subfloat[]{\includegraphics[width=0.495\textwidth]{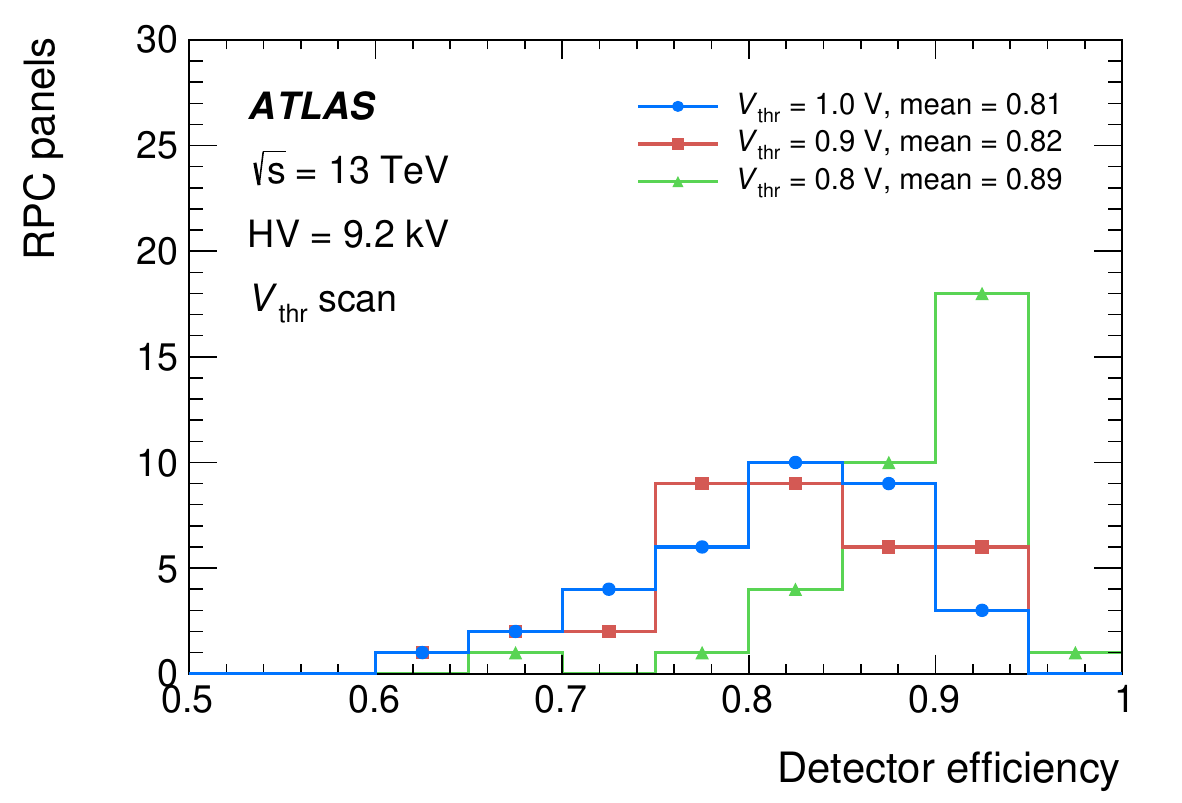}\label{fig:distributionsVth1}}
\subfloat[]{\includegraphics[width=0.495\textwidth]{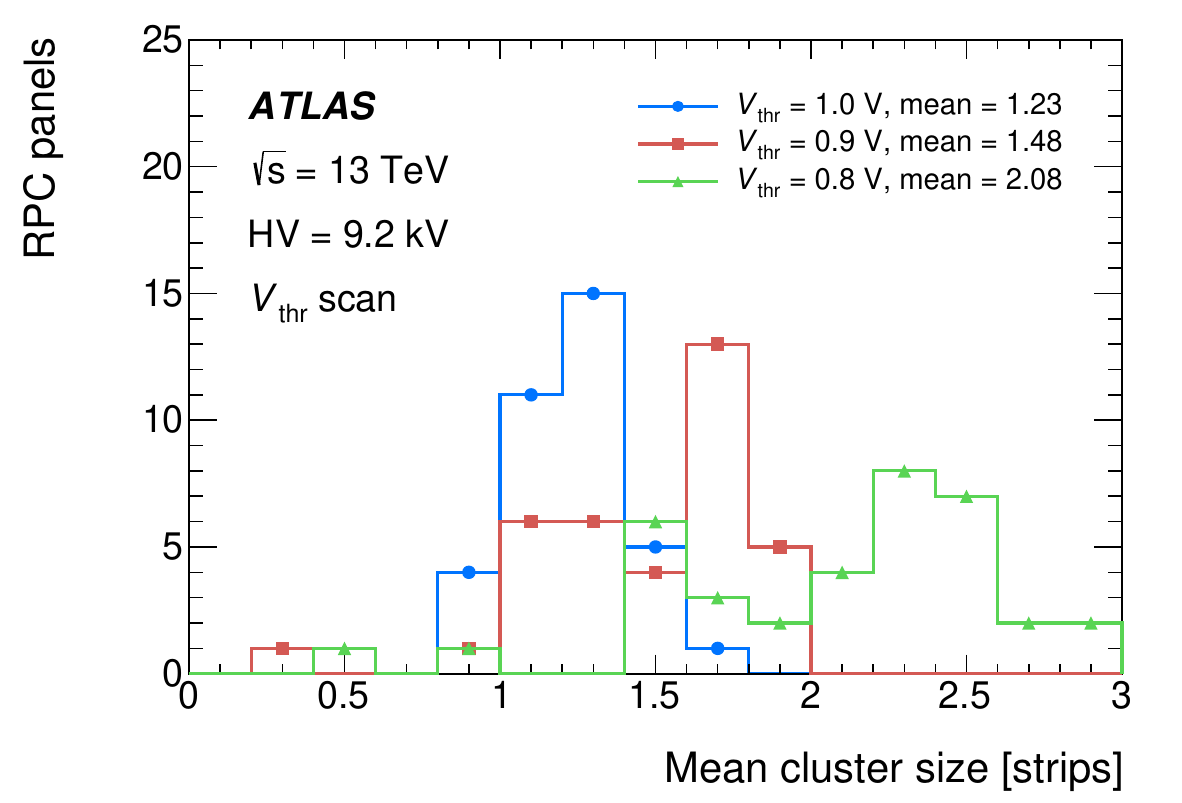}\label{fig:distributionsVth2}}
\caption{Distributions of (a) the panel efficiencies and (b) the mean cluster size measured at different {\vfe} threshold values with the operating voltage set to 9.2~kV.}
\label{fig:distributionsVth}
\end{center}
\end{figure}

Figure~\ref{fig:distributionsVth1} shows the distribution of the panel efficiencies at different {\vfe} threshold values for the 9.2~kV voltage setting. The average absolute efficiency gain is around 8\% when lowering the {\vfe} from 1.0~V to 0.8~V. The corresponding distribution for the mean cluster size is shown in Figure~\ref{fig:distributionsVth2}.
 
Despite the RPC efficiency decrease due to the lower applied voltage, the overall efficiency of the L1 muon barrel trigger is expected to increase after the ATLAS MS and trigger upgrades~\cite{CERN-LHCC-2017-017,CERN-LHCC-2017-020} planned for the HL-LHC. After these upgrades, three new layers of RPCs will increase the redundancy of the muon barrel trigger system and lead to a higher overall trigger efficiency. A study of the expected muon trigger performance after these upgrades is outside the scope of the present work, which instead aims to evaluate the expected performance of the RPC detector. The present studies demonstrate that lowering the operating voltage to 9.2~kV will allow safe RPC operation and provide an RPC detector efficiency similar to that obtained using the nominal voltage if the FE thresholds are lowered.
 
 
\section{Conclusions}
Measurements of the performance of the ATLAS RPC detector and L1 muon barrel trigger were obtained using data from \ilumi\ of $\sqrt{s}=13$~\TeV\ proton--proton collisions at the LHC recorded in 2018. The response of individual RPC modules was measured using muons produced in decays of $Z$ bosons, including measurements of efficiency, cluster hit multiplicity and time resolution. The RPC response was also measured as a function of time and as a function of the mean number of proton--proton collisions per bunch crossing. Performance of the L1 muon barrel trigger was also measured using muons produced in decays of $Z$ bosons, including measurements of trigger efficiency as a function of several variables. Acceptance rates were measured for the L1 muon trigger thresholds and the composition of accepted events was studied using reconstructed muon candidates. Measurements of RPC counting rates and currents were performed using zero-bias collisions, which are dominated by background events from neutron and photon interactions with the RPCs. The avalanche charge was measured for all RPC modules and was found to be consistent with earlier measurements in test beams. The RPC detector response was measured as a function of the instantaneous luminosity and several possible scenarios were tested for operating the RPCs for data-taking with the High Luminosity LHC. Overall, the RPC detector and L1 muon trigger show excellent performance.
 
\section*{Acknowledgements}

We thank CERN for the very successful operation of the LHC, as well as the
support staff from our institutions without whom ATLAS could not be
operated efficiently.
 
We acknowledge the support of ANPCyT, Argentina; YerPhI, Armenia; ARC, Australia; BMWFW and FWF, Austria; ANAS, Azerbaijan; SSTC, Belarus; CNPq and FAPESP, Brazil; NSERC, NRC and CFI, Canada; CERN; ANID, Chile; CAS, MOST and NSFC, China; COLCIENCIAS, Colombia; MSMT CR, MPO CR and VSC CR, Czech Republic; DNRF and DNSRC, Denmark; IN2P3-CNRS and CEA-DRF/IRFU, France; SRNSFG, Georgia; BMBF, HGF and MPG, Germany; GSRT, Greece; RGC and Hong Kong SAR, China; ISF and Benoziyo Center, Israel; INFN, Italy; MEXT and JSPS, Japan; CNRST, Morocco; NWO, Netherlands; RCN, Norway; MNiSW and NCN, Poland; FCT, Portugal; MNE/IFA, Romania; JINR; MES of Russia and NRC KI, Russian Federation; MESTD, Serbia; MSSR, Slovakia; ARRS and MIZ\v{S}, Slovenia; DST/NRF, South Africa; MICINN, Spain; SRC and Wallenberg Foundation, Sweden; SERI, SNSF and Cantons of Bern and Geneva, Switzerland; MOST, Taiwan; TAEK, Turkey; STFC, United Kingdom; DOE and NSF, United States of America. In addition, individual groups and members have received support from BCKDF, CANARIE, Compute Canada, CRC and IVADO, Canada; Beijing Municipal Science \& Technology Commission, China; COST, ERC, ERDF, Horizon 2020 and Marie Sk{\l}odowska-Curie Actions, European Union; Investissements d'Avenir Labex, Investissements d'Avenir Idex and ANR, France; DFG and AvH Foundation, Germany; Herakleitos, Thales and Aristeia programmes co-financed by EU-ESF and the Greek NSRF, Greece; BSF-NSF and GIF, Israel; La Caixa Banking Foundation, CERCA Programme Generalitat de Catalunya and PROMETEO and GenT Programmes Generalitat Valenciana, Spain; G\"{o}ran Gustafssons Stiftelse, Sweden; The Royal Society and Leverhulme Trust, United Kingdom.
 
The crucial computing support from all WLCG partners is acknowledged gratefully, in particular from CERN, the ATLAS Tier-1 facilities at TRIUMF (Canada), NDGF (Denmark, Norway, Sweden), CC-IN2P3 (France), KIT/GridKA (Germany), INFN-CNAF (Italy), NL-T1 (Netherlands), PIC (Spain), ASGC (Taiwan), RAL (UK) and BNL (USA), the Tier-2 facilities worldwide and large non-WLCG resource providers. Major contributors of computing resources are listed in Ref.~\cite{ATL-SOFT-PUB-2020-001}.
 

\printbibliography

@article{BarriusoPoy:2008zz,
      author         = "Barriuso Poy, A. and others",
      title          = "{The detector control system of the ATLAS experiment}",
      journal        = "JINST",
      volume         = "3",
      year           = "2008",
      pages          = "P05006",
      doi            = "10.1088/1748-0221/3/05/P05006",
      SLACcitation   = "%%CITATION = JINST,3,P05006;%%"
}

@article{Polini:2012zz,
      author         = "Polini, Alessandro",
      title          = "{Design and performance of the detector control system of
                        the ATLAS resistive-plate-chamber muon spectrometer}",
      booktitle      = "{Resistive plate chambers and related detectors.
                        Proceedings, 10th Workshop, RPC 2010, Darmstadt, Germany,
                        February 9-12, 2010}",
      collaboration  = "ATLAS Muon",
      journal        = "Nucl. Instrum. Meth. A",
      volume         = "661",
      year           = "2012",
      pages          = "S15-S18",
      doi            = "10.1016/j.nima.2010.08.006",
      SLACcitation   = "%%CITATION = NUIMA,A661,S15;%%"
}

@article{Aielli:2000rz,
      author         = "Aielli, G. and others",
      title          = "{Performance of a large-size RPC equipped with the final
                        ATLAS front end electronics at X5-GIF irradiation
                        facility}",
      booktitle      = "{Resistive plate chambers and related detectors
                        Proceedings, 5th International Workshop, RPC'99, Bari,
                        Italy, October 28-29, 1999}",
      journal        = "Nucl. Instrum. Meth. A",
      volume         = "456",
      year           = "2000",
      pages          = "77-81",
      doi            = "10.1016/S0168-9002(00)00966-9",
      SLACcitation   = "%%CITATION = NUIMA,A456,77;%%"
}

@article{Chiodini:2013rta,
      author         = "Chiodini, G. and Spagnolo, S.",
      title          = "{Off-line time calibration of the ATLAS RPC system}",
      booktitle      = "{Proceedings, 11th Workshop on Resistive Plate Chambers
                        and Related Detectors: Frascati, Rome, Italy, 5-10 Feb
                        2012}",
      collaboration  = "ATLAS",
      journal        = "JINST",
      volume         = "8",
      year           = "2013",
      pages          = "T02004",
      doi            = "10.1088/1748-0221/8/02/T02004",
      SLACcitation   = "%%CITATION = JINST,8,T02004;%%"
}

@article{Aad:2008zzm,
      author         = "{ATLAS Collaboration}",            
      title          = "{The ATLAS Experiment at the CERN Large Hadron Collider}",
      collaboration  = "ATLAS",
      journal        = "JINST",
      volume         = "3",
      year           = "2008",
      pages          = "S08003",
      doi            = "10.1088/1748-0221/3/08/S08003",
      SLACcitation   = "%%CITATION = JINST,3,S08003;%%"
}

@Report{ATLAS-TDR-2010-19,
    author         = "{ATLAS Collaboration}",
    title          = "{ATLAS Insertable B-Layer Technical Design Report}",
    type           = "ATLAS-TDR-19; CERN-LHCC-2010-013",
    year           = "2010",
    url            = "https://cds.cern.ch/record/1291633",
}

@Article{PIX-2018-001,
    author         = "Abbott, B. and others",
    title          = "{Production and integration of the ATLAS Insertable B-Layer}",
    journal        = "JINST",
    volume         = "13",
    year           = "2018",
    pages          = "T05008",
    doi            = "10.1088/1748-0221/13/05/T05008",
    eprint         = "1803.00844",
    archivePrefix  = "arXiv",
    primaryClass   = "physics.ins-det",
}

@article{Anulli:2009zz,
      author         = "Anulli, F. and others",
      title          = "{The Level-1 Trigger Muon Barrel System of the ATLAS
                        experiment at CERN}",
      journal        = "JINST",
      volume         = "4",
      year           = "2009",
      pages          = "P04010",
      doi            = "10.1088/1748-0221/4/04/P04010",
      SLACcitation   = "%%CITATION = JINST,4,P04010;%%"
}

@article{Riegler:2002vg,
      author         = "Riegler, W. and Lippmann, C. and Veenhof, R.",
      title          = "{Detector physics and simulation of resistive plate
                        chambers}",
      journal        = "Nucl. Instrum. Meth. A",
      volume         = "500",
      year           = "2003",
      pages          = "144-162",
      doi            = "10.1016/S0168-9002(03)00337-1",
      reportNumber   = "CERN-EP-2002-046",
      SLACcitation   = "%%CITATION = NUIMA,A500,144;%%"
}

@Booklet{CERN-LHCC-2017-017,
    author         = "{ATLAS Collaboration}",
    title          = "{Technical Design Report for the ATLAS Muon Spectrometer Phase-II Upgrade}",
    howpublished   = "{CERN-LHCC-2017-017}",
    url            = "http://cdsweb.cern.ch/record/2285580",
%    note           = {Available at \url{http://cdsweb.cern.ch/record/2285580}},    
    year           = "2017",
}

@Booklet{CERN-LHCC-2017-020,
    author         = "{ATLAS Collaboration}",
    title          = "{Technical Design Report for the Phase-II Upgrade of the ATLAS Trigger and Data Acquisition System}",
    howpublished   = "{CERN-LHCC-2017-020}",
    url            = "http://cdsweb.cern.ch/record/2285584",    
%    note           = {Available at \url{http://cdsweb.cern.ch/record/2285584}},
    year           = "2018",
}

@article{Aaboud:2016leb,
      author         = "{ATLAS Collaboration}",                  
      title          = "{Performance of the ATLAS Trigger System in 2015}",
      collaboration  = "ATLAS",
      journal        = "Eur. Phys. J. C",
      volume         = "77",
      year           = "2017",
      number         = "5",
      pages          = "317",
      doi            = "10.1140/epjc/s10052-017-4852-3",
      eprint         = "1611.09661",
      archivePrefix  = "arXiv",
      primaryClass   = "hep-ex",
      reportNumber   = "CERN-EP-2016-241",
      SLACcitation   = "%%CITATION = ARXIV:1611.09661;%%"
}

@article{Aad:2012xs,
      author         = "{ATLAS Collaboration}",            
      title          = "{Performance of the ATLAS Trigger System in 2010}",
      collaboration  = "ATLAS",
      journal        = "Eur. Phys. J. C",
      volume         = "72",
      year           = "2012",
      pages          = "1849",
      doi            = "10.1140/epjc/s10052-011-1849-1",
      eprint         = "1110.1530",
      archivePrefix  = "arXiv",
      primaryClass   = "hep-ex",
      reportNumber   = "CERN-PH-EP-2011-078",
      SLACcitation   = "%%CITATION = ARXIV:1110.1530;%%"
}

@article{Santonico:1981sc,
      author         = "Santonico, R. and Cardarelli, R.",
      title          = "{DEVELOPMENT OF RESISTIVE PLATE COUNTERS}",
      journal        = "Nucl. Instrum. Meth.",
      volume         = "187",
      year           = "1981",
      pages          = "377-380",
      doi            = "10.1016/0029-554X(81)90363-3",
      SLACcitation   = "%%CITATION = NUIMA,187,377;%%"
}

@article{Santonico:1988qi,
      author         = "Santonico, R. and Cardarelli, R. and Di Biagio, A. and
                        Lucci, A.",
      title          = "{PROGRESS IN RESISTIVE PLATE COUNTERS}",
      journal        = "Nucl. Instrum. Meth. A",
      volume         = "263",
      year           = "1988",
      pages          = "20-25",
      doi            = "10.1016/0168-9002(88)91011-X",
      SLACcitation   = "%%CITATION = NUIMA,A263,20;%%"
}

@article{Corradi:2016moz,
      author         = "Corradi, M.",
      title          = "{Performance of ATLAS RPC Level-1 muon trigger during the
                        2015 data taking}",
      booktitle      = "{Proceedings, 13th Workshop on Resistive Plate Chambers
                        and Related Detectors (RPC2016): Ghent, Belgium, February
                        22-26, 2016}",
      collaboration  = "ATLAS Muon",
      journal        = "JINST",
      volume         = "11",
      year           = "2016",
      number         = "09",
      pages          = "C09003",
      doi            = "10.1088/1748-0221/11/09/C09003",
      reportNumber   = "ATL-MUON-PROC-2016-003",
      SLACcitation   = "%%CITATION = JINST,11,C09003;%%"
}

@book{abbrescia2018resistive,
  author={Abbrescia, Marcello and Peskov, Vladimir and Fonte, Paulo},
  title={Resistive gaseous detectors: designs, performance, and perspectives},
  year={2018},
  publisher={John Wiley \& Sons}
}

@article{ATLASTDAQ:2016pov,
    author = "{ATLAS TDAQ Collaboration}",
    collaboration = "ATLAS TDAQ",
    title = "{The ATLAS Data Acquisition and High Level Trigger system}",
    doi = "10.1088/1748-0221/11/06/P06008",
    journal = "JINST",
    volume = "11",
    number = "06",
    pages = "P06008",
    year = "2016"
}

@article{Aad:2020uyd,
    author = "{ATLAS Collaboration}",
    collaboration = "ATLAS",
    title = "{Performance of the ATLAS muon triggers in Run 2}",
    eprint = "2004.13447",
    archivePrefix = "arXiv",
    primaryClass = "hep-ex",
    reportNumber = "CERN-EP-2020-031",
    doi = "10.1088/1748-0221/15/09/p09015",
    journal = "JINST",
    volume = "15",
    number = "09",
    pages = "P09015",
    year = "2020"
}

@article{Aielli:2002th,
    author = "Aielli, G. and Camarri, P. and Cardarelli, R. and de Asmundis, R. and Di Ciaccio, A. and Di Stante, L. and Liberti, B. and Paoloni, A. and Pastori, E. and Santonico, R.",
   %editor = "Jeitler, M. and Krammer, M. and Neuhofer, G. and Regler, M. and Wedenig, R.",
    title = "{RPC ageing studies}",
    doi = "10.1016/S0168-9002(01)01770-3",
    journal = "Nucl. Instrum. Meth. A",
    volume = "478",
    pages = "271--276",
    year = "2002"
}

@article{Aielli:2000sa,
    author = "Aielli, G. and Camarri, P. and Cardarelli, R. and Di Ciaccio, A. and Di Stante, L. and Liberti, B. and Paoloni, A. and Santonico, R.",
   %editor = "Grancagnolo, G. and Iaselli, G. and Ratti, S.L. and Santonico, R.",
    title = "{An RPC gamma irradiation test}",
    doi = "10.1016/S0168-9002(00)00967-0",
    journal = "Nucl. Instrum. Meth. A",
    volume = "456",
    pages = "82--86",
    year = "2000"
}

@article{Aad:2016jkr,
    author = "{ATLAS Collaboration}",
    collaboration = "ATLAS",
    title = "{Muon reconstruction performance of the ATLAS detector in proton--proton collision data at $\sqrt{s}$ =13 TeV}",
    eprint = "1603.05598",
    archivePrefix = "arXiv",
    primaryClass = "hep-ex",
    reportNumber = "CERN-EP-2016-033",
    doi = "10.1140/epjc/s10052-016-4120-y",
    journal = "Eur. Phys. J. C",
    volume = "76",
    number = "5",
    pages = "292",
    year = "2016"
}

@article{Aielli:2013nea,
    author = "Aielli, G. and Bindi, Marcello and Polini, Alessandro",
    title = "{Performance, operation and detector studies with the ATLAS Resistive Plate Chambers}",
    doi = "10.1088/1748-0221/8/02/P02020",
    journal = "JINST",
    volume = "8",
    pages = "P02020",
    year = "2013"
}

@article{Aad:2020wji,
    author = "{ATLAS Collaboration}",
    collaboration = "ATLAS",
    title = "{Operation of the ATLAS trigger system in Run 2}",
    eprint = "2007.12539",
    archivePrefix = "arXiv",
    primaryClass = "physics.ins-det",
    reportNumber = "CERN-EP-2020-109",
    doi = "10.1088/1748-0221/15/10/P10004",
    journal = "JINST",
    volume = "15",
    number = "10",
    pages = "P10004",
    year = "2020"
}

@Booklet{ATL-SOFT-PUB-2020-001,
    author         = "{ATLAS Collaboration}",
    title          = "{ATLAS Computing Acknowledgements}",
    howpublished   = "{ATL-SOFT-PUB-2020-001}",
    url            = "https://cds.cern.ch/record/2717821",
}

@article{Guida:2016qjq,
    author = "Guida, R. and Capeans, M. and Mandelli, B.",
    title = "{Characterization of RPC operation with new environmental friendly mixtures for LHC application and beyond}",
    doi = "10.1088/1748-0221/11/07/C07016",
    journal = "JINST",
    volume = "11",
    number = "07",
    pages = "C07016",
    year = "2016"
}

@Booklet{Ciapetti:2007zz,
    author         = "Ciapetti, G. and Pastore, F. and Petrolo, E. and Rosati, S. and Spila, F. and Vari, R. and Veneziano, S. and Zanello, Lucia",
    title          = "{The ATLAS barrel level-1 muon trigger calibration}",
    howpublished   = "{ATL-DAQ-CONF-2007-020}",
    url            = "http://cdsweb.cern.ch/record/1035895",
%    note           = {Available at \url{http://cdsweb.cern.ch/record/2285580}},    
    year           = "2007",
}

@article{Aaboud:2018tkc,
    author = "{ATLAS Collaboration}",
    collaboration = "ATLAS",
    title = "{Performance of missing transverse momentum reconstruction with the ATLAS detector using proton-proton collisions at $\sqrt{s}$ = 13 TeV}",
    eprint = "1802.08168",
    archivePrefix = "arXiv",
    primaryClass = "hep-ex",
    reportNumber = "CERN-EP-2017-274",
    doi = "10.1140/epjc/s10052-018-6288-9",
    journal = "Eur. Phys. J. C",
    volume = "78",
    number = "11",
    pages = "903",
    year = "2018"
}

@article{Amram:2011ue,
    author = "Amram, N. and others",
    title = "{Streamlined calibrations of the ATLAS precision muon chambers for initial LHC running}",
    eprint = "1103.0797",
    archivePrefix = "arXiv",
    primaryClass = "physics.ins-det",
    doi = "10.1016/j.nima.2011.12.086",
    journal = "Nucl. Instrum. Meth. A",
    volume = "671",
    pages = "40--50",
    year = "2012"
}

@article{Aielli:2003bx,
    author = "Aielli, G. and Camarri, P. and Cardarelli, R. and Di Ciaccio, A. and Di Stante, L. and Liberti, B. and Paoloni, A. and Pastori, E. and Santonico, R.",
   %editor = "Fonte, P. and Fraga, M. and Ratti, S.P. and Santonico, R.",
    title = "{Test of ATLAS RPCs Front-End electronics}",
    doi = "10.1016/S0168-9002(03)01349-4",
    journal = "Nucl. Instrum. Meth. A",
    volume = "508",
    pages = "189--193",
    year = "2003"
}

@article{Rigoletti:2020exj,
    author = "Rigoletti, G. and Mandelli, B. and Guida, R.",
   %editor = "Nessi, Marzio",
    title = "{Performance studies of RPC detectors with new environmentally friendly gas mixtures in presence of LHC-like radiation background}",
    doi = "10.1088/1748-0221/15/05/C05004",
    journal = "JINST",
    volume = "15",
    number = "05",
    pages = "C05004",
    year = "2020"
}

@article{Guida:2020yqs,
    author = "Guida, R. and Mandelli, B.",
   %editor = "Krammer, Manfred and Bergauer, Thomas and Dragicevic, Marko and Friedl, Markus and Jeitler, Manfred and Schieck, Jochen and Schwanda, Christoph",
    title = "{R\&D strategies for optimizing greenhouse gases usage in the LHC particle detection systems}",
    doi = "10.1016/j.nima.2019.04.089",
    journal = "Nucl. Instrum. Meth. A",
    volume = "958",
    pages = "162135",
    year = "2020"
}

@article{Acitelli:1995ft,
    author = "Acitelli, L. and others",
%    editor = "Bertolucci, E. and Cervelli, F. and Scribano, A.",
    title = "{Study of the efficiency and time resolution of an RPC irradiated with photons and neutrons}",
    doi = "10.1016/0168-9002(94)01219-9",
    journal = "Nucl. Instrum. Meth. A",
    volume = "360",
    pages = "42--47",
    year = "1995"
}

@article{Aad:2020gmm,
    author = "ATLAS Collaboration",
    collaboration = "ATLAS",
    title = "{Muon reconstruction and identification efficiency in ATLAS using the full Run 2 $pp$ collision data set at $\sqrt{s}=13$ TeV}",
    journal = "Eur. Phys. J. C",    
    volume = "81",
    pages = "578",
    doi = "10.1140/epjc/s10052-021-09233-2",
    eprint = "2012.00578",
    archivePrefix = "arXiv",
    primaryClass = "hep-ex",
    reportNumber = "CERN Preprint ID: CERN-EP-2020-199",
    year = "2021"
}

@article{Aad:2012tfa,
    author = "{ATLAS Collaboration}",
    collaboration = "ATLAS",
    title = "{Observation of a new particle in the search for the Standard Model Higgs boson with the ATLAS detector at the LHC}",
    eprint = "1207.7214",
    archivePrefix = "arXiv",
    primaryClass = "hep-ex",
    reportNumber = "CERN-PH-EP-2012-218",
    doi = "10.1016/j.physletb.2012.08.020",
    journal = "Phys. Lett. B",
    volume = "716",
    pages = "1--29",
    year = "2012"
}

@article{ATLAS:2020wny,
    author = "{ATLAS Collaboration}",
    collaboration = "ATLAS",
    title = "{Measurements of the Higgs boson inclusive and differential fiducial cross sections in the 4$\ell$ decay channel at $\sqrt{s}$ = 13 TeV}",
    eprint = "2004.03969",
    archivePrefix = "arXiv",
    primaryClass = "hep-ex",
    reportNumber = "CERN-EP-2020-035",
    doi = "10.1140/epjc/s10052-020-8223-0",
    journal = "Eur. Phys. J. C",
    volume = "80",
    number = "10",
    pages = "942",
    year = "2020"
}

@article{Aad:2020xfq,
    author = "{ATLAS Collaboration}",
    collaboration = "ATLAS",
    title = "{A search for the dimuon decay of the Standard Model Higgs boson with the ATLAS detector}",
    eprint = "2007.07830",
    archivePrefix = "arXiv",
    primaryClass = "hep-ex",
    reportNumber = "CERN-EP-2020-117",
    doi = "10.1016/j.physletb.2020.135980",
    journal = "Phys. Lett. B",
    volume = "812",
    pages = "135980",
    year = "2021"
}

@article{Aad:2020jvx,
    author = "{ATLAS Collaboration}",
    collaboration = "CMS, ATLAS",
    title = "{Combination of the W boson polarization measurements in top quark decays using ATLAS and CMS data at $\sqrt{s} =$ 8 TeV}",
    eprint = "2005.03799",
    archivePrefix = "arXiv",
    primaryClass = "hep-ex",
    reportNumber = "CMS-TOP-19-004, ATLAS-TOPQ-2018-02, CERN-EP-2020-012",
    doi = "10.1007/JHEP08(2020)051",
    journal = "JHEP",
    volume = "08",
    number = "08",
    pages = "051",
    year = "2020"
}

@article{Aad:2019wmn,
    author = "{ATLAS Collaboration}",
    collaboration = "ATLAS",
    title = "{Measurement of the transverse momentum distribution of Drell\textendash{}Yan lepton pairs in proton\textendash{}proton collisions at $\sqrt{s}=13$ TeV with the ATLAS detector}",
    eprint = "1912.02844",
    archivePrefix = "arXiv",
    primaryClass = "hep-ex",
    reportNumber = "CERN-EP-2019-223",
    doi = "10.1140/epjc/s10052-020-8001-z",
    journal = "Eur. Phys. J. C",
    volume = "80",
    number = "7",
    pages = "616",
    year = "2020"
}

@article{Aad:2019vnb,
    author = "{ATLAS Collaboration}",
    collaboration = "ATLAS",
    title = "{Search for electroweak production of charginos and sleptons decaying into final states with two leptons and missing transverse momentum in $\sqrt{s}=13$ TeV $pp$ collisions using the ATLAS detector}",
    eprint = "1908.08215",
    archivePrefix = "arXiv",
    primaryClass = "hep-ex",
    reportNumber = "CERN-EP-2019-106",
    doi = "10.1140/epjc/s10052-019-7594-6",
    journal = "Eur. Phys. J. C",
    volume = "80",
    number = "2",
    pages = "123",
    year = "2020"
}

@article{Aad:2019fac,
    author = "{ATLAS Collaboration}",
    collaboration = "ATLAS",
    title = "{Search for high-mass dilepton resonances using 139 fb$^{-1}$ of $pp$ collision data collected at $\sqrt{s}=$13 TeV with the ATLAS detector}",
    eprint = "1903.06248",
    archivePrefix = "arXiv",
    primaryClass = "hep-ex",
    reportNumber = "CERN-EP-2019-030",
    doi = "10.1016/j.physletb.2019.07.016",
    journal = "Phys. Lett. B",
    volume = "796",
    pages = "68--87",
    year = "2019"
}

@article{Aaboud:2018spl,
    author = "{ATLAS Collaboration}",
    collaboration = "ATLAS",
    title = "{Search for heavy Majorana or Dirac neutrinos and right-handed $W$ gauge bosons in final states with two charged leptons and two jets at $ \sqrt{s}=13 $ TeV with the ATLAS detector}",
    eprint = "1809.11105",
    archivePrefix = "arXiv",
    primaryClass = "hep-ex",
    reportNumber = "CERN-EP-2018-199",
    doi = "10.1007/JHEP01(2019)016",
    journal = "JHEP",
    volume = "01",
    pages = "016",
    year = "2019"
}

@article{Aaboud:2019nmv,
    author = "{ATLAS Collaboration}",
    collaboration = "ATLAS",
    title = "{Observation of Electroweak Production of a Same-Sign $W$ Boson Pair in Association with Two Jets in $pp$ Collisions at $\sqrt{s}=13$ TeV with the ATLAS Detector}",
    eprint = "1906.03203",
    archivePrefix = "arXiv",
    primaryClass = "hep-ex",
    reportNumber = "CERN-EP-2019-008",
    doi = "10.1103/PhysRevLett.123.161801",
    journal = "Phys. Rev. Lett.",
    volume = "123",
    number = "16",
    pages = "161801",
    year = "2019"
}

@article{Aaboud:2017svj,
    author = "{ATLAS Collaboration}",
    collaboration = "ATLAS",
    title = "{Measurement of the $W$-boson mass in pp collisions at $\sqrt{s}=7$ TeV with the ATLAS detector}",
    eprint = "1701.07240",
    archivePrefix = "arXiv",
    primaryClass = "hep-ex",
    reportNumber = "CERN-EP-2016-305",
    doi = "10.1140/epjc/s10052-017-5475-4",
    journal = "Eur. Phys. J. C",
    volume = "78",
    number = "2",
    pages = "110",
    year = "2018",
    note = "[Erratum: Eur.Phys.J.C 78, 898 (2018)]"
}

@article{Aaboud:2017jvq,
    author = "{ATLAS Collaboration}",
    collaboration = "ATLAS",
    title = "{Evidence for the associated production of the Higgs boson and a top quark pair with the ATLAS detector}",
    eprint = "1712.08891",
    archivePrefix = "arXiv",
    primaryClass = "hep-ex",
    reportNumber = "CERN-EP-2017-281",
    doi = "10.1103/PhysRevD.97.072003",
    journal = "Phys. Rev. D",
    volume = "97",
    number = "7",
    pages = "072003",
    year = "2018"
}

@article{Aad:2020tmz,
    author = "{ATLAS Collaboration}",
    collaboration = "ATLAS",
    title = "{Measurement of the $t\bar{t}$ production cross-section in the lepton+jets channel at $\sqrt{s}=13$ TeV with the ATLAS experiment}",
    eprint = "2006.13076",
    archivePrefix = "arXiv",
    primaryClass = "hep-ex",
    reportNumber = "CERN-EP-2020-096",
    doi = "10.1016/j.physletb.2020.135797",
    journal = "Phys. Lett. B",
    volume = "810",
    pages = "135797",
    year = "2020"
}

@Article{SOFT-2010-01,
    author         = "{ATLAS Collaboration}",
    title          = "{The ATLAS Simulation Infrastructure}",
    journal        = "Eur. Phys. J. C",
    volume         = "70",
    year           = "2010",
    pages          = "823",
    doi            = "10.1140/epjc/s10052-010-1429-9",
    eprint         = "1005.4568",
    archivePrefix  = "arXiv",
    primaryClass   = "physics.ins-det",
}

@Report{ATLAS-TDR-10,
    author         = "{ATLAS Collaboration}",
    title          = "{ATLAS Muon Spectrometer: Technical Design Report}",
    type           = "ATLAS-TDR-10; CERN-LHCC-97-022",
    institution    = "CERN",
    year           = "1997",
    url            = "https://cds.cern.ch/record/331068",
}
 
\clearpage
 
\begin{flushleft}
\hypersetup{urlcolor=black}
{\Large The ATLAS Collaboration}

\bigskip

\AtlasOrcid[0000-0002-6665-4934]{G.~Aad}$^\textrm{\scriptsize 101}$,    
\AtlasOrcid[0000-0002-5888-2734]{B.~Abbott}$^\textrm{\scriptsize 128}$,    
\AtlasOrcid[0000-0002-7248-3203]{D.C.~Abbott}$^\textrm{\scriptsize 102}$,    
\AtlasOrcid[0000-0002-2788-3822]{A.~Abed~Abud}$^\textrm{\scriptsize 36}$,    
\AtlasOrcid[0000-0002-1002-1652]{K.~Abeling}$^\textrm{\scriptsize 53}$,    
\AtlasOrcid[0000-0002-2987-4006]{D.K.~Abhayasinghe}$^\textrm{\scriptsize 93}$,    
\AtlasOrcid[0000-0002-8496-9294]{S.H.~Abidi}$^\textrm{\scriptsize 29}$,    
\AtlasOrcid[0000-0002-8279-9324]{O.S.~AbouZeid}$^\textrm{\scriptsize 40}$,    
\AtlasOrcid{N.L.~Abraham}$^\textrm{\scriptsize 156}$,    
\AtlasOrcid[0000-0001-5329-6640]{H.~Abramowicz}$^\textrm{\scriptsize 161}$
% ...
\bigskip
\\

% ...
$^{29}$Physics Department, Brookhaven National Laboratory, Upton NY; United States of America.\\
% ...
$^{36}$CERN, Geneva; Switzerland.\\
% ...
$^{40}$Niels Bohr Institute, University of Copenhagen, Copenhagen; Denmark.\\
% ...
$^{53}$II. Physikalisches Institut, Georg-August-Universit\"{a}t G\"ottingen, G\"ottingen; Germany.\\
% ...
$^{93}$Department of Physics, Royal Holloway University of London, Egham; United Kingdom.\\
% ...
$^{101}$CPPM, Aix-Marseille Universit\'e, CNRS/IN2P3, Marseille; France.\\
$^{102}$Department of Physics, University of Massachusetts, Amherst MA; United States of America.\\
% ...
$^{128}$Homer L. Dodge Department of Physics and Astronomy, University of Oklahoma, Norman OK; United States of America.\\
% ...
$^{156}$Department of Physics and Astronomy, University of Sussex, Brighton; United Kingdom.\\
% ...
$^{161}$Raymond and Beverly Sackler School of Physics and Astronomy, Tel Aviv University, Tel Aviv; Israel.\\
% ...

\end{flushleft}


\end{document}